\documentclass[aps,prd,groupedaddress,amsmath,amssymb,amsfonts,nofootinbib,preprint]{revtex4-1}

\usepackage{amssymb,amsfonts}
\usepackage{mathrsfs}
\usepackage{tikz}
\usetikzlibrary{external}
\usepackage{amsthm}

\newtheorem{thm}{Theorem}
\newtheorem{cor}{Corollary}
\newtheorem{rem}{Remark}

\newtheorem{prop}{Proposition}
\usepackage[colorlinks, allcolors=blue!70!black, linktocpage]{hyperref}
\usepackage{accents}
\usepackage{graphicx}
\usepackage[utf8x]{inputenc}
\usepackage{float}
\usepackage{amsmath}
\DeclareMathAlphabet{\mathpzc}{OT1}{pzc}{m}{it}
\usetikzlibrary{patterns}
\usepackage{titlesec}
\usepackage{cases}
\usepackage{mathtools}
\usepackage{color}
\usepackage{cleveref}
\crefname{lem}{lemma}{lemmas}
\crefname{thm}{theorem}{theorems}
\crefname{cor}{corollary}{corollaries}
\crefname{rem}{remark}{remarks}
\crefname{prop}{proposition}{propositions}

\begin{document}

\title{The Asymptotic Behavior of Massless Fields and the Memory Effect}

\author{Gautam Satishchandran}
\email{gautamsatish@uchicago.edu}
\author{Robert M. Wald}
\email{rmwa@uchicago.edu}
\affiliation{Enrico Fermi Institute and Department of Physics \\
  The University of Chicago \\
  5640 South Ellis Avenue, Chicago, Illinois 60637, USA.}

\begin{abstract}
We investigate the behavior of massless scalar, electromagnetic, and linearized gravitational perturbations near null infinity in $d \geq 4$ dimensional Minkowski spacetime (of both even and odd dimension) under the assumption that these fields admit a suitable expansion in $1/r$. For even $d$ with $d > 4$, our $1/r$ expansion ansatz is equivalent to smoothness at ${\mathscr I}^+$, whereas for $d=4$ it is slightly weaker, so all solutions that are smooth at ${\mathscr I}^+$ are encompassed by our analysis. We also analyze the solutions to the full nonlinear Einstein equation in $d \geq 4$ dimensions near null infinity, assuming a similar $1/r$ expansion. We show that for $d > 4$ the Lorenz gauge condition can be imposed for electromagnetic and gravitational perturbations in a manner compatible with our assumed $1/r$ expansion. However, for $d=4$ the Lorenz gauge can be imposed if and only if there is no flux of charge-current (in the electromagnetic case) or stress-energy (in the linearized gravitational case) to null infinity. Similarly, in the nonlinear gravitational case, the harmonic gauge condition can be imposed for $d > 4$ but cannot be imposed for $d=4$ if there either is a flux of stress-energy at null infinity or if the Bondi news is nonvanishing. We explicitly obtain the recursion relations on the coefficients of the $1/r$ expansion implied by the wave equation as well as the ``constraints'' in the electromagnetic and gravitational cases arising from the Lorenz/harmonic gauge condition. We also characterize the ``free data'' needed to determine a solution. We then consider the memory effect in fully nonlinear general relativity, i.e., the permanent displacement of test particles near null infinity following a burst of gravitational radiation. We show that in even dimensions, the memory effect first arises at Coulombic order---i.e., order $1/r^{d-3}$---and can naturally be decomposed into ``null memory'' and ``ordinary memory.'' Null memory is associated with an energy flux to null infinity. We show that ordinary memory is associated with the metric failing to be stationary at one order faster fall-off than Coulombic in the past and/or future, as will typically be the case if matter (on timelike inertial trajectories) comes in or goes out to infinity. In odd dimensions, we show that the total memory effect at Coulombic order and slower fall-off always vanishes. It is easily seen that null memory is always of ``scalar type'' with regard to its behavior on spheres, but the ordinary memory can be of any (i.e., scalar, vector, or tensor) type. In $4$-spacetime dimensions, we give an explicit example in linearized gravity of an expanding shell with vector stresses which gives rise to a nontrivial vector (i.e., magnetic parity) ordinary memory effect at order $1/r$. We show that scalar memory is described by a diffeomorphism, which is an asymptotic symmetry (a supertranslation) in $d=4$ and a gauge transformation for $d > 4$. Vector and tensor memory cannot be described by diffeomorphisms. In $d=4$ dimensions, we show that there is a close relationship between memory and the charge and flux expressions associated with supertranslations. Similar formulas are given in higher dimensions. We analyze the behavior of solutions that are stationary at Coulombic order and show how these suggest ``antipodal matching'' between future and past null infinity, which gives rise to conservation laws. The relationship between memory and infrared divergences of the ``out'' state in quantum gravity is analyzed, and the nature of the ``soft theorems'' is explained.

\end{abstract}
\maketitle
\tableofcontents
\section{Introduction}
In the early 1960's, Bondi and collaborators \cite{BMS1,BMS2,BMS3} performed a general analysis of the asymptotic behavior of the metric near ``null infinity'' ($r \to \infty$ at fixed retarded time $u$) for asymptotically flat spacetimes. They assumed an expansion of the metric in powers of $1/r$ and obtained a recursive algorithm for solving the Einstein equations near null infinity. Several years later, Penrose \cite{Penrose} gave an elegant, geometric reformulation of the Bondi ansatz via conformal compactification. A similar analysis of higher even-dimensional, asymptotically flat spacetimes can be given using conformal compactification \cite{HI}. However, such a conformal compactification is not possible for odd dimensional  spacetimes with gravitational radiation \cite{HW}. 

In \Cref{GenBehavFields} of 
this paper, we will analyze the asymptotic behavior of massless scalar, electromagnetic, and linearized gravitational fields near null infinity in Minkowski spacetimes with $d \geq 4$. We will then analyze asymptotically flat, nonlinear general relativity near null infinity. Since we wish to treat odd dimensions as well as even dimensions, we will not use conformal compactification but, instead, will assume an expansion in powers of $1/r$ as an ansatz. For $d$ even with $d > 4$, our ansatz is precisely equivalent to smoothness\footnote{It should be noted that
our analysis will be primarily concerned with behavior of fields at $1/r^{d-3}$ and slower fall off, so for our main results, ``smoothness'' can be replaced by differentiability to the corresponding order.} at $\mathscr{I}^{+}$ in the conformally compactified spacetime, whereas we will see in \Cref{smoothscri} that for $d=4$ it is slightly weaker, i.e., we allow a small class of additional solutions that would not be allowed by smoothness at $\mathscr{I}^{+}$.
Our fields will be allowed to have arbitrary interior sources, i.e., only the field equations near null infinity will be used. Near null infinity
the fall-off of the sources is required to be rapid enough to ensure that there is a finite flux through spheres near null infinity.  

In \Cref{Memeffect} of this paper, we will give a thorough analysis of the memory effect in nonlinear general relativity in all dimensions $d \geq 4$.
An important aim of our analysis is to extend and clarify the work of Strominger and collaborators \cite{Strom1:2014,Strom2:2014,Strom1:2015,Strom2:2015,Strom:2017,Strom:2018}.

We begin our analysis in \Cref{scalaran} by considering a massless scalar field, $\phi$, in $d$-dimensional Minkowski spacetime. We show that the wave equation gives a recursion relation that relates different coefficients in an expansion of the field in powers of $1/r$. This recursion relation motivates an expansion in integer steps, with the slowest fall-off being $1/r^{d/2 -1}$ (``radiative order''). In odd dimensions, integer powers starting at $1/r^{d-3}$ (``Coulombic order'') must also be allowed. The ``free data'' needed to specify a solution is characterized in \Cref{scalarsol}.

We then consider an electromagnetic field, $A_a$, in \Cref{subsecMaxeqa}. It is very convenient to put $A_a$ in Lorenz gauge,
$\partial^a A_a = 0$, since then many of the results for the scalar field can be directly taken over.
In order to put the electromagnetic field in Lorenz gauge, we need to solve the scalar wave equation with a source. We show that when $d > 4$, this can be done in a manner compatible with our $1/r$ expansion ansatz. However, when $d=4$ we cannot do this if there is a nonvanishing flux of charge to null infinity. In Lorenz gauge, each Cartesian component of $A_a$ satisfies the same recursion relations as the scalar wave equation, but there also are additional conditions (``constraints'') arising from the Lorenz gauge condition itself. It is convenient to write the recursion relations and constraints in terms of the components $A_u, A_r, A_A$ in coordinates $(u,r,x^A)$ where $u$ is the retarded time and $x^A$ denotes coordinates on the $(d-2)$-sphere. We do this explicitly in \Cref{subsecMaxeqa}. The ``free data'' is then characterized.

Gravitational perturbations, $h_{ab}$, are considered in \Cref{Einsteinseqs}. In order to put $h_{ab}$ in Lorenz gauge, $\partial^a \bar{h}_{ab} = 0$ (with $\bar{h}_{ab} \equiv h_{ab} - 1/2 \eta_{ab} h$ and $h\equiv \eta^{ab}h_{ab}$), we need to solve the vector wave equation with a source. Again, we find that when $d > 4$, this can be done in a manner compatible with our $1/r$ expansion ansatz. However, when $d=4$ we cannot do this if there is a nonvanishing flux of matter stress-energy to null infinity. We give the recursion relations and constraints explicitly in terms of the components $h_{uu}, h_{ur}, h_{rr}, h_{uA}, h_{rA}, h_{AB}$ and identify the ``free data.''

It might be thought that the full, nonlinear Einstein equation would be much more difficult to analyze. However, as we shall see in \Cref{nleins}, the nonlinear terms first enter Einstein's equation at order $1/r^{d-2}$ and they first affect the behavior of the metric at Coulombic order $1/r^{d-3}$. Similarly, the nonlinear terms in the harmonic gauge condition first affect the metric at Coulombic order. Thus, under our ansatz concerning the expansion of the metric in powers of $1/r$, the analysis of the nonlinear Einstein equation coincides with the linearized analysis until Coulombic order, and the differences at Coulombic order can be taken into account in a relatively straightforward manner.

In \Cref{Memeffect}, we turn our attention to the memory effect, i.e., the permanent relative displacement of an arrangement of test particles near null infinity that are initially at rest. We assume that the metric initially is stationary to Coulombic order, goes through a non-stationary epoch, and again becomes stationary to Coulombic order. The precise stationarity assumptions and the motivation for them are spelled out in \Cref{statcond1}. We obtain general properties of the memory tensor in \Cref{coulmem1}. In \Cref{coulmem2}, we calculate the memory tensor for all $d \geq 4$. We show that the memory tensor vanishes at all fall-off slower than Coulombic, i.e., it vanishes at order $1/r^n$ for all $n < d-3$. In even dimensions, the memory tensor at Coulombic order can be nonvanishing \cite{Strom:2018,Mao} and  we also show that it naturally decomposes into ``null memory'' and ``ordinary memory,'' in a manner similar to the known decomposition in $4$-dimensions \cite{BG}. ``Null memory'' is associated with a flux of energy to null infinity, whereas we show that ``ordinary memory'' is associated with the metric being non-stationary at one order faster fall-off than Coulombic, as will generically occur if there is a flux of matter stress-energy moving inertially in from infinity or out to infinity at less than the speed of light. In odd dimensions, we show that the total memory effect vanishes near null infinity at Coulombic order. 

As discussed in \Cref{nonscal}, in all dimensions, the memory effect can be decomposed into scalar, vector and tensor parts on the $(d-2)$-sphere. Null memory is always of scalar type, but ordinary memory can be of any type. We give an explicit example in linearized gravity in $d=4$ dimensions involving a shell of matter with vector stresses that gives rise to vector (i.e., ``magnetic parity'') ordinary memory at order $1/r$. In \Cref{memdiff}, we show that scalar memory can be characterized by a diffeomorphism. This diffeomorphism is an asymptotic symmetry in $d=4$ dimensions, but it is gauge for $d>4$. Vector 
and tensor memory cannot be described by a diffeomorphism. 

We then consider the relationship of memory to charges and conservation laws in \Cref{chcons}. In $d=4$ dimensions, we show in \Cref{chmem} how the charges and fluxes associated with supertranslations can be used to derive the formula for scalar memory. Although memory cannot be associated with an asymptotic symmetry when $d > 4$, similar expressions are obtained from our general formulas for memory in \Cref{coulmem2}. In \Cref{claws} we provide some arguments in favor of ``antipodal matching'' of solutions between future and past null infinity, and show that under the assumption of antipodal matching, we obtain expressions that can be interpreted as representing conservation laws relating charges and fluxes at past and future null infinity.

Finally, in \Cref{softthm} we show that in $d=4$ dimensions, the presence of a nontrivial memory effect at future null infinity is intimately related to infrared divergences in the ``out'' state in quantum field theory. The factorization of the ``out'' state vector into a product of ``hard'' and ``soft'' parts is shown for the case of quantum linearized gravity with a classical source, and is argued to hold generally.

We work in geometrized units $(G=c=1)$ and will use the notation and sign conventions of \cite{Wald-book}. In particular, our metric signature is ``mostly positive'' and our sign convention for curvature is such that the scalar curvature of a round sphere is positive. Latin indices from the early alphabet $(a,b,c,\dots)$ denote abstract spacetime indices. Greek indices $(\mu,\nu,\dots)$ denote spacetime components of tensors. Throughout the paper, Latin and Greek indices are raised and lowered with respect to the ``background'' Minkowski metric $\eta_{ab}$. Capital latin indices $(A,B,C,\dots)$ will be used to denote tensors on the $(d-2)-$sphere. We will also use capital latin indices to denote coordinates, $x^A$, on the sphere and components in this coordinate basis. (We do not feel that the potential confusion resulting from using the same notation for a tensor on a sphere and its components in a coordinate basis is sufficient to justify introducing another alphabet into our notation.)
When we expand a scalar field $\phi$ in powers of $1/r$, $\phi^{(n)}$ will denote the coefficient of $1/r^n$. When we expand a tensor field $t_{a_1 \dots a_k}$ in powers of $1/r$, the quantity $t^{(n)}_{a_1 \dots a_k}$ will denote the coefficient of $1/r^n$ in a {\em normalized} basis. In particular, for a co-vector field, $t_a$, the quantity $t^{(n)}_A$ is such that its action on the normalized basis element $\frac{1}{r} \frac{\partial}{\partial x^A}$ falls as $1/r^n$. This differs from a much more common convention \cite{Strom:2018,BG,Eanna} where $t^{(n)}_A$ would be such that its action on $\frac{\partial}{\partial x^A}$ falls as $1/r^n$. Our conventions thereby avoid a spurious mixing of orders, and the orders we assign to components do not depend on whether we are using Cartesian or spherical coordinates.


\section{The General Behavior of Fields near Near Null Infinity}
\label{GenBehavFields}

Consider $d$-dimensional Minkowski spacetime with $d \geq 4$. In terms of global inertial coordinates $(t, x^1, \dots, x^{d-1})$, the metric takes the form
\begin{equation}
\eta = -dt^{2} + \sum_{\mu = 1}^{d-1} (dx^\mu)^2.
\end{equation}
Let $r=(\sum (x^{\mu})^{2})^{1/2}$, let $u\equiv t-r$, and let $x^A$ be arbitrary coordinates on the spheres of constant $r$ and $u$. 
In the coordinates $(u,r,x^A)$, the Minkowski metric $\eta$ takes the form
\begin{equation}
\eta = -du^{2}-2dudr+r^2 q_{AB}dx^{A}dx^{B}
\label{metric}
\end{equation}
where $q_{AB}$ is the metric on the round unit $(d-2)$-sphere. 
Let 
\begin{equation}
K^{a}= (\partial/\partial r)^a
\label{K}
\end{equation}
\begin{equation}
l^a=(\partial/\partial u)^a-\frac{1}{2}(\partial/\partial r)^a
\end{equation}
so that $K^a$ and $l^{a}$ are the future-directed, radially outgoing and ingoing null vector fields, which satisfy
\begin{equation}
K^{a}l_{a}=-1\, .
\end{equation}
Let $q_{ab}$ denote the spacetime tensor field whose pullback to spheres of constant $u$ and $r$ is $q_{AB}$
and $K^{a}q_{ab}=0=l^{a}q_{ab}$. The metric can be written as 
\begin{equation}
\eta_{ab}=-2K_{(a}l_{b)}+r^2 q_{ab}.
\end{equation}

We will be concerned in the following with the behavior of fields near ``null infinity'' in this spacetime, i.e., the limit as $r \to \infty$ at fixed $(u, x^A)$.


\subsection{Ansatz for the Massless Scalar Field}
\label{scalaran}

Consider a massless Klein-Gordon field $\phi$ satisfying
\begin{equation}
\label{scalar1}
{\Box} \phi = 0
\end{equation}
where ${\Box} \equiv \eta^{ab}{\partial}_{a}{\partial}_{b}$. (In the next subsection, we will allow a source term $S$, i.e., we will consider ${\Box} \phi = S$.) We assume, as a {\em preliminary ansatz}, that near null infinity, $\phi$ can be expanded as a series in $1/r$ as follows:
\begin{equation}
\label{preansatz}
\phi \sim \sum_{j=0}^\infty \frac{1}{r^{\alpha +j}} \phi^{(j)}(u, x^A)
\end{equation}
where $\alpha \in (0,1]$. Here, the meaning of the ``$\sim$'' in \cref{preansatz} is as follows: We do {\em not} require that the series on the right side of this equation converges (even for large $r$) but require that for any $N \geq 0$ we have
\begin{equation}
\label{series}
\phi - \sum_{j=0}^N \frac{1}{r^{\alpha +j}} \phi^{(j)}(u, x^A) = O(1/r^{\alpha + N +1})
\end{equation}
as $r \to \infty$, i.e., we require this series to be asymptotic.
We further require that all partial derivatives of the left side of \cref{series} with respect to $u$ and $x^A$ are also $O(1/r^{\alpha + N +1})$, whereas $k$ partial derivatives with respect to $r$ are $O(1/r^{\alpha + N +1 + k})$. For convenience, we have taken the upper limit in the sum in \cref{preansatz} to be $\infty$, but all of our results will require \cref{series} to hold only for finite $N$ (with the precise value of $N$ needed depending on the result).

We now substitute \cref{preansatz} into 
\cref{scalar1} and collect the terms that fall off as $1/r^{\alpha + j +1}$. We thereby obtain the following recursion relations for the coefficients appearing in \cref{preansatz}
\begin{equation}
\label{scawavew/osourc(1/r^j+1)}
[\mathcal{D}^{2}+(\alpha +j-1)(\alpha +j-d+2)]\phi^{(j-1)}+(2\alpha + 2j-d+2)\partial_{u}\phi^{(j)}=0
\end{equation}
Here, $\mathcal{D}^{2} = \mathcal{D}_A \mathcal{D}^A$ is the Laplacian on the unit sphere, where $\mathcal{D}_A$ is the derivative operator associated with $q_{AB}$ and sphere indices are lowered and raised with $q_{AB}$ and $q^{AB}$. 

It follows immediately from \cref{scawavew/osourc(1/r^j+1)} that if, for some $i \geq 0$, $\phi^{(i)}$ has nonpolynomial dependence on $u$, then for even $d$, no solution of the form \cref{preansatz}
exists unless $\alpha = 1$, whereas for odd $d$, no solution of the form \cref{preansatz}
exists unless $\alpha = 1/2$. To see this, we note that unless the coefficient of the $\partial_{u}\phi^{(j)}$ term vanishes for some $j$, the nonpolynomial dependence of $\phi^{(i)}$ will propagate to $\phi^{(i-1)}$ and thence to $\phi^{(i-2)}$, etc. This will result in an inconsistency in 
\cref{scawavew/osourc(1/r^j+1)} at the lowest nontrivial order, $j=0$, since the first term in that equation is then absent. Thus, the coefficient of
$\partial_{u}\phi^{(j)}$ in \cref{scawavew/osourc(1/r^j+1)} must vanish for some $j$.
For $d$ even, this requires $\alpha = 1$, in which case the coefficient vanishes for $j = d/2 -2$. For $d$ odd, this requires $\alpha = 1/2$, in which case the coefficient vanishes for $j = (d-3)/2$. 

However, in the odd dimensional case, \cref{preansatz}
with $\alpha = 1/2$ is not adequate for several reasons. First, \cref{preansatz}
 with $\alpha=1/2$ does not admit static solutions, since static solutions satisfy Laplace's equation and fall off as integral powers of $1/r$, starting at order, $1/r^{d-3}$. Second, when a source term $S$ is considered in \cref{scalar1}, it is natural to allow $S$ to fall off with integral powers of $1/r$. In particular, in order to have a nonvanishing, finite source flux at null infinity, it will be necessary to have $S$ fall off as $1/r^{d-2}$. Such source terms will generate terms in $\phi$ that fall off as integral powers of $1/r$, again starting at order $1/r^{d-3}$. Third, even if one does not consider sources, for nonlinear equations such as Einstein's equation, quadratic and higher order even powers of the field will generate terms that fall off as integral powers of $1/r$. This will lead to inconsistencies unless one also includes integral powers of $1/r$ in the fall-off of the field, again starting at order $1/r^{d-3}$.

Thus, in odd dimensions, we must allow integral powers of $1/r$ starting at least at order $1/r^{d-3}$. However, in odd dimensions, the coefficient of a term that falls as $1/r^p$ for integer $p < d-3$ must have polynomial dependence in $u$ of degree $<p$ in order for the recursion relations to terminate. (Source terms and nonlinear terms will not enter the recursion relations at these orders.) Such solutions do not appear to be of any physical interest, and we will exclude them from our ansatz. 

Thus, we adopt the following as the final form of our ansatz:
\begin{eqnarray}
\phi &\sim& \sum_{n=d/2 -1}^\infty \frac{1}{r^{n}} \phi^{(n)}(u, x^A) \, \quad \quad \quad \quad \quad \quad \quad \quad \quad \quad \quad d \,\, {\rm even} \label{devenan} \\
\phi &\sim& \sum_{n=d/2 -1}^\infty \frac{1}{r^{n}} \phi^{(n)}(u, x^A) + \sum_{p=d-3}^\infty \frac{1}{r^{p}} \tilde{\phi}^{(p)}(u, x^A) \quad \quad d \,\, {\rm odd} \label{doddan} 
\end{eqnarray}
where the meaning of ``$\sim$'' is as explained below \cref{preansatz}. 
Note that in eq.~(\ref{doddan}), $n$ runs over half-integer values rather than integer values (as in eq.~(\ref{devenan})). We have done this (rather than insert $\alpha = 1/2$ and keep integer values) so that the superscript ``$(n)$'' is always associated with $1/r^n$ fall-off and so that we can write the recursion in the same form 
\begin{equation}
\label{scawavew/osourc(1/r^n+1)}
[\mathcal{D}^{2}+(n-1)(n-d+2)]\phi^{(n-1)}+(2n-d+2)\partial_{u}\phi^{(n)}=0
\end{equation}
in both even and odd dimensions. In both even and odd dimensions, we refer to the leading (slowest fall-off) term $n = d/2 -1$ as {\em radiative order}, and we refer to the term with $1/r^{d-3}$ fall-off as {\em Coulombic order}. 
In odd dimensions, the $\tilde{\phi}^{(p)}$ satisfy separate recursion relations of the same form
\begin{equation}
\label{tildephi}
[\mathcal{D}^{2}+(p-1)(p-d+2)]\tilde{\phi}^{(p-1)}+(2p-d+2)\partial_{u}\tilde{\phi}^{(p)}=0.
\end{equation}
In the source free case, $\tilde{\phi}^{(p)}$ must have polynomial dependence in $u$ with degree no higher than $p - d +3$ in order for the expansion to terminate at order $d-3$. However, this restriction will not apply when source terms or nonlinear terms are present.

\begin{rem}
\label{termin}
{\em Note that the lower limit of the sum in (\ref{devenan}) was taken to be radiative order, $n = d/2 -1$. However, the ansatz would not be changed if we allowed the lower limit of the sum to extend to $n=1$ for $d > 4$ because the recursion relation \cref{scawavew/osourc(1/r^n+1)} at $n=d/2 - 1$ yields
\begin{equation}
[\mathcal{D}^{2}-(d/2-2)(d/2-1)]\phi^{(d/2-2)} = 0
\end{equation}
which implies $\phi^{(d/2-2)} = 0$. The recursion relations at smaller $n$ then successively yield $\phi^{(n)} = 0$ for all $n < d/2 -1$. Similarly, the lower limit of the first sum in (\ref{doddan}) could be taken to be $n=1/2$ without affecting the ansatz.
The upper limit of the sums appearing in (\ref{devenan}) and (\ref{doddan}) were taken to be $\infty$ for convenience. Most of our analysis will concern the behavior of fields at Coulombic order and slower fall-off and only a small number of derivatives will be taken, so the asymptotic expansion need hold only to the corresponding order.}
\end{rem}

Finally, we address the issue of the reasonableness of our ansatz, i.e., what classes of solutions to \cref{scalar1} satisfy our ansatz. In Minkowski spacetime of both even and odd\footnote{Future null infinity does not exist for an odd dimensional radiating spacetime \cite{HW}, but it exists for odd dimensional Minkowski spacetime.} 
dimensions, there is an alternative criterion of smoothness of the conformally rescaled field $\bar{\phi} = \Omega^{-(d/2 -1)}\phi$ at future null infinity, $\mathscr{I}^+$, in the conformally completed spacetime. Since $\Omega = 1/r$ is a suitable conformal factor for Minkowski spacetime, it is easily seen that smoothness of $\bar{\phi}$ at $\Omega = 0$ is equivalent to our asymptotic expansion eq.~(\ref{devenan}) in even dimensions and our asymptotic expansion
eq.~(\ref{doddan}) without the integer power terms in odd dimensions.
By the argument\footnote{Prop. 11.1.1 of \cite{Wald-book} is stated for $d=4$ but is easily generalized to Minkowski spacetime of arbitrary dimension} of Prop. 11.1.1 of \cite{Wald-book}, smoothness at $\mathscr{I}^+$ holds for all solutions to \cref{scalar1} with smooth initial data of compact support. Thus, all solutions with initial data of compact support satisfy our ansatz.
Furthermore, static, asymptotically flat solutions satisfy the asymptotic expansion eq.~(\ref{devenan}) in even dimensions and the asymptotic expansion
eq.~(\ref{doddan}) with only the integer power terms in odd dimensions. It follows that in both even and odd dimensions, all solutions to \cref{scalar1} with smooth initial that corresponds to a static asymptotically flat solution outside of a compact region satisfy our ansatz.


\subsection{Solutions to the Scalar Wave Recursion Relations}
\label{scalarsol}

We now consider the scalar wave equation with smooth source $S$
\begin{equation}
\label{scalar2}
{\Box} \phi = S.
\end{equation}
We assume that $S$ also has an expansion in powers of $1/r$. In order that the flux of $S$ through a sphere near null infinity be finite in the limit as $r \to \infty$, we must have $S = O(1/r^{d-2})$. We take as our ansatz for $S$
\begin{equation}
S\sim \sum_{n= d-2}^\infty \frac{1}{r^n} S^{(n)}(u, x^A).
\label{sexpand}
\end{equation}
In even dimensions, the sum ranges over integer $n$. In odd dimensions, we could also allow half-integral powers of $1/r$ in the expansion of $S$, beginning at order $1/r^{d-5/2}$. Indeed, for nonlinear equations, half-integral powers would appear as an effective source generated by cubic and higher order terms in the field, although these terms would first enter only at order $1/r^{3(d/2 - 1)}$. However, we will be primarily interested in the behavior of solutions $\phi$ at fall-off ranging from radiative ($1/r^{d/2 - 1}$) to Coulombic ($1/r^{d-3}$) orders.
In odd dimensions, only the leading order source term $S^{(d-2)}/r^{d-2}$ will enter our analysis. Therefore, for notational simplicity, we will take the sum in \cref{sexpand} to range only over integer values of $n$ in both even and odd dimensions. Note that our asymptotic expansion takes account only of sources ``near null infinity.'' Sources that go out to infinity along, e.g., timelike inertial trajectories do not contribute at all to the asymptotic expansion of $S$.

In even dimensions, under the ansatz eq.~(\ref{devenan}), the recursion relations \cref{scawavew/osourc(1/r^n+1)} are modified by the source term to become
\begin{equation}
\label{scawavew/sourc(1/r^n+1)}
[\mathcal{D}^{2}+(n-1)(n-d+2)]\phi^{(n-1)}+(2n-d+2)\partial_{u}\phi^{(n)}=S^{(n+1)}.
\end{equation}
In odd dimensions, under the ansatz eq.~(\ref{doddan}), \cref{scawavew/osourc(1/r^n+1)} is unmodified, but \cref{tildephi} is modified to become
\begin{equation}
\label{tildephisource}
[\mathcal{D}^{2}+(p-1)(p-d+2)]\tilde{\phi}^{(p-1)}+(2p-d+2)\partial_{u}\tilde{\phi}^{(p)}= S^{(p+1)}.
\end{equation}

It should be noted that when $d=4$, \cref{scawavew/sourc(1/r^n+1)} for $n=1$ yields $S^{(2)} = 0$. Thus, for $d=4$ there is an inconsistency with our ansatz eq.~(\ref{devenan}) when $S^{(2)} \neq 0$, i.e., when there is nonvanishing flux of the source through spheres near null infinity. This could be accommodated by modifying the ansatz in $d=4$ to allow an additional series of terms that fall as $\ln r/r^n$. This issue will arise in the next subsections when we consider whether the Lorenz gauge condition can be imposed on electromagnetic fields and linearized gravitational perturbations, and we will see that a non-vanishing flux of charge current or stress energy will provide an obstruction to imposing the Lorenz gauge in $d=4$ in a manner compatible with our ansatz. Similarly, in full, nonlinear general relativity, we will find that a non-vanishing flux of stress energy or Bondi news will provide an obstruction to imposing the harmonic gauge in $d=4$ in a manner compatible with our ansatz. Rather than include any such additional $\ln r$ terms in these cases, we will simply not impose the Lorenz and harmonic gauges in $d=4$ when these obstructions exist.
For the analysis of this subsection, we will simply restrict consideration to the case that $S^{(2)} = 0$ when $d=4$, so that our ansatz can be imposed. 

We now consider two procedures for solving the above recursion relations.
The first procedure is as follows: Consider, first, the even dimensional case, where we must solve \cref{scawavew/sourc(1/r^n+1)} with integral $n$. By our ansatz for $\phi$ and $S$, this equation automatically holds for $n=d/2 -1$, since $\phi^{(d/2 -2)} = S^{(d/2)} = 0$ and the coefficient of $\partial_{u}\phi^{(d/2 -1)}$ vanishes. (Here, when $d=4$, we have assumed that $S^{(2)} = 0$.) Thus, we may specify $\phi^{(d/2 -1)}(u,x^A)$ arbitrarily. The $n=d/2$ equation then yields
\begin{equation}
\label{phiwaved/2}
2\partial_{u}\phi^{(d/2)}=S^{(d/2+1)} - [\mathcal{D}^{2}+(n-1)(n-d+2)]\phi^{(d/2-1)}.
\end{equation}
The right side is ``known,'' so this equation can be straightforwardly integrated to obtain $\phi^{(d/2)}$. The solution is unique up to the arbitrary specification of $\phi_0^{(d/2)} (x^A)= \phi^{(d/2)}(u_0, x^A)$ at the retarded time $u=u_0$. This procedure can then be iterated indefinitely to solve for $\phi^{(n)}$ for all $n > d/2 -1$ up to the arbitrary specification of $\phi_0^{(n)} (x^A)= \phi^{(n)}(u_0, x^A)$.

In odd dimensions, we must solve \cref{scawavew/osourc(1/r^n+1)} with half-integral $n$ as well as \cref{tildephisource}. To solve \cref{scawavew/osourc(1/r^n+1)}, we may again, specify $\phi^{(d/2 -1)}(u,x^A)$ arbitrarily. We may then again uniquely solve for $\phi^{(n)}$ for all $n > d/2 -1$ up to the arbitrary specification of $\phi_0^{(n)} (x^A)= \phi^{(n)}(u_0, x^A)$. Similarly, we can uniquely solve \cref{tildephisource} with $p=d-3$ for $\tilde{\phi}^{(d-3)}$, up to the arbitrary specification of $\tilde{\phi}_0^{(d-3)} (x^A) = \tilde{\phi}^{(d-3)}(u_0, x^A)$. We can then perform a similar iteration to obtain $\tilde{\phi}^{(p)}$ for all $p > d-3$, up to the arbitrary specification of $\tilde{\phi}_0^{(p)} (x^A) = \tilde{\phi}^{(p)}(u_0, x^A)$.

We summarize these results in the following proposition.
\begin{prop}
\label{first}
Let $\phi$ be given by the asymptotic expansion eq.~(\ref{devenan})-(\ref{doddan}) and let $S$ be given by the asymptotic expansion \cref{sexpand}. Assume further that for $d=4$ we have $S^{(2)} = 0$. Then, in even dimensions, a unique solution to the recursion relations \cref{scawavew/sourc(1/r^n+1)} is obtained by arbitrarily specifying $\phi^{(d/2 -1)}(u,x^A)$ (i.e., specifying $\phi$ at ``radiative order'') and arbitrarily specifying $\phi^{(n)}(u_0,x^A)$ for all $n > d/2 -1$ at some initial time $u_0$. Similarly, in odd dimensions, a unique solution to the recursion relations \cref{scawavew/osourc(1/r^n+1)} and \cref{tildephisource} is obtained by arbitrarily specifying $\phi^{(d/2 -1)}(u,x^A)$ (i.e., specifying $\phi$ at ``radiative order'') and arbitrarily specifying both $\phi^{(n)}(u_0,x^A)$ for all $n > d/2 -1$ and $\tilde{\phi}^{(p)}(u_0, x^A)$ for all $p \geq d-3$
at some initial time $u_0$.
\end{prop}

The second procedure involves solving the recursion relations in the reverse order. Suppose that, for some $n > d/2 - 1$, we specify $\phi^{(n)}(u,x^A)$ arbitrarily. We can then try to solve \cref{scawavew/sourc(1/r^n+1)} for $\phi^{(n-1)}$. In order to do so, we must invert the angular operator $\mathcal{D}^{2}+(n-1)(n-d+2)$. A unique inverse of this operator exists whenever $-(n-1)(n-d+2)$ is not an eigenvalue of the Laplacian, $\mathcal{D}^{2}$. Since the eigenvalues of $\mathcal{D}^{2}$ are $-\ell(\ell + d - 3)$ for $\ell = 0, 1, \dots$, it can be seen that this operator is invertible at every order in odd dimensions, where $n$ is half-integer. On the other hand in even dimensions, this operator is invertible when $n \leq d-3$, but it is not invertible when $n > d-3$. Thus, in even dimensions, we can specify $\phi^{(d-3)}(u,x^A)$ arbitrarily and then uniquely solve for $\phi^{(d-4)}(u,x^A)$ by inverting the angular operator in \cref{scawavew/sourc(1/r^n+1)}. Iterating this process, we uniquely obtain $\phi^{(n)}(u,x^A)$ for all $n < d-3$. We then can solve for $\phi^{(n)}(u,x^A)$ for all $n > d-3$ as before, with the freedom to arbitrarily specify $\phi^{(n)}(u_0,x^A)$. In odd dimensions, we can similarly arbitrarily specify $\phi^{(n_0)}(u,x^A)$ for {\em any} half-integer $n_0 \geq d/2 -1$. We can then uniquely solve for $\phi^{(n)}(u,x^A)$ for all $n < n_0$ by inversion of the angular operators, and then solve for $\phi^{(n)}(u,x^A)$ for all $n > n_0$ as before, with the freedom to arbitrarily specify $\phi^{(n)}(u_0,x^A)$. This can be summarized as follows:

\begin{prop}
\label{second}
Let $\phi$ be given by the asymptotic expansion eq.~(\ref{devenan})-(\ref{doddan}) and let $S$ be given by the asymptotic expansion \cref{sexpand}. Assume further that for $d=4$ we have $S^{(2)} = 0$. Then, in even dimensions, a unique solution to the recursion relations \cref{scawavew/sourc(1/r^n+1)} is obtained by arbitrarily specifying $\phi^{(d -3)}(u,x^A)$ (i.e., specifying $\phi$ at ``Coulombic order'') and arbitrarily specifying $\phi^{(n)}(u_0,x^A)$ for all $n > d - 3$ at some initial time $u_0$. Similarly, in odd dimensions, a unique solution to the recursion relations \cref{scawavew/osourc(1/r^n+1)} and \cref{tildephisource} is obtained by arbitrarily specifying $\phi^{(n_0)}(u,x^A)$ for any half-integral $n_0$, and, for some initial time $u_0$, arbitrarily specifying $\phi^{(n)}(u_0,x^A)$ for all $n > n_0$ and $\tilde{\phi}^{(p)}(u_0, x^A)$ for all $p \geq d-3$.
\end{prop}

An important corollary of the argument leading to Proposition \ref{second} is the following:

\begin{cor}
\label{statcoulordcorr}
Suppose for $d$ even we have $\partial_{u}\phi^{(n_0)} = 0$ for some $n_0 < d-3$. Then $\phi^{(n)} = 0$ for all $n < n_0$. Similarly, if $\partial_{u}\phi^{(d-3)} = 0$ and $S^{(d-2)} = 0$, then $\phi^{(n)} = 0$ for all $n < d-3$.
For $d$ odd, if $\partial_{u}\phi^{(n_0)} = 0$ for some half-integral $n_0$ (without restriction), then $\phi^{(n)} = 0$ for all $n < n_0$.
\end{cor}

Finally, it is worth noting that for $n > d - 3$, the spherical harmonic $Y_{n-d+2,m}$ is in the kernel of $\mathcal{D}^{2}+(n-1)(n-d+2)$. It follows immediately that in the source-free case, for $d$ even we have that
\begin{equation}
\label{NPalpha}
\alpha_{nm}^{d}\equiv\int  Y_{n-d+2,m}\phi^{(n)} d\Omega
\end{equation}
is a constant of motion for all $n > d - 3$ \cite{NPconst,JezNPconst}, i.e., $\partial_{u}\alpha_{nm}^{d}=0$, where $d\Omega$ is the measure on the $(d-2)-$sphere. Similarly, in the source free case, for $d$ odd we have that 
\begin{equation}
\label{NPalphatilde}
\tilde{\alpha}_{pm}^{d}\equiv\int  Y_{p-d+2,m}\tilde{\phi}^{(p)} d\Omega
\end{equation}
is a constant of motion for all $p > d - 3$.


\subsection{Maxwell's Equations}
\label{subsecMaxeqa}

Consider Maxwell's Equations with vector potential ${A}_{a}$ and charge-current $j_{a}$ on $d-$dimensional Minkowski spacetime
\begin{equation}
\label{MaxwellsEqs}
{\Box} {A}_{a}-{\partial}_{a}{\partial}^{b}{A}_{b}=-4\pi j_{a}
\end{equation}
where $\partial^a j_{a} = 0$. In analogy with the scalar field ansatz (\ref{devenan}) and (\ref{doddan}), we assume as an ansatz that there exists a choice of gauge for $A_a$ such that it admits an asymptotic expansion of the form
\begin{eqnarray}
A_a &\sim& \sum_{n=d/2 -1}^\infty \frac{1}{r^{n}} A_a^{(n)}(u, x^A) \, \quad \quad \quad \quad \quad \quad \quad \quad \quad \quad \quad d \,\, {\rm even} \label{Adevenan} \\
A_a &\sim& \sum_{n=d/2 -1}^\infty \frac{1}{r^{n}} A_a^{(n)}(u, x^A) + \sum_{p=d-3}^\infty \frac{1}{r^{p}} \tilde{A}_a^{(p)}(u, x^A) \quad \quad d \,\, {\rm odd.} \label{Adoddan} 
\end{eqnarray}

We further assume, in analogy with \cref{sexpand} that $j_a$ admits an asymptotic expansion of the form
\begin{equation}
j_a \sim \sum_{n= d-2}^\infty \frac{1}{r^n} j_a^{(n)}(u, x^A).
\label{jexpand}
\end{equation}
In addition, we require that $j_{a}^{(d-2)}(u,x^{A})\rightarrow 0$ as $u\rightarrow -\infty$, i.e. there is no current flux to future null infinity at asymptotically early times. Here, as already mentioned at the end of the Introduction, $A_a^{(n)}$, $\tilde{A}_a^{(n)}$, and $j_a^{(n)}$ are defined so that their {\em normalized} basis components are independent of $r$---in contrast to a more common convention where the orders of the expansion would denote the powers of $1/r$ occurring in the expansion of coordinate basis components of $A_a$ in the coordinates of \cref{metric}. Thus, in our convention, $A_r^{(n)}$, $A_u^{(n)}$, and $A_A^{(n)}$ all contribute to the {\em physical} fall off rate of $1/r^n$, i.e., $A_A^{(n)}$ is the $1/r^n$ part of $1/r (\partial/\partial x^A)^a A_a$, not the $1/r^n$ part of $(\partial/\partial x^A)^a A_a$.
Our convention avoids a spurious ``mixing of orders'' in equations due to the different behavior of the coordinate basis elements. Again, our assumption that upper limits in the above asymptotic expansions run to $\infty$ is for convenience, as only finitely many orders will be needed for our main results.

In even\footnote{We are not aware of any smoothness at $\mathscr{I}^+$ criterion for $A_a$ that can be formulated in odd dimensions, since $A_a$ itself cannot be smooth at $\mathscr{I}^+$ for radiating solutions and giving $A_a$ a conformal weight would not appear to be of any use since Maxwell's equations are not conformally invariant when $d \neq 4$.} dimensions, we now compare our ansatz eq.~(\ref{Adevenan}) to what would be obtained by requiring that $A_a$ (with no conformal weight) be smooth at $\mathscr{I}^+$. Since $\Omega = 1/r$ is a suitable conformal factor for Minkowski spacetime, the necessary and sufficient condition for smoothness of $A_a$ at $\mathscr{I}^+$ is that its components, $(A_u, A_\Omega, A_A)$, defined by
\begin{equation}
A = A_u du + A_\Omega d \Omega + A_A d x^A
\end{equation}
be smooth functions of $(u, \Omega, x^A)$ at $\Omega = 0$. For $d=4$, it is easily seen that this smoothness criterion differs from the asymptotic expansion eq.~(\ref{Adevenan}) only in that the smoothness criterion (i) allows a $0$th order term, $A^{(0)}_u$, in $A_u$ and (ii) requires $A^{(1)}_r = 0$. It is easily seen that $A^{(0)}_u$ can be set to zero by a gauge transformation, so smoothness at $\mathscr{I}^+$ implies that our ansatz eq.~(\ref{Adevenan}) holds. Conversely, we show  in \Cref{appelec} that starting from our ansatz eq.~(\ref{Adevenan}), one can set $A^{(1)}_r = 0$ by a gauge transformation if and only if\footnote{$j_{r}^{(3)}$ must be independent of $u$ by conservation of current. A nonvanishing $j_{r}^{(3)}$ would correspond to having an {\em ingoing} null current near $\mathscr{I}^+$.} $j_{r}^{(3)}=0$. Thus, for $d=4$ our ansatz eq.~(\ref{Adevenan}) is slightly weaker than smoothness at $\mathscr{I}^+$ in that it admits additional solutions with $j_{r}^{(3)} \neq 0$. 

In higher even dimensional spacetimes, eq.~(\ref{Adevenan}) requires strictly faster fall-off than needed for smoothness of $A_a$ (with no conformal weighting) at $\mathscr{I}^+$. Thus, eq.~(\ref{Adevenan}) is nominally stronger than the condition of smoothness of $A_a$ at $\mathscr{I}^+$.
However, we show in \cref{appelectlorgauge} that the Lorenz gauge condition can be imposed when $d > 4$ within a slower fall-off ansatz. As explained in \Cref{hg}, the slower fall-off solutions excluded by eq.~(\ref{Adevenan}) are therefore pure gauge. Thus, in even dimensional spacetimes with $d > 4$, our ansatz is exactly equivalent to smoothness of $A_a$ (in some gauge) at $\mathscr{I}^+$. 

In the following, we will focus on the even dimensional case, and then indicate how the arguments can be modified to accommodate the odd dimensional case.
Just as in the scalar case, Maxwell's equations  give rise to recursion relations for the coefficients of the asymptotic expansions eq.~(\ref{Adevenan}) and \cref{jexpand}. In the even dimensional case, these recursion relations are explicitly
\begin{equation}
\label{Maxwellseqsu-comparb}
\big[\mathcal{D}^{2} +(n-1)(n-d+2)\big]A_{u}^{(n-1)}+(2n-d+2)\partial_{u}A_{u}^{(n)}-\partial_{u}\psi^{(n+1)}=-4\pi j_{u}^{(n+1)}
\end{equation}
\begin{equation}
\label{Maxwellseqsr-comparb}
\big[\mathcal{D}^{2}+n(n-d+1)\big]A_{r}^{(n-1)}+(d-2)A_{u}^{(n-1)}+(2n-d+2)\partial_{u}A_{r}^{(n)}-2\mathcal{D}^{A}A_{A}^{(n-1)}+n\psi^{(n)}=-4\pi j_{r}^{(n+1)} 
\end{equation}
\begin{equation}
\label{MaxwellseqsA-comparb}
\big[\mathcal{D}^{2}+(n-1)(n-d+2)-1\big]A_{A}^{(n-1)}- 2\mathcal{D}_{A}(A_{u}^{(n-1)} - A_{r}^{(n-1)})+(2n-d+2)\partial_{u}A_{A}^{(n)}-\mathcal{D}_{A}\psi^{(n)}=-4\pi j_{A}^{(n+1)}
\end{equation}
where $n$ takes integer values.
Here, we have defined 
\begin{equation}
\psi\equiv {\partial}^{a}{A}_a
\end{equation}
so
\begin{equation}
\label{psiatorder1/r^n}
\psi^{(n)}=\mathcal{D}^{A}A_{A}^{(n-1)}+(d-n-1)(A_{r}^{(n-1)}-A_{u}^{(n-1)})-\partial_{u}A_{r}^{(n)}.
\end{equation}

It would be very convenient to put $A_a$ in Lorenz gauge, $\psi = 0$.
On general grounds, we know that $A_a$ can always be put in the Lorenz gauge, but it is not obvious {\em a priori} whether it can be put in Lorenz gauge in such a way that the form of the asymptotic expansions, eq.~(\ref{Adevenan}) is maintained. We now investigate this issue. 

Under a gauge transformation, we have
\begin{equation}
\label{gaugetranseq}
{A}_{a} \to A_a - {\partial}_{a} \phi.
\end{equation}
Thus, in order to put $A_a$ in Lorenz gauge, we must solve
\begin{equation}
\label{waveeqphi}
{\Box} \phi =\psi.
\end{equation}
Thus, the equation that we must solve is of the same form as \cref{scalar2}, which we analyzed in the previous section. However, there are two key differences: (i) From its definition, {\em a priori}, $\psi$ may fall off as slowly as $1/r^{d/2 - 1}$ rather than $1/r^{d-2}$. (ii) We do not require that $\phi$ satisfy the ansatz eq.~(\ref{devenan}) but rather that $\partial_a \phi$ satisfy the ansatz eq.~(\ref{Adevenan}). Therefore, we may take the ansatz for $\phi$ to be
\begin{equation}
\label{phiansnew}
    \phi \sim  \sum_{n=d/2 -2}^{\infty}\frac{1}{r^{n}}\phi^{(n)}(u,x^{A})
\end{equation}
where $\partial_u \phi^{(d/2 -2)} = 0$. In $d=4$ dimensions, we may also add the term $c \ln r$ to the ansatz for $\phi$, where $c$ is a constant.

We first note that it follows immediately from $\partial^a j_{a} = 0$ that $\partial_u j_r^{(d-2)} = 0$. Hence, if $j_a^{(d-2)} \to 0$ as $u \to - \infty$ as we have assumed in our ansatz above, we have
\begin{equation}
j_r^{(d-2)} = 0. 
\end{equation}
Thus, the $r$-component of $j_a$ falls off at least one power of $1/r$ faster than required by the ansatz \cref{jexpand}.
Since $d/2 \leq d- 2$ for all $d \geq 4$, it follows 
immediately from \cref{Maxwellseqsr-comparb} with $n = d/2 - 1$ that 
\begin{equation}
\label{psid/2-1}
\psi^{(d/2 - 1)} = 0, 
\end{equation}
i.e., Maxwell's equations require $\psi$ to fall off at least one power of $1/r$ faster than implied by the ansatz (\ref{Adevenan}). To proceed further, we must separately consider the cases $d > 4$ and $d=4$.

When $d > 4$ all components of $j_a$ vanish at order $n=d/2$. It follows from \cref{Maxwellseqsu-comparb} with $n = d/2 -1$ that 
\begin{equation}
\partial_u \psi^{(d/2)} = 0. 
\end{equation}
We now can solve the scalar recursion relation \cref{scawavew/sourc(1/r^n+1)} at order $n= d/2 -1$ by allowing a nonvanishing $\phi^{(d/2 - 2)}$ given by
\begin{equation}
\phi^{(d/2 - 2)} = [\mathcal{D}^{2}-(d/2 - 2)^2]^{-1} \psi^{(d/2)}.
\end{equation}
Although $\phi^{(d/2 - 2)}$ falls off more slowly than allowed by the ansatz eq.~(\ref{devenan}), since $\partial_u \phi^{(d/2 - 2)} = 0$ the gradient of $\phi^{(d/2 - 2)}/r^{d/2 -2}$ will be compatible with the ansatz eq.~(\ref{Adevenan}). Furthermore, since $\partial_u \phi^{(d/2 - 2)} = 0$, the scalar recursion relations imply that all slower fall-off terms vanish. We may now specify $\phi^{(d/2 -1)}$ arbitrarily and solve the recursion relations for the faster fall-off terms in the same manner as in Proposition \ref{first}. Thus, when $d > 4$, there is no difficulty in putting $A_a$ in the Lorenz gauge in a manner compatible with the ansatz eq.~(\ref{Adevenan}).

When $d=4$, we still have $\psi^{(1)} = 0$ but we now have
\begin{equation}
\partial_u \psi^{(2)} = 4 \pi j_u^{(2)}.
\end{equation}
The scalar recursion relation \cref{scawavew/sourc(1/r^n+1)} at order $n= 1$ (with the term $c \ln r$ added to the ansatz for $\phi$) yields
\begin{equation}
c + {\mathcal D}^2 \phi^{(0)} =  \psi^{(2)}. 
\label{phi0}
\end{equation}
However, $\phi^{(0)}$ has to be $u$-independent in order that $\partial_a \phi$ satisfy the ansatz eq.~(\ref{Adevenan}). This requires $\partial_u \psi^{(2)}$ to vanish and hence $j_u^{(2)}=0$, i.e., there can be no flux of charge to infinity.\footnote{The Lorenz gauge can be imposed with $j_{u}^{(2)}\neq 0$ by adding a series with terms of the form $\ln r/r^{n}$ \cite{Himwich:2019}.} Conversely, if $j_u^{(2)}=0$, then $\psi^{(2)}$ is $u$-independent. We can choose $c$ to cancel the $\ell = 0$ part of $ \psi^{(2)}$. We can then invert ${\mathcal D}^2$ to solve for $\phi^{(0)}$. 
Thus, for $d=4$, we can solve \cref{phi0} if and only if $j_u^{(2)}=0$. We may then choose $\phi^{(1)}$ arbitrarily and solve the remaining recursion relations for the faster fall-off terms in the same manner as in Proposition \ref{first}. Thus, for $d=4$, $A_a$ can be put in the Lorenz gauge in a manner compatible with the ansatz eq.~(\ref{Adevenan}) if and only if $j_u^{(2)}=0$.

We now describe the modifications to the above results for odd dimensions. The recursion relations for $A_a^{(n)}$ take the form eqs. (\ref{Maxwellseqsu-comparb})-(\ref{MaxwellseqsA-comparb}) with $n$ half-integral and with the current source terms absent, whereas the recursion relations for $\tilde{A}_a^{(p)}$ take the same form as eqs. (\ref{Maxwellseqsu-comparb})-(\ref{MaxwellseqsA-comparb}) with $n$ replaced by $p$, with $p$ an integer. The ansatz for $\phi$ is taken to be
\begin{equation}
    \phi \sim \sum_{n= d/2 -2}^{\infty}\frac{1}{r^{n}}\phi^{(n)}(u,x^{A}) + \sum_{p=d-3}^{\infty}\frac{1}{r^{p}}\tilde{\phi}^{(p)}(u,x^{A})
\end{equation}
with $\partial_u \phi^{(d/2 -2)} = 0$.
The analysis of imposing the Lorenz gauge then proceeds in close parallel to the even dimensional case for $d>4$. We find that the Lorenz gauge can always be imposed in a manner compatible with the ansatz eq.~(\ref{Adoddan}).

We summarize our above results on the imposition of the Lorenz gauge in the following proposition:

\begin{prop}
\label{lorenz}
In Minkowski spacetime of dimension $d \geq 4$, suppose that in some gauge the vector potential  $A_a$ satisfies our ansatz eq.~(\ref{Adevenan}) (for $d$ even) or our ansatz eq.~(\ref{Adoddan}) (for $d$ odd). Suppose further that the charge-current $j_a$ satisfies \cref{jexpand} 
and that $j_{a}^{(d-2)}(u,x^{A})\rightarrow 0$ as $u\rightarrow -\infty$.
Then for all $d>4$, $A_a$ can be put in the Lorenz gauge in such a way that it continues to satisfy our ansatz. In $d=4$ the Lorenz gauge condition can be imposed within the ansatz eq.~(\ref{Adevenan}) if and only if $j_u^{(2)}=0$, i.e., if and only if the flux of charge to null infinity vanishes.
\end{prop}

\begin{rem}
\label{hg}
{\em We show in \cref{appelectlorgauge} that if, for $d > 4$, we had allowed the sum in eq.~(\ref{Adevenan}) to extend to $n = 1$ and the sum in eq.~(\ref{Adoddan}) to extend to $n = 1/2$, our proof that the Lorenz gauge condition can be imposed within the revised ansatz would still go through. Since $\Box A_a = -4\pi j_{a}$ in the Lorenz gauge and $j_{a}^{(n)} = 0$ for $n < d-2$, it follows from \Cref{termin} that in Lorenz gauge, we have $A_{a}^{(n)} = 0$ for all $n < d/2 -1$. Thus, the only solutions excluded by starting the sums at $n = d/2 -1$ in eqs.~(\ref{Adevenan}) and (\ref{Adoddan}) (rather than at $n= 1$ and $n = 1/2$) are pure gauge.}
\end{rem}

\begin{rem}
\label{station}
{\em Suppose that $A_a$ satisfies the ansatz eq.~(\ref{Adevenan}) and is stationary at all orders $n \leq m$ where $m \leq d-2$. Suppose further that $j_a^{(n)} = 0$ for all $n \leq m +1$. By conservation of $j_a$, we obtain $\partial_u j_r^{(n)} = 0$ for all $n \leq m + 2$.
It follows directly from its definition, \cref{psiatorder1/r^n}, that $\psi^{(n)}$ must be stationary for all $n \leq m$. However, using \cref{Maxwellseqsr-comparb} and the stationarity of $j_r^{(n)}$ for all $n \leq m + 2$, we obtain the stronger result that $\psi^{(n)}$ actually must be stationary for all $n \leq m+1$. We then may solve the recursion relation \cref{scawavew/sourc(1/r^n+1)} for all $n \leq m-1$ by setting 
\begin{equation}
\phi^{(n-1)}=[\mathcal{D}^{2}+(n-1)(n-d+2)]^{-1} \psi^{(n+1)}.
\end{equation}
We may then set $\phi^{(m-1)} = 0$ and solve the recursion relations for $\phi^{(n)} = 0$ for $n \geq m$ as in Proposition \ref{first}. The resulting gauge transformation will put $A_a$ in the Lorenz gauge satisfying the ansatz eq.~(\ref{Adevenan}) and maintaining stationarity at all orders $n \leq m$. In particular, if a solution with $j_a = 0$ is stationary in some gauge to order $m \leq d-2$, then it is stationary in a Lorenz gauge to the same order.}
\end{rem}

\begin{rem}
{\em Let $d=4$ and suppose $j_{u}^{(2)}=0$. Suppose, further, that $j_{r}^{(3)}=0$ so that, as shown in \Cref{appelec}, our ansatz is equivalent to smoothness of $A_a$ at $\mathscr{I}^{+}$ in some gauge. Although, by \Cref{lorenz}, the Lorenz gauge can be imposed
within our ansatz eq.~(\ref{Adevenan}), it need not be the case that $A_{r}^{(1)}=0$ in the Lorenz gauge, in which case $A_a$ in the Lorenz gauge will not be smooth at $\mathscr{I}^{+}$. In other words, in $d=4$ when $j_{u}^{(2)}=0$, the Lorenz gauge is compatible with our ansatz but it need not be compatible with smoothness of $A_a$ at  $\mathscr{I}^{+}$.}
\end{rem}

When $A_a$ is in Lorenz gauge---as, by Proposition \ref{lorenz} we may assume for $d>4$ and for $d=4$ when $j_{u}^{(2)}=0$---it satisfies
\begin{eqnarray}
\Box A_a = - 4 \pi j_a    \label{mel} \\
\partial^a A_a = 0. \label{gc}
\end{eqnarray}
The recursion relations arising from $\Box A_a = - 4 \pi j_a$ are just eqs.~ (\ref{Maxwellseqsu-comparb})-(\ref{MaxwellseqsA-comparb}) with $\psi = 0$ in even dimensions. (They are modified as described above in odd dimensions.) The recursion relations arising from $\partial^a A_a = 0$ are just $\psi^{(n)} = 0$ where $\psi^{(n)}$ is given by \cref{psiatorder1/r^n}. However, it is more convenient to work with a linear combination of this equation and the other equations so as to eliminate all $u$-derivatives. This can be achieved by defining
\begin{equation}
\label{constraintscalarfield}
\omega=K^{a} [\Box A_a+4\pi j_{a}]- 2K^a {\partial}_{a}\psi - (d-2) \psi/r
\end{equation}
where $K^a = (\partial/\partial r)^a$.
When eq.~(\ref{mel}) holds, the vanishing of $\omega$ is equivalent to the vanishing of $\psi$. The relation $\omega^{(n+2)} = 0$ yields
\begin{equation}
\label{Lorenzgaugeconstraintequationdgeq4}
[\mathcal{D}^{2}-(n-d+2)(n-d+3)]A_{r}^{(n)}+(2n-d+2)(n-d+3)A_{u}^{(n)}+(2n-d+2)\mathcal{D}^{A}A_{A}^{(n)}=-4\pi j_{r}^{(n+2)}
\end{equation}
which contains no $u$-derivatives (and therefore also does not mix different orders). 

We now consider the analogs of Propositions \ref{first} and \ref{second} for Maxwell's equations in Lorenz gauge. By eq.~(\ref{mel}) each Cartesian component of $A_a$ satisfies the scalar wave equation. Therefore, we may directly apply Propositions \ref{first} and \ref{second} to determine the data needed to uniquely determine a solution to eq.~(\ref{mel}) alone. Thus, the remaining task is to specify this data in such a way that eq.~(\ref{gc}) holds. However, if eq.~(\ref{mel}) holds we have
\begin{equation}
\Box \psi = \Box \partial^a A_a = \partial^a  \Box A_a = -4 \pi \partial^a j_a = 0.
\end{equation}
Thus, $\psi$ satisfies the homogeneous scalar wave equation, and we can ensure that $\psi = 0$ by choosing data for $A_a$ so as to ensure that the corresponding data for $\psi$ yields the solution $\psi = 0$. Again, we can determine this using Propositions \ref{first} and \ref{second}, and also using the fact that when eq.~(\ref{mel}) holds, the vanishing of $\psi^{(n)}$ is equivalent to the vanishing of $\omega^{(n+1)}$.  Putting all of the above statements together, it follows using Proposition \ref{first} that a unique solution to Maxwell's equations in Lorenz gauge can be determined by specifying $A^{(d/2-1)}_a$ subject to \cref{Lorenzgaugeconstraintequationdgeq4} for $n=d/2 -1$, and then specifying $A^{(n)}_a(u_0)$ for all $n > d/2 -1$ subject to \cref{Lorenzgaugeconstraintequationdgeq4} holding at $u=u_0$ (see exercise $2$ of \cite{Strom:2017} for the case $d=4$ with $j_{a}=0$). In odd dimensions, we also must similarly specify data for $\tilde{A}_{a}^{(p)}$ at $u=u_{0}$ subject to the constraint for all $p$. 

Alternatively, in even dimensions, using Proposition \ref{second}, a solution can be uniquely determined by specifying data at Coulombic order, $A^{(d-3)}_a$. However, in this case, the constraint \cref{Lorenzgaugeconstraintequationdgeq4} at $n=d-3$ ensures that $\psi^{(d-2)} = 0$ but this does not quite suffice to ensure that $\psi $ vanishes at all slower fall-off. This is because the recursion relation \cref{scawavew/osourc(1/r^n+1)} for $n=d-2$ yields
\begin{equation}
\mathcal{D}^{2}\psi^{(d-3)} = -(d-4)\partial_{u}\psi^{(d-2)}=0
\end{equation}
which does not imply that the $\ell = 0$ part of $\psi^{(d-3)}$ must vanish. Hence, the condition 
\begin{equation}
[\psi^{(d-3)}]|_{\ell = 0} =0
\end{equation}
must be imposed separately. Using eq.~(\ref{mel}), we may write this condition purely in terms of the Coulombic order data as
\begin{equation}
\label{chargeconservation}
\partial_{u}\mathcal{Q}(u)=-\mathcal{A}_{d} j_{u}^{(d-2)}\vert_{\ell=0}
\end{equation}
where 
\begin{equation}
\label{chargeaspect}
\mathcal{Q}(u)=\frac{\mathcal{A}_{d}}{4\pi}[A_{r}^{(d-3)}+(d-4)A_{u}^{(d-3)}]\vert_{\ell=0} \quad \quad \textrm{$d$ even (in Lorenz gauge)}
\end{equation}
and 
$\mathcal{A}_{d}$ is the area of a unit $(d-2)-$sphere 
\begin{equation}
\label{areaunitsphere}
   \mathcal{A}_{d}=\frac{2\pi^{\frac{d-1}{2}}}{\Gamma(\frac{d-1}{2})}.
\end{equation}
Using the Lorenz gauge condition, it can be verified that
$\mathcal{Q}(u)$ is the total electric charge at time $u$, defined by 
\begin{equation}
    \mathcal{Q}(u)\equiv \frac{1}{4\pi}\int F_{ur}^{(d-2)}d\Omega 
\end{equation}
with $F_{ab}=2\partial_{[a}A_{b]}$.
Thus, \cref{chargeconservation} expresses conservation of charge. Note that the formula \cref{chargeaspect} for $\mathcal{Q}(u)$ holds only in the Lorenz gauge and thus cannot be used in $d=4$ when $j_{u}^{(2)} \neq 0$. In odd dimensions, we do not obtain a similar additional constraint, but \cref{chargeconservation} follows directly from the recursion relation for $\tilde{A}_a^{(d-3)}$ corresponding to \cref{Maxwellseqsu-comparb} with $p=d-3$ as well as the Lorenz gauge condition given by \cref{psiatorder1/r^n} with $p=d-3$, where the charge is now given by
\begin{equation}
    \mathcal{Q}(u)=\frac{\mathcal{A}_{d}}{4\pi}[\tilde{A}_{r}^{(d-3)}+(d-4)\tilde{A}_{u}^{(d-3)}]\vert_{\ell=0} \quad \quad \textrm{ $d$ odd (in Lorenz gauge).}
\end{equation}

We summarize our results as follows:

\begin{thm}
\label{theoremcharinitprblmMaxeqs}
Suppose $d>4$ or $d=4$ and $j_u^{(2)} = 0$, so that the Lorenz gauge condition can be imposed. Then a unique solution to the recursion relations and constraints for Maxwell's equations in the Lorenz gauge is obtained by specifying data in either of the following two ways:
\begin{enumerate}
\item {\em Radiative Order Data:} Specify $A^{(d/2-1)}_a(u, x^A)$ subject to the constraint \cref{Lorenzgaugeconstraintequationdgeq4} at $n = d/2 - 1$. Specify $A^{(n)}_a (u=u_0, x^A)$ for all $n > d/2 -1$ subject to the constraint \cref{Lorenzgaugeconstraintequationdgeq4} at $u=u_0$. In odd dimensions, also specify $\tilde{A}^{(p)}_a (u=u_0, x^A)$ for all $p \geq d-3$, subject to the constraint \cref{Lorenzgaugeconstraintequationdgeq4} at $u=u_0$.

\item {\em Coulombic Order Data:} In even dimensions, specify $A^{(d-3)}_a(u, x^A)$ subject to the constraint \cref{Lorenzgaugeconstraintequationdgeq4} at $n = d-3$ and the additional constraint 
\cref{chargeconservation}; specify $A^{(n)}_a (u=u_0, x^A)$ for all $n > d-3$ subject to the constraint \cref{Lorenzgaugeconstraintequationdgeq4} at $u=u_0$. In odd dimensions, specify $A^{(m)}_a(u, x^A)$ for any half-integer $m \geq d/2 -1$, subject to the constraint \cref{Lorenzgaugeconstraintequationdgeq4} at $n = m$, specify $A^{(n)}_a (u=u_0, x^A)$ for all $n > m$ subject to the constraint \cref{Lorenzgaugeconstraintequationdgeq4} at $u=u_0$; specify $\tilde{A}^{(p)}_a (u=u_0, x^A)$ for all $p \geq d-3$, subject to the constraint \cref{Lorenzgaugeconstraintequationdgeq4} at $u=u_0$.

\end{enumerate}

\end{thm}

\subsection{Linearized Einstein Equation}
\label{Einsteinseqs}
We consider the linearized Einstein equation on $d$-dimensional Minkowski spacetime for a metric perturbation $h_{ab}$ with stress-energy source $T_{ab}$
\begin{equation}
\label{lineineqarbgauge}
-2\delta G_{ab}\equiv \Box {\bar{h}}_{ab}-2\partial_{(a}\partial^{c}{\bar{h}}_{b)c}+\eta_{ab}\partial^{c}\partial^{d}{\bar{h}}_{cd}=-16\pi T_{ab}
\end{equation}
where $\delta G_{ab}$ is the linearized Einstein tensor and
\begin{equation}
{\bar{h}}_{ab}\equiv {h}_{ab}-\frac{1}{2}{h}\eta_{ab}
\end{equation}
with $h \equiv \eta^{ab} h_{ab}$.
Our ansatz for $h_{ab}$ is 
\begin{eqnarray}
h_{ab} &\sim& \sum_{n=d/2 -1}^\infty \frac{1}{r^{n}} h_{ab}^{(n)}(u, x^A) \, \quad \quad \quad \quad \quad \quad \quad \quad \quad \quad \quad d \,\, {\rm even} \label{hdevenan} \\
h_{ab} &\sim& \sum_{n=d/2 -1}^\infty \frac{1}{r^{n}} h_{ab}^{(n)}(u, x^A) + \sum_{p=d-3}^\infty \frac{1}{r^{p}} \tilde{h}_{ab}^{(p)}(u, x^A) \quad \quad d \,\, {\rm odd} \label{hdoddan} 
\end{eqnarray}
where our conventions for labeling the orders in this expansion is as in the electromagnetic case.
Our ansatz for $T_{ab}$ is
\begin{equation}
T_{ab} \sim \sum_{n= d-2}^\infty \frac{1}{r^n} T_{ab}^{(n)}(u, x^A).
\label{Texpand}
\end{equation}
In addition, we require that $T_{ab}$ satisfy the dominant energy condition and that $T_{ab}^{(d-2)}(u,x^{A})\rightarrow 0$ as $u\rightarrow -\infty$, i.e. there is no stress energy flux to future null infinity at asymptotically early times. In odd dimensions, it would be reasonable to also allow terms in the expansion of $T_{ab}$ that fall as half-integral powers of $1/r$---and when we consider the full Einstein's equation, nonlinearities will effectively generate such terms in the equations. However, our analysis will mainly be concerned with the terms in $h_{ab}$ with fall-off ranging from radiative ($1/r^{d/2 -1}$) to Coulombic ($1/r^{d-3}$) orders, for which only the leading order terms in the expansion of $T_{ab}$ will contribute, so for simplicity, we do not include half-integral powers of $1/r$ in the ansatz for $T_{ab}$ in odd dimensions.

In even dimensions, we can compare our ansatz eq.~(\ref{hdevenan}) to what would be obtained by requiring that $\Omega^2 h_{ab}$ with $\Omega = 1/r$ be smooth at $\mathscr{I}^{+}$, i.e., at $\Omega = 0$. For $d=4$, if one assumes smoothness at $\mathscr{I}^{+}$ in some gauge, then, by a further choice of gauge (see \cite{Hollands_Thorne} or p.280 of \cite{Wald-book}), one can ensure that $h_{ab}$ satisfies our ansatz eq.~(\ref{hdevenan}). Conversely, if $h_{ab}$ satisfies our ansatz, then $\Omega^2 h_{ab}$ will be smooth at $\mathscr{I}^{+}$ if and only if $h_{rr}^{(1)}$ vanishes.
In \Cref{applingrav}, we show that we can set $h_{rr}^{(1)}=0$ by a gauge transformation provided that\footnote{$T_{ur}^{(3)}$, $T_{rr}^{(3)}$, and $T_{rA}^{(3)}$ are independent of $u$ by conservation and the dominant energy condition. These quantities vanish identically if the stress-energy is produced by a scalar or electromagnetic field satisfying our ansatz for those fields.}
$T_{ur}^{(3)}=T_{rr}^{(3)}=T_{rA}^{(3)}=0$. Thus, for $d=4$, our ansatz is slightly weaker than smoothness of $\Omega^2 h_{ab}$ at $\mathscr{I}^{+}$ in that we allow additional solutions with 
$T_{ra}^{(3)} \neq 0$.

For even dimensional spacetimes with $d > 4$, our ansatz eq.~(\ref{hdevenan}) 
requires faster fall-off than what is needed for smoothness of $\Omega^2 h_{ab}$ at $\mathscr{I}^{+}$. However, starting with smoothness at $\mathscr{I}^{+}$ and choosing
the conformal Gaussian null gauge, it was shown in \cite{Hollands_Thorne} that the fall-off given by our ansatz holds; we also will show in \Cref{applingravlorgauge} that, starting with smoothness of 
$\Omega^2 h_{ab}$ at $\mathscr{I}^{+}$, the Lorenz gauge condition can be imposed, which also implies the faster fall-off given by our ansatz. Thus, in even dimensional spacetimes with $d > 4$, our ansatz is precisely equivalent to smoothness of $\Omega^2 h_{ab}$ at $\mathscr{I}^{+}$ in some gauge.

In even dimensions, where $n$ is integer, Einstein's equation gives rise to the following system of recursion relations: 
\begin{align}
\label{linEinsteqsuu-comparb}
&\big[\mathcal{D}^{2} +(n-1)(n-d+2)\big]\bar{h}_{uu}^{(n-1)}+(2n-d+2)\partial_{u}\bar{h}_{uu}^{(n)}-2\partial_{u}\chi_{u}^{(n+1)}-\mathcal{D}^{A}\chi_{A}^{(n)}-(d-n-2)(\chi_{r}^{(n)}-\chi_{u}^{(n)})\nonumber\\
&+\partial_{u}\chi_{r}^{(n+1)}=-16\pi T_{uu}^{(n+1)}
\end{align}
\begin{align}
\label{linEinsteqsur-comparb}
&\big[\mathcal{D}^{2}+n(n-d+1)\big]\bar{h}_{ur}^{(n-1)}+(d-2)\bar{h}_{uu}^{(n-1)}+(2n-d+2)\partial_{u}\bar{h}_{ur}^{(n)}-2\mathcal{D}^{A}\bar{h}_{uA}^{(n-1)}-\mathcal{D}^{A}\chi_{A}^{(n)}\nonumber \\
&-(d-n-2)\chi_{r}^{(n)}+(d-2)\chi_{u}^{(n)}=-16\pi T_{ur}^{(n+1)}
\end{align}
\begin{equation}
\label{linEinsteqsuA-comparb}
\big[\mathcal{D}^{2}+(n-1)(n-d+2)-1\big]\bar{h}_{uA}^{(n-1)}-2\mathcal{D}_{A}(\bar{h}_{uu}^{(n-1)} - \bar{h}_{ur}^{(n-1)})+(2n-d+2)\partial_{u}\bar{h}_{uA}^{(n)}-\mathcal{D}_{A}\chi_{u}^{(n)}-\partial_{u}\chi_{A}^{(n+1)}=-16\pi T_{uA}^{(n+1)}
\end{equation}
\begin{align}
\label{linEinsteqsrr-comparb}
&\big[\mathcal{D}^{2}+(n-1)(n-d+2)-2(d-2)]\bar{h}_{rr}^{(n-1)}+2(d-2)\bar{h}_{ur}^{(n-1)}+2q^{AB}\bar{h}^{(n-1)}_{AB}+(2n-d+2)\partial_{u}\bar{h}_{rr}^{(n)}\nonumber \\
&-4\mathcal{D}^{A}\bar{h}^{(n-1)}_{Ar}+2n\chi_{r}^{(n)}=-16\pi T_{rr}^{(n+1)}
\end{align}
\begin{align}
\label{linEinsteqsrA-comparb}
&\big[\mathcal{D}^{2}+(n-1)(n-d+2)-d-1\big]\bar{h}_{rA}^{(n-1)}+d\bar{h}_{uA}^{(n-1)}-2\mathcal{D}_{A}\bar{h}^{(n-1)}_{ur}+2\mathcal{D}_{A}\bar{h}^{(n-1)}_{rr}+(2n-d+2)\partial_{u}\bar{h}_{rA}^{(n)}\nonumber \\
&-2\mathcal{D}^{B}\bar{h}_{BA}^{(n-1)}-\mathcal{D}_{A}\chi_{r}^{(n)}+(n+1)\chi_{A}^{(n)}=-16\pi T_{rA}^{(n+1)}
\end{align}
\begin{align}
\label{linEinsteqsAB-comparb}
&\big[\mathcal{D}^{2}+(n-1)(n-d+2)-2\big]\bar{h}_{AB}^{(n-1)}-4\mathcal{D}_{(A}\bar{h}_{B)u}^{(n-1)} +4\mathcal{D}_{(A}\bar{h}_{B)r}^{(n-1)}+ 2\big(\bar{h}_{rr}^{(n-1)}-2\bar{h}_{ur}^{(n-1)}+\bar{h}_{uu}^{(n-1)}\big)q_{AB} \nonumber \\
&+(2n-d+2)\partial_{u}\bar{h}_{AB}^{(n)}-2\bigg(\mathcal{D}_{(A}\chi_{B)}^{(n)}-\frac{q_{AB}}{2}\mathcal{D}^{C}\chi_{C}^{(n)}\bigg)+(d-n-4)(\chi_{r}^{(n)}-\chi_{u}^{(n)})q_{AB}-q_{AB}\partial_{u}\chi_{r}^{(n+1)}\nonumber \\
&=-16\pi T_{AB}^{(n+1)}.
\end{align}
Here we have defined 
\begin{equation}
\chi_a = \partial^b \bar{h}_{ab}
\end{equation}
so that
\begin{equation}
\label{chiequcomp}
\chi_{u}^{(n)}=\mathcal{D}^{A}\bar{h}_{Au}^{(n-1)} + (d-n-1)\big(\bar{h}_{ur}^{(n-1)}-\bar{h}_{uu}^{(n-1)}\big)-\partial_{u}\bar{h}^{(n)}_{ur}
\end{equation}
\begin{equation}
\label{chieqrcomp}
\chi_{r}^{(n)}=\mathcal{D}^{A}\bar{h}_{Ar}^{(n-1)}+ (d-n-1)\big(\bar{h}_{rr}^{(n-1)}-\bar{h}_{ur}^{(n-1)}\big)-q^{AB}\bar{h}_{AB}^{(n-1)} - \partial_{u}\bar{h}_{rr}^{(n)}
\end{equation}
\begin{equation}
\label{chieqAcomp}
\chi_{A}^{(n)}=\mathcal{D}^{B}\bar{h}_{AB}^{(n-1)}+(d-n)\big(\bar{h}_{rA}^{(n-1)}-\bar{h}_{uA}^{(n-1)}\big)-\partial_{u}\bar{h}_{rA}^{(n)}.
\end{equation}
In odd dimensions, where $n$ is half-integral, eqs.~(\ref{linEinsteqsuu-comparb})-(\ref{linEinsteqsAB-comparb}) hold with $T_{ab} = 0$, whereas the recursion relations for $\tilde{h}_{ab}^{(p)}$ are the same as eqs.~(\ref{linEinsteqsuu-comparb})-(\ref{linEinsteqsAB-comparb}).

In the electromagnetic case, the current $j_a$ is subject only to the conservation law $\partial^a j_a = 0$. This gave rise to the condition $K^a j_a^{(d-2)} = 0$, where $K^a = (\partial/\partial r)^a$. The stress-energy tensor $T_{ab}$ is also subject to the conservation law $\partial^a T_{ab} = 0$. This gives rise to the condition
\begin{equation}
K^a T_{ab}^{(d-2)} = 0.
\label{kdott}
\end{equation}
However, in the gravitational case, we have the further requirement that the stress-energy tensor satisfy the dominant energy condition. The only way \cref{kdott} can be compatible with the dominant energy condition is if
\begin{equation}
T_{ab}^{(d-2)} = \alpha K_a K_b
\label{dec}
\end{equation}
for some function $\alpha(u,x^A)$. Thus, all components of $T^{(d-2)}_{ab}$ must vanish except for $T^{(d-2)}_{uu}$.

It is of interest to examine the gauge dependence of the radiative order metric $h^{(d/2-1)}_{ab}$ and the gauge invariant quantities that can be constructed from $h^{(d/2-1)}_{ab}$. Under a gauge transformation, we have 
\begin{equation}
h_{ab} \to h_{ab} -\partial_{(a} \xi_{b)}
\label{gtrans}
\end{equation}
so
\begin{equation}
h^{(d/2-1)}_{ab} \to h^{(d/2-1)}_{ab} - [\partial_{(a}\xi_{b)}]^{(d/2-1)}
\end{equation}
where
\begin{align}
\label{radgaugetrans}
    [\partial_{(a}\xi_{b)}]^{(d/2-1)}=&\mathcal{D}_{(A}\xi_{B)}^{(d/2-2)}-K_{(a}\mathcal{D}_{B)}\xi_{u}^{(d/2-2)}+r_{(a}\mathcal{D}_{B)}\xi_{r}^{(d/2-2)}-r_{(a}\xi_{B)}^{(d/2-2)}+(\xi_{r}^{(d/2-2)}-\xi_{u}^{(d/2-2)})q_{AB}\nonumber \\
    &-\bigg(\frac{d}{2}-2\bigg)r_{(a}\xi_{b)}^{(d/2-2)}-K_{(a}\partial_{u}\xi_{b)}^{(d/2-1)}.
\end{align}
Here $\xi_{a}^{(d/2-2)}$ must be stationary in order to maintain our ansatz eqs.(\ref{hdevenan}) and (\ref{hdoddan}). It is clear from \cref{radgaugetrans} that $\xi_{a}^{(d/2-1)}$ can always be used to set $h_{uu}^{(d/2-1)},h_{ur}^{(d/2-1)}$ and $h_{uA}^{(d/2-1)}$ to zero. It also is clear from \cref{radgaugetrans} that the remaining components can be changed only by a stationary transformation. It follows immediately that $\partial_u h_{\mu \nu}^{(d/2-1)}$ is gauge invariant for all $\mu, \nu \neq u$.
However, using the linearized Einstein equation, it can be shown\footnote{The vanishing of $\partial_u h_{rr}^{(d/2-1)}$, $\partial_u h_{rA}^{(d/2-1)}$ and $\partial_u (q^{AB} h_{AB}^{(d/2-1)})$ follows from \cref{chi1} below, together with eqs.~(\ref{chiequcomp})-(\ref{chieqAcomp}) for $n=d/2 -1$.} that $\partial_u h_{rr}^{(d/2-1)} = \partial_u h_{rA}^{(d/2-1)} = \partial_u (q^{AB} h_{AB}^{(d/2-1)}) = 0$. Therefore, the only nontrivial gauge invariant quantity that can be constructed from $\partial_u h^{(d/2-1)}_{ab}$ is 
\begin{equation}
\label{news}
    N_{ab}\equiv \bigg(q_{a}{}^{c}q_{b}{}^{d}-\frac{1}{d-2}q_{ab}q^{cd}\bigg)\partial_{u}h_{ab}^{(d/2-1)}.
\end{equation}
We may view $N_{ab}$ as a tensor on the sphere, denoted $N_{AB}$. $N_{AB}$ is called the {\em Bondi news tensor}.

We now seek to put $h_{ab}$ in Lorenz gauge, 
\begin{equation}
 \partial^b \bar{h}_{ab} = 0
\end{equation}
while preserving the form of the ansatz eqs.~(\ref{hdevenan}) or (\ref{hdoddan}). Under a gauge transformation, $h_{ab}$ changes by \cref{gtrans}.
Thus, we can put $h_{ab}$ into Lorenz gauge if and only if we can solve
\begin{equation}
\label{xigaugeeq}
{\Box} \xi_{a}=2\chi_{a}.
\end{equation}
Thus, the equations we must solve take the same basic form as the scalar wave equation, and we can analyze them in close parallel to the electromagnetic case. 
We take our ansatz for $\xi_a$ to be 
\begin{eqnarray}
\xi_a &\sim&  \sum_{n=d/2 -2}^\infty \frac{1}{r^{n}} \xi_a^{(n)}(u, x^A) \, \quad \quad \quad \quad \quad \quad \quad \quad \quad \quad \quad \quad d \,\, {\rm even} \label{xidevenannew} \\
\xi_a &\sim& \sum_{n=d/2 -2}^\infty \frac{1}{r^{n}} \xi_a^{(n)}(u, x^A) + \sum_{p=d-3}^\infty \frac{1}{r^{p}} \tilde{\xi}_a^{(p)}(u, x^A) \quad \quad \quad d \,\, {\rm odd} \label{xidoddannew} 
\end{eqnarray}
where it is required in both of these expressions that $\partial_u \xi_a^{(d/2 -2)} = 0$.
When $d=4$, we may also add a term $c(\partial/\partial u)_a \ln r$ to $\xi_a$, where $c$ is a constant.

When $d>4$, the stress-energy terms in eqs.~(\ref{linEinsteqsuu-comparb})-(\ref{linEinsteqsAB-comparb}) do not enter at radiative order $n=d/2-1$. The $ur$,$rr$ and $rA$ components of these equations yield, respectively
\begin{equation}
-(d/2 -1)\chi_{r}^{(d/2 -1)}+2(d/2 -1)\chi_{u}
^{(n)}-\mathcal{D}^{A}\chi_{A}^{(d/2 -1)}=0
\end{equation}
\begin{equation}
(d-2)\chi_{r}^{(d/2-1)}=0
\end{equation}
\begin{equation}
(d/2)\chi_{A}^{(d/2-1)}-\mathcal{D}_{A}\chi_{r}^{(d/2-1)}=0.
\end{equation}
Thus, we have
\begin{equation}
\chi_{a}^{(d/2-1)}=0.
\label{chi1}
\end{equation}
The $uu$, $uA$ and $AB$ components yield, respectively
\begin{equation}
-2\partial_{u}\chi_{u}^{(d/2)}+\partial_{u}\chi_{r}^{(d/2)}= 0
\end{equation}
\begin{equation}
\partial_{u}\chi_{A}^{(d/2)}= 0
\end{equation}
\begin{equation}
q_{AB}\partial_{u}\chi_{r}^{(d/2)}= 0
\end{equation}
which implies
\begin{equation}
\partial_u \chi_{a}^{(d/2)}=0.
\label{chi2}
\end{equation}
As in the electromagnetic case, equations (\ref{chi1}) and (\ref{chi2}) ensure that we can solve \cref{xigaugeeq} within the ansatz.

However, when $d=4$, we still have that $\chi_{a}^{(1)}$ vanishes but \cref{linEinsteqsuu-comparb} for $n=1$ yields
\begin{equation}
\partial_{u}\chi_{u}^{(2)}=8\pi T_{uu}^{(2)}.
\end{equation}
As in the electromagnetic case, this will give rise to an obstruction to solving \cref{xigaugeeq} within the ansatz if and only if $T_{uu}^{(2)}$ is nonvanishing. Thus, for $d=4$, the necessary and sufficient condition for imposing the Lorenz gauge within our ansatz is that $T_{uu}^{(2)}$ vanish identically. 

We summarize these results in the following proposition:

\begin{prop}
\label{lorenzgrav}
For all $d>4$,
any $h_{ab}$ that satisfies our ansatz eq.~(\ref{hdevenan}) (for $d$ even) or ansatz eq.~(\ref{hdoddan}) (for $d$ odd)
can be put in the Lorenz gauge in such a way that it continues to satisfy our ansatz. In $d=4$ the Lorenz gauge condition can be imposed within the ansatz if and only if $T_{uu}^{(2)}=0$.
\end{prop}

\begin{rem}
\label{lgstat}
{\em As in the electromagnetic case, for $d > 4$ we show in \cref{applingravlorgauge} that the Lorenz gauge condition could still be imposed if we weakened the fall-off conditions to $1/r$ fall-off in even dimensions and $1/\sqrt{r}$ fall-off in odd dimensions. As in \cref{hg}, this justifies our taking the lower limit of the sum in eq.~(\ref{hdevenan}) and eq.~(\ref{hdoddan}) to start at $n=d/2 -1$.
Also, as in the electromagnetic case, it follows that if a solution is stationary in some gauge for all $n \leq m$ with $m \leq d-2$ and if $T^{(n)}_{ab} = 0$ for all $n \leq m+1$, then it is stationary in a Lorenz gauge for all $n \leq m$.}
\end{rem}

\begin{rem}
{\em Let $d=4$ and $T_{uu}^{(2)}=0$. Suppose further that $T_{ra}^{(3)}=0$ so that our ansatz is equivalent to smoothness of $\Omega^2 h_{ab}$ at $\mathscr{I}^{+}$ in some gauge.
Although, by \cref{lorenzgrav}, the Lorenz gauge can be imposed within our ansatz eq.~(\ref{hdevenan}), it need not be the case that $h_{rr}^{(1)}=0$ in the Lorenz gauge, in which case $\Omega^2 h_{ab}$ in the Lorenz gauge will not be smooth at $\mathscr{I}^{+}$, i.e., the Lorenz gauge need not be compatible with smoothness at $\mathscr{I}^{+}$.}
\end{rem}

When $h_{ab}$ is in Lorenz gauge---as, by Proposition \ref{lorenzgrav} we may assume for $d>4$ and for $d=4$ when $T_{uu}^{(2)}=0$---it satisfies
\begin{eqnarray}
\Box \bar{h}_{ab} = - 16 \pi T_{ab}    \label{gel} \\
\partial^a \bar{h}_{ab} = 0. \label{ggc}
\end{eqnarray}
The recursion relations for eq.~(\ref{gel}) are eqs.~(\ref{linEinsteqsuu-comparb})-(\ref{linEinsteqsAB-comparb}) with $\chi_a = 0$. The recursion relations arising from eq.~(\ref{ggc}) are just eqs.~(\ref{chiequcomp})-(\ref{chieqAcomp}) with $\chi_a = 0$. Again, it is useful to eliminate the terms in eqs.~(\ref{chiequcomp})-(\ref{chieqAcomp}) with $u$-derivatives using eqs.~(\ref{linEinsteqsuu-comparb})-(\ref{linEinsteqsAB-comparb}). This can be achieved by defining
\begin{equation}
\label{linconstraint}
{\tau}_{a}=K^{b}[\Box\bar{h}_{ab}+16\pi T_{ab}]-2K^{b}\partial_{b}\chi_{a}-(d-2)\chi_{a}/r.
\end{equation}
When eq. \ref{gel} holds, the vanishing of $\tau_{a}$ is equivalent to the vanishing of $\chi_{a}$. The relation $\tau_{a}^{(n+2)}=0$ yields
\begin{equation}
[\mathcal{D}^{2}-(n-d+2)(n-d+3)]\bar{h}_{ru}^{(n)}+(n-d+3)(2n-d+2)\bar{h}_{uu}^{(n)}-(2n-d+2)\mathcal{D}^{A}\bar{h}_{uA}^{(n)} =-16\pi T_{ru}^{(n+2)} \label{cu}
\end{equation}
\begin{align}
&[\mathcal{D}^{2}-((n-d+2)^{2}+n)]\bar{h}_{rr}^{(n)}+(d-2+(n-d+3)(2n-d+2))\bar{h}_{ur}^{(n)}-(2n-d+2)q^{AB}\bar{h}_{AB}^{(n)}\nonumber\\ 
&+(2n-d)\mathcal{D}^{A}\bar{h}_{Ar}^{(n)}=-16\pi T_{rr}^{(n+2)} \label{cr}
\end{align}
\begin{align}
&[\mathcal{D}^{2}-(n-d+3)(n-d+2)+(2n-d+1)]\bar{h}_{rA}^{(n)}+(2n-d+2)(n-d+2)\bar{h}_{uA}^{(n)}\nonumber \\
&+2\mathcal{D}_{A}(\bar{h}_{rr}^{(n)}-\bar{h}_{ur}^{(n)})+(2n-d+2) \mathcal{D}^{B}\bar{h}_{AB}^{(n)}=-16\pi T_{rA}^{(n+2)}. \label{cA}
\end{align}
\Cref{cu,cr,cA} reduce to the ``constraint equations'' given by \cite{Strom:2018} if one applies the additional gauge conditions that they impose. 

The analysis of the appropriate data for solutions to eq.~(\ref{gel}) and (\ref{ggc}) follows in exact parallel with the electromagnetic case. We solve the wave equation given by eqs.~(\ref{linEinsteqsuu-comparb})-(\ref{linEinsteqsAB-comparb}) with $\chi_a = 0$, subject to the constraints eqs.~(\ref{cu})-(\ref{cA}). We can specify data at radiative order subject to the constraints and solve for the faster fall-off terms exactly as in the electromagnetic case. We also can specify data at Coulombic order and solve for slower fall-off terms. In exact parallel with the electromagnetic case, in even dimensions, in addition to the Coulombic order constraints, the Coulombic order data must satisfy
\begin{equation}
\label{bondimassbalancelawlin}
    \partial_{u}\mathcal{M}=-\mathcal{A}_{d} T_{uu}^{(d-2)}\vert_{\ell=0}
\end{equation}
where 
    \begin{equation}
    \label{Bondimassevend}
    \mathcal{M}=\frac{1}{16\pi}\mathcal{A}_{d}[\bar{h}_{ur}^{(d-3)}+(d-4)\bar{h}_{uu}^{(d-3)}]\vert_{\ell=0} \quad \quad \textrm{ $d$ even (in Lorenz gauge)}
\end{equation}
with $\mathcal{A}_{d}$ given by \cref{areaunitsphere}.
Thus, $\mathcal M$ satisfies the same flux relation as the linearized Bondi mass in linearized gravity, and thus it can differ from the linearized Bondi mass only by a constant. To show that $\mathcal M$ is, indeed, the linearized Bondi mass, it suffices to show that it agrees with the Bondi mass in the stationary case, where $t^a = (\partial/\partial u)^a$, is a Killing field. In the stationary case, it can be verified that $\mathcal M$ agrees with the Komar mass formula
\begin{equation}
\label{Komar}
    \mathcal{M}=-\frac{1}{16\pi}\frac{(d-2)}{(d-3)}\int _{\infty}\boldsymbol{\epsilon}_{abcd}\nabla^{a}t^{b}
\end{equation}
where $\boldsymbol{\epsilon}_{abcd}$ is the volume form and the integral is taken over a sphere near infinity. Since the Komar mass agrees with the Bondi mass in the stationary case \cite{Myers-Perry}, it follows that $\mathcal M$ is, indeed, the linearized Bondi mass\footnote{We caution the reader that \cref{Bondimassevend} holds only in the Lorenz gauge, which cannot be imposed for $d=4$ when $T_{uu}^{(2)}\neq0$.}. In odd dimensions, we do not obtain a similar additional constraint, but  the recursion relation \cref{linEinsteqsuu-comparb} with $p=d-3$ as well as the Lorentz gauge constraint \cref{chiequcomp} with $p=d-3$ implies that \cref{bondimassbalancelawlin} holds where the linearized Bondi mass is given by
\begin{equation}
\label{Bondimassoddd}
    \mathcal{M}=\frac{1}{16\pi}\mathcal{A}_{d}[\bar{\tilde{h}}_{ur}^{(d-3)}+(d-4)\bar{\tilde{h}}_{uu}^{(d-3)}]\vert_{\ell=0} \quad \quad \textrm{ $d$ odd (in Lorenz gauge).}
\end{equation}

We summarize our results on solutions to the linearized Einstein equation in Lorenz gauge with the following theorem:

\begin{thm}
\label{theoremcharinitprblmgrav}
Suppose $d>4$ or $d=4$ and $T_{uu}^{(2)}= 0$, so that the Lorenz gauge condition can be imposed. Then a unique solution to the recursion relations and constraints for the linearized Einstein equation in Lorenz gauge is obtained by specifying data in either of the following two ways:
\begin{enumerate}
\item {\em Radiative Order Data:} Specify $h^{(d/2-1)}_{ab}(u, x^A)$ subject to the constraints eqs.~(\ref{cu})-(\ref{cA}) at $n = d/2 - 1$. Specify $h^{(n)}_{ab} (u=u_0, x^A)$ for all $n > d/2 -1$ subject to the constraints eqs.~(\ref{cu})-(\ref{cA}) at $u=u_0$. In odd dimensions, also specify $\tilde{h}^{(p)}_{ab} (u=u_0, x^A)$ for all $p \geq d-3$, subject to the constraint eqs.~(\ref{cu})-(\ref{cA}) at $u=u_0$.

\item {\em Coulombic Order Data:} In even dimensions, specify $h^{(d-3)}_{ab}(u, x^A)$ subject to the constraints eqs.~(\ref{cu})-(\ref{cA}) at $n = d-3$ and the additional constraint 
\cref{bondimassbalancelawlin}; specify $h^{(n)}_{ab} (u=u_0, x^A)$ for all $n > d-3$ subject to the constraints eqs.~(\ref{cu})-(\ref{cA}) at $u=u_0$. In odd dimensions, specify $h_{ab}^{(m)}(u, x^A)$ for any $m \geq d/2 -1$, subject to the constraints eqs.~(\ref{cu})-(\ref{cA}) at $n = m$, specify $h^{(n)}_{ab} (u=u_0, x^A)$ for all $n > m$ subject to the constraints eqs.~(\ref{cu})-(\ref{cA}) at $u=u_0$; specify $\tilde{h}^{(p)}_{ab} (u=u_0, x^A)$ for all $p \geq d-3$, subject to the constraints eqs.~(\ref{cu})-(\ref{cA}) at $u=u_0$.

\end{enumerate}

\end{thm}

\subsection{Nonlinear Einstein Equation}
\label{nleins}

For the nonlinear Einstein equation, we write $g_{ab} = \eta_{ab} + h_{ab}$ and we assume the same ansatz for $h_{ab}$ as in linearized gravity (see \Cref{Einsteinseqs}). For $d$ even with $d > 4$, our ansatz eq.~(\ref{hdevenan}) is equivalent to smoothness of $\Omega^2 h_{ab}$ (and, therefore, smoothness of $\Omega^2 g_{ab} = \Omega^2 \eta_{ab} + \Omega^2 h_{ab}$) at $\mathscr{I}^{+}$ by the same arguments as for the linearized case. For $d=4$ our ansatz eq.~(\ref{hdevenan}) in linearized gravity was slighter weaker than smoothness of $\Omega^2 h_{ab}$ at $\mathscr{I}^{+}$ in that it admitted additional solutions for which 
$h_{rr}^{(1)}$ cannot be set to zero by a gauge transformation within our ansatz. However, we show in \Cref{appnonlin} that, if the Bondi news is nonvanishing at all angles at any time, such additional solutions do not exist in the nonlinear theory. Thus, for $d=4$ our ansatz eq.~(\ref{hdevenan}) is also equivalent to smoothness at $\mathscr{I}^{+}$ for spacetimes in which $N_{AB}$ is nonvanishing everywhere on some cross-section.

The nonlinear Einstein equation is far more complex than the linearized Einstein equation. However, since the slowest fall-off of $h_{ab}$ is $1/r^{d/2 -1}$, the nonlinear terms first enter at order $(1/r^{d/2 -1})^2 = 1/r^{d-2}$. Consequently, for $n < d -3$, the recursion relations for the full Einstein equation are identical to eqs.~(\ref{linEinsteqsuu-comparb})-(\ref{linEinsteqsAB-comparb}) in the linearized case.
For $n= d-3$, the equations are modified by terms of the form $(\partial_u h^{(d/2 -1)}_{ab})^2$
and $h_{ab}^{(d/2 -1)} \partial^2_u h_{cd}^{(d/2 -1)}$, which are the only types of nonlinear terms that can contribute at this order. At higher orders, the nonlinear correction terms are far more complicated, but they always involve adding terms arising from metric components of slower fall-off. 

We define the non-linear part of the Einstein tensor $\mathcal{G}_{ab}$ as 
\begin{equation}
    \mathcal{G}_{ab}\equiv G_{ab}-\delta G_{ab}
\end{equation}
where $G_{ab}$ is the Einstein tensor and $\delta G_{ab}$ is the linearized Einstein tensor defined in \cref{lineineqarbgauge}.  Our ansatz then implies an asymptotic expansion of $\mathcal{G}_{ab}$ in integer powers of $1/r$ in even dimensions and both integer and half-integer powers in odd dimensions. In both even and odd dimensions, the expansion starts at order $1/r^{d-2}$. In all dimensions, Einstein's equations give rise to the same set of recursion relations as in the linearized case with the replacement
\begin{equation}
\label{nonlinsub}
    8\pi T_{ab}^{(n)}\rightarrow 8\pi T_{ab}^{(n)} - \mathcal{G}_{ab}^{(n)} \textrm{ for $n\geq d-2$}
\end{equation}
where $n$ is an integer in even dimensions and takes on both integer and half integer values in odd dimensions. By a direct calculation, we find that the leading order contribution to $\mathcal{G}_{ab}$ is given by
\begin{equation}
     \mathcal{G}_{ab}^{(d-2)}=-\frac{1}{4}N^{cd}N_{cd}K_{a}K_{b}+\frac{1}{2}\partial_{u}\bigg(q^{cd}q^{ef}c_{ce}N_{df}K_{a}K_{b}+q^{cd}c_{rc}N_{d(a}K_{b)}+c_{rr}N_{ab}\bigg) \label{nonlineinstd-2}\\
\end{equation}
where $c_{ab}\equiv h_{ab}^{(d/2-1)}$ and $N_{ab}$ is the Bondi news tensor as defined in \cref{news}. In writing \cref{nonlineinstd-2}, we have used the fact that, as in the linearized case (see \cref{chi1}), the recursion relations imply that $\chi_{a}^{(d/2-1)} = 0$, where $\chi_a \equiv \partial^b \bar{h}_{ab}$.

We wish to determine whether the metric $g_{ab}$ can be put in the harmonic gauge while maintaining our $1/r$ expansion ansatz. To put the metric in harmonic gauge, we must find coordinate functions $x^{\mu}$ such that 
\begin{equation}
\label{Harmgauge1}
    \Box_{g}x^{\mu}=0
\end{equation}
where $\Box_{g}\equiv g^{ab}\nabla_{a}\nabla_{b}$ and $\nabla_{a}$ is the derivative operator compatible with $g_{ab}$. Let 
\begin{equation}
\label{coordinatesnearscri}
    x^{\mu}=\accentset{\circ}{x}^{\mu}+\xi^{\mu}
\end{equation}
where $\accentset{\circ}{x}^{\mu}$ are global inertial coordinates of $\eta_{ab}$, satisfying $\partial_{\alpha}\accentset{\circ}{x}^{\mu}=\delta_{\alpha}{}^{\mu}$. Applying $\Box_{g}$ to \cref{coordinatesnearscri} we obtain
\begin{equation}
\label{Harmgaugewaveeq}
    \Box \xi^{\mu}=-\frac{1}{\sqrt{-g}}\partial_{\alpha}(\sqrt{-g}g^{\alpha\mu})-H^{\alpha\beta}\partial_{\alpha}\partial_{\beta}\xi^{\mu}-\frac{1}{\sqrt{-g}}\partial_{\alpha}(\sqrt{-g}g^{\alpha\beta})\partial_{\beta}\xi^{\mu}
\end{equation}
where, again, $\Box\equiv \eta^{ab}\partial_{a}\partial_{b}$, and
\begin{equation}
    H^{\alpha\beta}\equiv g^{\alpha\beta}-\eta^{\alpha\beta} \, .
\end{equation}
Here we have used the fact that, for any function $f$, 
\begin{equation}
\Box_{g}f=\frac{1}{\sqrt{-g}}\partial_{\alpha}(\sqrt{-g}g^{\alpha\beta}\partial_{\beta}f).
\end{equation}

In parallel with the analysis of imposition of the Lorenz gauge condition in linearized gravity, we will be able to put the metric in harmonic gauge in nonlinear gravity while maintaining our expansion ansatz if we can solve  \cref{Harmgaugewaveeq} via the ansatz 
\begin{equation}
\xi^{\mu} \sim \sum_{n= d/2-2}^\infty \frac{1}{r^n} \xi^{\mu \, (n)}(u, x^A),
\label{xiexpand}
\end{equation}
with $\partial_u \xi^{\mu \, (d/2-2)} = 0$. In odd dimensions, the sum in \cref{xiexpand} is allowed to run over integer values (starting at $d-3$) as well as half-integer values. For $d=4$, in the case of a stationary spacetime with Killing field $\partial/\partial u$, we may also add a term $c (\partial/\partial u)^ \mu \ln r$ to $\xi^\mu$, where $c$ is a constant.

To analyze existence of solutions to \cref{Harmgaugewaveeq} of the form \cref{xiexpand}, we note that \cref{Harmgaugewaveeq} is of the form
 \begin{equation}
 \label{Harmgaugedgreater4split}
     \Box \xi^{\mu}=\chi^{\mu}+L^{\mu}(h,\xi)
 \end{equation}
where, again, $\chi^{\mu}\equiv\partial_{\alpha}\bar{h}^{\alpha\mu}$, and where $L^{\mu}$ is composed of terms that are (i) quadratic and higher order in $h_{\mu \nu}$ or (ii) linear in $\xi^\mu$ and linear or higher order in $h_{\mu \nu}$. The leading order contribution of $L^\mu$ to this equation arises at order $1/r^{d-2}$. 

Consider, first, the case $d>4$. As noted previously, the non-linear contributions to Einstein's equation enter at order $1/r^{d-2}$, so the recursion relations derived for the linearized Einstein's equation given by \cref{linEinsteqsuu-comparb,linEinsteqsur-comparb,linEinsteqsrr-comparb,linEinsteqsuA-comparb,linEinsteqsrA-comparb,linEinsteqsAB-comparb} are equivalent to the recursion relations for the full, non-linear Einstein's equation for $n\leq d-3$. As already noted above, these equations imply that $\chi_{a}^{(d/2-1)}$ must vanish. It also follows that $\chi_{a}^{(d/2)}$ is stationary. It then follows that we can solve \cref{Harmgaugedgreater4split} at order $1/r^{d/2}$ by a choice of $\xi^{\mu \, (d/2-2)}$ that is stationary. We may then specify $\xi_{\mu}^{(d/2-1)}$ arbitrarily and recursively solve \cref{Harmgaugewaveeq} with the ansatz \cref{xiexpand} for all of the faster fall-off terms, in the same manner as in \Cref{first}. The source $L^{\mu}$ plays an innocuous role in this procedure since it is obtained from $\xi^{\mu}$ at orders that have already been solved for and thus is a ``known'' source term.

For the case $d=4$, we still have that $\chi_{a}^{(1)}=0$. In addition, since $\partial_u h_{rA}^{(1)} = 0$ (as follows from eqs.~(\ref{chi1}) and (\ref{chieqAcomp})), we may perform a gauge transformation of the form \cref{radgaugetrans} to set $h_{rA}^{(1)} = 0$. We then find that  $\chi_{r}^{(2)}$ and $\chi^{(2)}_{A}$ are stationary. However, $\chi^{(2)}_{u}$ now satisfies
\begin{equation}
\label{chiu2nonlin}
    \partial_{u}\chi_{u}^{(2)}=8\pi T_{uu}^{(2)} - {\mathcal G}^{(2)}_{uu}.
\end{equation} 
Using \cref{nonlineinstd-2} together with $h_{rA}^{(1)} = 0$, we obtain
\begin{equation}
\label{leadingordnonlinEindis4}
    \mathcal{G}^{(2)}_{uu}= -\frac{1}{4}N^{CD}N_{CD}+\frac{1}{2}\partial_{u} \left(C^{CD}N_{CD} \right)
\end{equation}
where $C_{AB}$ is the trace free part of the projection of $h_{ab}^{(1)}$ onto the sphere. 
However, \cref{Harmgaugedgreater4split} implies the $u$-component of the leading order term $\xi^{\mu \, (0)}$ satisfies
\begin{equation}
\label{xi0nonlin}
    \mathcal{D}^{2}\xi_{u}^{(0)}=\chi_{u}^{(2)}+N_{AB}C^{AB}
\end{equation}
and hence
\begin{equation}
    \mathcal{D}^{2}\left(\partial_u \xi_{u}^{(0)}\right)= 8\pi T_{uu}^{(2)} + \frac{1}{4}N^{CD}N_{CD} + \frac{1}{2} \partial_u \left(N_{AB}C^{AB} \right).
\end{equation}
Since $T^{(2)}_{uu} \geq 0$, if we assume that $N_{AB}$ vanishes as $u \to \pm \infty$, it is easily seen that we cannot have $\partial_u \xi_{u}^{(0)} = 0$ at all $u$ as required unless both $T_{uu}^{(2)}$ and $N_{AB}$ vanish identically. Thus, we cannot impose the harmonic gauge condition 
within our ansatz\footnote{We could impose the harmonic gauge condition for nonvanishing 
$T_{uu}^{(2)}$ or $N_{AB}$ for $d=4$
if we modified our ansatz to allow additional series involving terms of the form $(\ln r)^k/r^n$ \cite{Fock-book}.} if $T_{uu}^{(2)} \neq 0 $ or $N_{AB} \neq 0$.
On the other hand, if the spacetime is stationary---i.e. if it admits a timelike Killing field $t^{a}$---then $T_{uu}^{(2)} = 0$ and $N_{AB}=0$. Using the fact that the equations for $\xi_{r}^{(0)}$ and $\xi_{A}^{(0)}$ contain only ``source terms'' that are stationary---it can be seen that we can solve \cref{xi0nonlin} by choosing $\xi^{\mu(0)}$ to be stationary (provided that we again add the term $cg_{ab}t^{b}\ln(r)$ to our ansatz to solve the $\ell=0$ part of \cref{xi0nonlin}). The recursion relations for all faster fall-off can then be solved as in the case $d > 4$ so the harmonic gauge condition can be imposed\footnote{If the spacetime is non-stationary and $T_{uu}^{(2)} = 0=N_{AB}$ then we do not believe that the metric can be put in harmonic gauge within our ansatz, but we have not proven this.}.

We summarize these results in the following proposition:
\begin{prop}
\label{harmonicgrav}
For all $d>4$,
any $g_{ab}=\eta_{ab}+h_{ab}$ that satisfies our ansatz eq.~(\ref{hdevenan}) (for $d$ even) or our ansatz eq.~(\ref{hdoddan}) (for $d$ odd)
can be put in the harmonic gauge in such a way that it continues to satisfy our ansatz. In $d=4$ the harmonic gauge condition cannot be imposed within the ansatz if $T_{uu}^{(2)}\neq 0$ or $N_{AB}\neq 0$.
\end{prop}

\begin{rem}
{\em In linearized gravity, the restriction $T_{uu}^{(2)}=0$ in $d=4$ allows all vacuum solutions as well as all solutions with a stress-energy source that has vanishing flux at null infinity. Thus, the Lorenz gauge can be imposed in linearized gravity within our ansatz in a wide variety of circumstances of interest. However, in nonlinear general relativity, the harmonic gauge cannot be imposed within our ansatz in $d=4$ if---in addition to $T_{uu}^{(2)}\neq0$---the Bondi news is also nonvanishing, i.e., in $d=4$ the harmonic gauge cannot be imposed within our ansatz in any spacetime with gravitational radiation. In particular, for $d=4$ we cannot use the harmonic gauge when considering the memory effect in the next section, so we will have to treat the case $d=4$ separately.}
\end{rem}

When $g_{ab}$ is in the harmonic gauge, it satisfies
\begin{eqnarray}
G_{ab}^{H} = 8\pi T_{ab}    \label{EinEqs} \\
H^{b}\equiv\frac{1}{\sqrt{-g}}\partial_{b}[\sqrt{-g}g^{ab}] = 0 \label{Hgauge}
\end{eqnarray}
where $G_{ab}^{H}$ is the Einstein tensor in the harmonic gauge
\begin{equation}
\label{EinsteinisHarmplusgarmgauge}
  G_{ab}^{H} = G_{ab} + g_{c(a}\partial_{b)}H^{c}-\frac{1}{2}g_{ab}\partial_{c}H^{c}.
\end{equation}
We now turn to the issue of whether these equations can be solved recursively within our ansatz. We restrict consideration to $d >4$, since, as just remarked above, the harmonic gauge can be imposed only in trivial cases when $d=4$.

Taking the divergence of \cref{EinsteinisHarmplusgarmgauge} with respect to $\nabla_{a}$ and using the Bianchi identity we find that when $G_{ab}^{H} = 8\pi T_{ab}$, we have
\begin{equation}
\label{Harmgaugepropeq}
   \Box H_{a}=W_{a}(h,H)
\end{equation}
where $W_{a}$ is linear in $H_a$ and its first derivative and is quadratic and higher order in $h_{ab}$ and its first derivative. It follows that if $H^{(d/2 -1)}_a  = 0$ for all $u$ and $H^{(n)}_a  = 0$ for $n > d/2 -1$ at some $u=u_0$, then $H_a = 0$. Namely, if we inductively assume that $H^{(n)}_a  = 0$ for all $n \leq k$, then the source term arising from $W_a$ that appears in the recursion equation for $H^{(k+1)}_a$ will vanish. It then follows from the same arguments as used to prove \Cref{first} that $H^{(k+1)}_a =0$.

It is convenient to replace $H^a$ by
\begin{equation}
  {\tau'}_{a}=K^{b}[-2G^{(H)}_{ab}+16\pi T_{ab}]+2K^{b}\partial_{b}H_{a}+(d-2)H_{a}/r
\end{equation}
where the form of ${\tau'}_a$ has been chosen so that, for $n < d-3$, ${\tau'}_{a}^{(n+2)}$ can be expressed purely in terms of $h_{ab}^{(n)}$, with no $u$-derivatives of $h_{ab}$ appearing.
When eq. (\ref{EinEqs}) holds, the vanishing of ${\tau'}^{(n+1)}_{a}$ implies the vanishing of $H^{(n)}_{a}$. Thus we obtain a solution to eqs. (\ref{EinEqs}) and (\ref{Hgauge}) 
if we can solve eq.~(\ref{EinEqs}) in such a way that we also obtain $\tau'_a = 0$.

The recursion relations for eq. (\ref{EinEqs}) for $n<d-3$ are identical to \cref{linEinsteqsuu-comparb,linEinsteqsur-comparb,linEinsteqsrr-comparb,linEinsteqsuA-comparb,linEinsteqsrA-comparb,linEinsteqsAB-comparb} with $\chi_{a}=0$. In addition, we have ${\tau'}_{a}^{(n+2)}={\tau}^{(n+2)}_{a}$ for $n<d-3$, where ${\tau}_a$ is the corresponding quantity in linearized gravity given by \cref{linconstraint}. Thus, for $n < d-3$, the recursion relations and constraints are identical to the linearized case. It follows that if one specifies data at radiative order, one may solve the recursion relations for $h^{(n)}_{ab}$ for all $n<d-3$ exactly as in the linearized case. The recursion relations and constraints needed to solve for $h^{(n)}_{ab}$ for $n \geq d-3$ receive nonlinear corrections relative to the linearized equations. However, the nonlinear terms entering the equations will be of the form of products of metric perturbations arising at lower orders. 
Consequently, the non-linear terms can be effectively treated as source terms in our recursive analysis and they pose no difficulties in solving for $h^{(n)}_{ab}$ for $n \geq d-3$. We thereby obtain the following theorem:

\begin{thm}
\label{theoremcharinitprblmnonlingrav}
Suppose $d>4$ so that, by \Cref{harmonicgrav}, the harmonic gauge condition can be imposed. Then a unique solution to the recursion relations and constraints for the Einstein's equation in the harmonic gauge is obtained by the following specification of data: Specify $h_{ab}^{(d/2-1)}(u,x^{A})$ subject to the constraints $\tau'^{(d/2 +1)}_a = 0$ (which are identical to eqs.~(\ref{cu})-(\ref{cA}) at $n = d/2 - 1$). Specify $h_{ab}^{(n)}(u=u_{0},x^{A})$ for all $n>d/2-1$ subject to the constraints ${\tau'}^{(n+2)}_{a}=0$ at $u=u_{0}$. In odd dimensions, also specify $\tilde{h}_{ab}^{(p)}(u=u_{0},x^{A})$ for all $p\geq d-3$ subject to the constraint $\tau'^{(p+2)}_{a}=0$ at $u=u_{0}$. 
\end{thm}

Note that there is no analog of the ``Coulombic order data specification'' method for getting a solution of the recursion relations in nonlinear general relativity, since the Bondi news enters the equations for the metric at Coulombic order. Thus, we need to know the solution at radiative order before we can determine whether $h_{ab}^{(d-3)}(u,x^{A})$ is a solution to the recursion relations and constraints. 

Finally, it is worth noting that the analog of \cref{bondimassbalancelawlin} in nonlinear general relativity for $d>4$ is
\begin{equation}
\label{Bondimassbalancenonlin}
    \partial_{u}\mathcal{M}=- \mathcal{A}_{d}T_{uu}^{(d-2)}\vert_{\ell=0}-\frac{1}{32\pi}\mathcal{A}_{d}N^{AB}N_{AB}\vert_{\ell=0}
\end{equation}
where in even dimensions
\begin{equation}
\label{Bondimassnonlineven}
    \mathcal{M}= \frac{1}{16\pi}\mathcal{A}_{d}[\bar{h}_{ur}^{(d-3)}+(d-4)\bar{h}_{uu}^{(d-3)}-C^{AB}N_{AB}]\vert_{\ell=0} \quad \textrm{$d$ even (in harmonic gauge),}
\end{equation}
and in odd dimensions
\begin{equation}
\label{Bondimassnonlinodd}
    \mathcal{M}= \frac{1}{16\pi}\mathcal{A}_{d}[\bar{\tilde{h}}_{ur}^{(d-3)}+(d-4)\bar{\tilde{h}}_{uu}^{(d-3)}-C^{AB}N_{AB}]\vert_{\ell=0} \quad\textrm{$d$ odd (in harmonic gauge)}
\end{equation}
where $\mathcal{A}_{d}$ is the area of a unit $(d-2)-$sphere given by \cref{areaunitsphere}. By the same arguments as given in the linearized case, $\mathcal M$ is the Bondi mass. Again, the above formulas for $\mathcal M$ apply only in harmonic gauge and thus cannot be applied when $d=4$ if $T_{uu}^{(2)}\neq 0$ or $N_{AB}\neq 0$. A gauge invariant expression for the Bondi mass in all even dimensions $d\geq 4$ was given in \cite{HI}. Positivity of the Bondi mass in even dimensions was proven in \cite{Hollands_Thorne}.

\newpage

\section{The Memory Effect}
\label{Memeffect}
We now turn our attention to the analysis of the memory effect in nonlinear general relativity in $d \geq 4$ dimensions. 
In physical terms, the memory effect can be described as the permanent relative displacement resulting from the passage of a ``burst of gravitational radiation'' of a system of test particles that are initially at rest. 
The relative displacement of test particles is governed by the geodesic deviation equation
\begin{equation}
\label{geodev}    (v^{a}\nabla_{a})^{2}\xi^{b}=-R_{acd}{}^{b}v^{a}v^{d}\xi^{c}
\end{equation}
where $v^{a}$ is the tangent the worldline of the test particle, $\xi^{a}$ is the deviation vector and $R_{abcd}$ is the Riemann tensor. In our case, we will be interested in test particles near future null infinity and wish to determine the leading order memory effect in a $1/r$ expansion.

We note that there are closely analogous ``memory effects'' for electromagnetic and scalar fields \cite{BG_Scalar_Mem,GHITWeven,SWodd}. For the electromagnetic field or the scalar field, the memory effect would correspond to a charged particle with electric or scalar charge, originally at rest, getting a momentum kick after the passage of a burst of electromagnetic or scalar radiation. However, since we now have fully developed the machinery for the gravitational case, we will bypass the analysis of these other cases and go directly to the analysis of the memory effect in general relativity.

\subsection{Stationarity Conditions at Early and Late Retarded Times}
\label{statcond1}

Our first task in analyzing the memory effect is to define more precisely what we mean by a ``burst of gravitational radiation,'' i.e., to specify the stationarity conditions that we will assume hold at early and late retarded times.

We wish to consider spacetimes where there is significant gravitational radiation near future null infinity only over some finite range of retarded time. We envision this radiation as arising from ``localized event'' in the interior of the spacetime involving the interaction of matter and/or black holes and/or gravitational waves---although our entire analysis will be done near future null infinity and will not make any assumptions about the source of the gravitational radiation. Thus, we wish to consider a situation where the metric is (nearly) stationary at early retarded times and again becomes (nearly) stationary at late retarded times. 

However, it would be much too strong a condition to demand that the metric becomes stationary at early and late retarded times at all orders in $1/r$. This is because we wish to allow for the presence of bodies of matter (or black holes) that move inertially from/towards infinity at early/late retarded times. To see the implications of this, we note that a static multipole of angular order $\ell$ will decay as $r \to \infty$ at fixed global inertial time $t$ as $1/r^{\ell + d-3}$. However, for inertially moving bodies, the $\ell$th multipole moment will grow with time as $t^{\ell}$. Thus, near future null infinity, there will be contributions from the $\ell$th multipole solution that result in $h_{\mu \nu}$ behaving as\footnote{The $\ell$th multipole solution with leading order time dependence \cref{hcoul} will also have  terms that behave as $t^{\ell-2k}/r^{\ell-2k+d-3}$ with $k$ integer and $2k \leq \ell$, which also will contribute to the field at future null infinity in the same manner as indicated in \cref{hcoul}.}
\begin{equation}
h_{\mu \nu} \sim \frac{t^\ell}{r^{\ell + d-3}} = \frac{(u+r)^\ell}{r^{\ell + d-3}} = \frac{1}{r^{d-3}} + \frac{\ell u}{r^{d-2}}  + \dots
\label{hcoul}
\end{equation}
Thus, the leading order behavior of $h_{\mu \nu}$ is Coulombic---but note that $h_{\mu \nu}$ is {\em not} spherically symmetric near null infinity at Coulombic order. Although $h_{\mu \nu}$ is stationary at Coulombic order, it is, in general, non-stationary for $\ell \geq 1$ at order $1/r^{d-2}$. This non-stationarity can be removed for $\ell = 1$ by Lorentz boosting to a frame where the center of mass of the matter is at rest, but $h_{\mu \nu}$ will, in general, be genuinely non-stationary at order $1/r^{d-2}$ for $\ell \geq 2$.

The late time behavior near null infinity in curved spacetime with matter (or black holes) inertially moving to infinity along timelike trajectories cannot be expected to satisfy a stronger stationarity condition than would hold for inertially moving bodies in Minkowski spacetime. Indeed, as we shall see in the next subsection, if we were to require stationarity at order $1/r^{d-2}$ at both late and early retarded times, we would entirely exclude the ``ordinary memory'' effect. On the other hand, we do not believe that we would exclude any interesting phenomena by assuming that the metric becomes stationary at Coulombic order at early and late retarded times. 

We will therefore adopt as our stationarity condition that, in some gauge within our ansatz, the metric becomes stationary at Coulombic order and slower fall-off at early and late retarded times. More precisely, in even dimensions we require that there exist a gauge in which 
\begin{equation}
\label{evendstatcond}
\partial_u h^{(n)}_{\mu \nu} \to 0 \quad {\rm as} \quad u \to \pm \infty \textrm{ for $n\leq d-3$}, 
\end{equation}
and in odd dimensions we require that there exist a gauge in which
\begin{align}
\label{odddstatcond1}
    &\partial_u h^{(n)}_{\mu \nu} \to 0 \quad {\rm as} \quad u \to \pm \infty \textrm{ for $n< d-3$}\\
    &\partial_{u}\tilde{h}^{(d-3)}_{\mu \nu} \to 0 \quad {\rm as} \quad u \to \pm \infty.
    \label{odddstatcond2}
\end{align}

It follows immediately from these conditions that in the stationary eras, the nonlinear terms in Einstein's equation are $O(1/r^{2(d-2)})$ and will not enter the equations to the orders to which we will work. In addition, stationarity at Coulombic order implies that the Bondi mass (which can be defined in any gauge) is time independent, which implies that $T_{uu}^{(d-2)}\vert_{\ell=0}$  vanish in the stationary eras. However, positivity of $T_{uu}^{(d-2)}$ then implies that $T_{uu}^{(d-2)} = 0$ and the dominant energy condition then implies that $T_{\mu \nu}^{(d-2)} = 0$. We can then apply \Cref{lgstat} to conclude that, for $d > 4$, without loss of generality, the fall-off conditions \cref{evendstatcond} or \cref{odddstatcond1,odddstatcond2} can be assumed to hold in a harmonic gauge, as we shall assume in the following. It then follows from 
\Cref{statcoulordcorr} that in even dimensions we have $h^{(n)}_{\mu \nu} = 0$ for all $n < d-3$, and in odd dimensions, $h^{(n)}_{\mu \nu} = 0$ for all $n < d-2$.

Finally, we note Madler and Winicour \cite{Mad_Winic} have imposed a ``weak stationarity condition'' in their treatment of the memory effect in linearized gravity in $4$ dimensions. Their condition effectively requires the metric to be stationary at order $1/r^2$, i.e., one order faster fall-off than Coulombic.
Thus, their condition is stronger than ours. As we shall see in the next subsection, this stronger condition rules out all ``ordinary memory'' effects.

\subsection{The Memory Tensor and its Properties at Coulombic Order and Slower Fall-Off}
\label{coulmem1}

As discussed in the previous subsection, we wish to consider a spacetime where the metric near future null infinity is stationary at Coulombic order, $1/r^{d-3}$, at early and late retarded times. We 
consider an array of test particles near null infinity whose tangents $v^a$ initially point in the $(\partial/\partial u)^a$ direction. We wish to compute the memory effect for such test particles at all orders $n \leq d-3$. Since the metric differs from the Minkowski metric only at order $1/r^{d/2-1}$, the geodesic determined by $v^a$ will differ from the corresponding integral curve of $(\partial/\partial u)^a$ beginning only at order $1/r^{d/2-1}$, and $u$ will differ from an affine parametrization also beginning only at this order. Since the curvature also falls off as $r^{d/2-1}$, it can be seen that the deviations of $v^a$ from $(\partial/\partial u)^a$ in \cref{geodev} can affect $\xi^\mu$ only at order $r^{d-2}$ and faster fall-off. Since we consider only the memory effect at orders $n \leq d-3$, we may therefore replace $v^a$ in \cref{geodev} with $(\partial/\partial u)^a$, i.e., we may replace \cref{geodev} with
\begin{equation}
\label{geodev2}    \frac{\partial^2}{\partial u^2} \xi^{\mu}=-R_{u\nu u}{}^{\mu} \xi^{\nu}.
\end{equation}
Since, by our ansatz, $T_{ab} = O(1/r^{d-2})$, it follows immediately from Einstein's equation that the Ricci tensor vanishes at Coulombic order and slower fall off. Consequently, we may replace the Riemann tensor in \cref{geodev2} with the Weyl tensor. We also may replace $\xi^\nu$ on the right side of \cref{geodev2} with its initial value, $\xi^\nu_0$, since $\xi^\nu - \xi^\nu_0 = O(1/r^{d/2-1})$, so this difference cannot contribute to the right side at Coulombic and slower fall-off. Thus, at Coulombic and slower fall-off, we have
\begin{equation}
\label{geodev3}    \frac{\partial^2}{\partial u^2} \xi^{\mu}=-C_{u\nu u}{}^{\mu} \xi^{\nu}_0.
\end{equation}

Now suppose that the metric is stationary at Coulombic order and slower fall-off for $u \to \pm \infty$, as discussed in the previous subsection.
Integrating \cref{geodev3} twice, we obtain
\begin{equation}
\label{deltaxi}
\xi^{(n)\mu}\Big|_{u=-\infty}^{u=\infty}={\Delta^{(n)\mu}}_\nu \, \xi^\nu_0 \quad \quad \textrm{ for $n\leq d-3$}
\end{equation}
where
\begin{equation}
\label{memtensornthord}
    \Delta^{(n)}_{\mu \nu} \equiv - \int_{-\infty}^{\infty}du^{\prime}\int_{-\infty}^{u^{\prime}}du^{\prime\prime}C^{(n)}_{u\nu u \mu}.
\end{equation}
We refer to $ \Delta^{(n)}_{\mu \nu}$ as the $n$-th order {\em memory tensor}. It characterizes the memory effect at order $1/r^n$. We note that the Weyl tensor at these orders is equivalent to the linearized Weyl tensor and is gauge invariant. Therefore, the memory effect at these orders is manifestly gauge invariant.

It follows immediately from its definition, \cref{memtensornthord}, that for all $n \leq d-3$ the memory tensor, $ \Delta^{(n)}_{\mu \nu}$, is symmetric, trace-free, and has vanishing $u$-components, 
\begin{equation}
 \Delta^{(n)}_{\mu \nu} = {\Delta^{(n)}_{\nu \mu} \, , \quad \Delta^{(n)\mu}}_\mu = 0 \, , \quad
 \Delta^{(n)}_{u \nu} = 0 \quad \quad  {\rm for \, all} \, n \leq d-3.
\end{equation}
Obviously, from its definition, $\Delta^{(n)}_{\mu \nu}$ does not depend on $u$, so we also have $\partial_u  \Delta^{(n)}_{\mu \nu} = 0$. 

Additional properties of $ \Delta^{(n)}_{\mu \nu}$ follow from the Bianchi identity. We remind the reader that the uncontracted Bianchi identity is 
\begin{equation}
\label{bianchi}
    \nabla_{[a}R_{bc]de}=0.
\end{equation}
Contracting over $a$ and $d$ yields
\begin{equation}
\label{divofRiem}
    g^{ad}\nabla_{a}R_{bcde}=2\nabla_{[b}R_{c]e}.
\end{equation}
Applying $g^{af}\nabla_{f}$ to \cref{bianchi} we obtain 
\begin{equation}
\label{wave1}
    \Box_{g}R_{bcde}+g^{fa}\nabla_{f}\nabla_{b}R_{cade}+g^{fa}\nabla_{f}\nabla_{c}R_{abde}=0.
\end{equation}
Commuting the derivatives in the second and third terms of \cref{wave1} and using \cref{divofRiem} we obtain 
\begin{align}
\label{waveRiem}
    \Box_{g} R_{bcde}=&4\nabla_{[b}\nabla_{|[d}R_{e]|c]}-2g^{af}g^{mn}R_{f[bc]m}R_{nade}-2g^{mn}R_{m[b}R_{c]nde}-2g^{af}g^{mn}R_{dmf[b}R_{c]ane}\nonumber \\
    &-2g^{af}g^{mn}R_{mef[b}R_{c]adn}.
    \end{align}
We also remind the reader that the Riemann tensor is related to the Weyl tensor by 
\begin{equation}
\label{RiemtoWeyl}
    R_{abcd}=C_{abcd}+\frac{4}{d-2}g_{[a|[c}R_{d]|b]}-\frac{2}{(d-1)(d-2)}Rg_{a[c}g_{d]b}.
\end{equation}

In {\em linearized gravity} with $R_{ab} = 0$, the above relations imply
\begin{equation}
\label{lingravdivweyl}
\partial^a C_{abcd} = 0 \quad \quad \quad \quad \textrm{(linearized gravity)}
\end{equation}
and
\begin{equation}
\label{lingravboxweyl}
\Box C_{abcd} = 0 \quad \quad \quad \quad \textrm{(linearized gravity).}
\end{equation}
These relations, of course, do not hold in nonlinear general relativity, and they also do not hold in linearized gravity when 
 $R_{ab} \neq 0$. However, let 
\begin{equation}
    E_{\mu \nu}\equiv C_{u\nu u \mu}
\end{equation}
so that $E_{\mu\nu}$ is the ``electric part'' of the Weyl tensor. Define $\mathcal{T}_{\mu}$ by
\begin{equation}
\label{constraintel}
    \mathcal{T}_{\mu}=K^{\nu}\Box E_{\mu\nu}-2K^{\nu}\partial_{\nu}\partial^\alpha E_{\alpha \mu}-(d-2)\partial^\alpha E_{\alpha \mu}/r
\end{equation}
where $K^{a} = (\partial/\partial r)^{a}$. In linearized gravity with $R_{ab} = 0$, we have $\mathcal{T}_{\mu} = 0$. 
Remarkably, we find\footnote{The peeling properties of the Weyl tensor \cite{Reall} (which are a consequence of our ansatz) were used to show this.} that in nonlinear general relativity with our ansatz for $h_{ab}$ and $T_{ab}$, we have 
$\mathcal{T}_{\mu}^{(n+2)} = 0$ for all $n\leq d-3$. Now, the formula \cref{constraintel} defining $\mathcal{T}_{\mu}$ is exactly the same as the formula \cref{linconstraint} defining $\tau_{\mu}^{(n+2)}$ under the substitution $h_{\mu \nu} \to E_{\mu \nu}$ and $T_{\mu \nu} \to 0$. Thus, $E_{\mu \nu}$ satisfies eqs.~(\ref{cr}) and (\ref{cA}) with vanishing right side for all $n \leq d-3$. (Equation (\ref{cu}) is trivial since the $u$-components of $E_{\mu \nu}$ vanish.) 
Integrating this equation twice with respect to $u$, we find that for $n\leq d-3$, $\Delta_{\mu\nu}^{(n)}$ satisfies 
\begin{align}
&[\mathcal{D}^{2}-(n-d+1)(n-d+2)]\Delta_{rr}^{(n)}+(2n-d)\mathcal{D}^{A}\Delta_{Ar}^{(n)}=0 \textrm{ for $n\leq d-3$}
\label{Mr}
\end{align}
\begin{align}
&[\mathcal{D}^{2}-(n-d+3)(n-d+2)+(2n-d+1)]\Delta_{rA}^{(n)}+2\mathcal{D}_{A}\Delta_{rr}^{(n)}+(2n-d+2)\mathcal{D}^{B}\Delta_{AB}^{(n)}=0 \textrm{ for $n\leq d-3$}
\label{MA}
\end{align}
where we used the fact that the trace of $\Delta_{\mu\nu}$ vanishes to relate $\Delta_{rr}$ to $q^{AB}\Delta_{AB}$. We note that \cref{Mr,MA} have nothing to do with the harmonic gauge condition and hold for $d = 4$ as well as $d > 4$.

These relations will be used in \Cref{nonscal} below. They also have the following important consequence. The spherically symmetric $(\ell =0)$ part of $\Delta_{\mu\nu}$ automatically has $\Delta_{rA}=0$ and $\Delta_{AB}\propto q_{AB}$, since no vector on the sphere can be spherically symmetric and $q_{AB}$ is the only tensor of this index type that is spherically symmetric. Consequently,  \cref{Mr} implies that the spherically symmetric part of $\Delta_{\mu\nu}^{(n)}$ vanishes for $n\leq d-3$. Similar arguments also show that \cref{Mr,MA} imply that the $\ell=1$ part of $\Delta_{\mu\nu}^{(n)}$ vanishes for all $n\leq d-3$. This implies that 
\begin{equation}
\label{lgreat2}
[\Delta^{(n)}_{\mu \nu}] \lvert_{\ell=0,1} = 0 \quad \quad {\rm for \, all} \, n \leq d-3.
\end{equation}
In addition, in $d=4$ dimensions, \cref{Mr,MA} imply 
\begin{equation}
\label{4dcomp}
    \Delta^{(1)}_{r \nu} = 0
\end{equation}
and, similarly, in $d=6$ dimensions, we obtain
\begin{equation}
\label{6drr}
    \Delta^{(3)}_{rr} = 0.
\end{equation}
However, in higher dimensions, all components of the Coulombic order memory tensor (other than $u$ components and the trace) may be nonvanishing. These results in $d=4$ and $d=6$ dimensions also follow directly from the peeling properties of the Weyl tensor in these dimensions \cite{Reall}.

\subsection{Evaluation of the Memory Tensor at Coulombic Order and Slower Fall-Off}
\label{coulmem2}

We now evaluate $\Delta^{(n)}_{\mu \nu}$ for all $n \leq d-3$. We separately consider the cases (1) $d > 4$ and even, (2) $d$ odd, and (3) $d=4$. For $d > 4$, we impose the harmonic gauge condition to greatly simplify the analysis.

\subsubsection{$d$ even, $d>4$}

For $n \leq d -3$, the relevant components of the $n$th order Weyl tensor take the form
\begin{equation}
\label{nthorderweyl}
    C_{uaub}^{(n)}=\alpha^{(n)}_{ab}{}^{cd}h_{cd}^{(n-2)}+\beta^{(n)}_{ab}{}^{cd}\partial_{u}h_{cd}^{(n-1)}+\gamma^{(n)}_{ab}{}^{cd}\partial_{u}^{2}h_{cd}^{(n)}
\end{equation}
Here $\alpha^{(n)}_{ab}{}^{cd},\beta^{(n)}_{ab}{}^{cd},\gamma^{(n)}_{ab}{}^{cd}$ are given by
\begin{equation}
    \alpha^{(n)}_{ab}{}^{cd}=-\frac{1}{2}(n-1)(n-2)r_{a}r_{b}n^{c}n^{d}+(n-2)n^{c}n^{d}r_{(a}\mathcal{D}_{b)}-\frac{1}{2}n^{c}n^{d}\mathcal{D}_{a}\mathcal{D}_{b}+\frac{1}{2}(n-2)q_{ab}n^{c}n^{d},
\end{equation}
\begin{align}
    \beta^{(n)}_{ab}{}^{cd}=&-(n-1)r_{a}r_{b}n^{(c}K^{d)}+n^{(c}K^{d)}r_{(a}\mathcal{D}_{b)}-nr_{(a}n^{(c}q_{b)}{}^{d)}+n^{c}q_{(b}{}^{d}\mathcal{D}_{a)} \nonumber \\
    &+\frac{1}{2}q_{ab}(2n^{(c}K^{d)}-n^{c}n^{d}) ,
\end{align}
\begin{equation}
    \gamma^{(n)}_{ab}{}^{cd}=-\frac{1}{2}r_{a}r_{b}K^{c}K^{d}-r_{(a}K^{(c}q_{b)}{}^{d)}-\frac{1}{2}q_{a}{}^{c}q_{b}{}^{d}
\label{gamma}
\end{equation}
where $K^{a}=(\partial/\partial r)^{a}$, $n^{a}=(\partial/\partial u)^{a}$ and $r_{a}=(dr)_{a}$.

We now use the recursion relations to eliminate $h_{ab}^{(n-2)}$ and $h_{ab}^{(n-1)}$ in favor of $h_{ab}^{(n)}$ in \cref{nthorderweyl}. We consider, first, the case $n < d -3$; we will treat the case $n = d-3$ after we have completed the analysis for $n < d-3$.

For $n < d-3$, the relevant recursion relations do not contain any nonlinear terms in $h_{ab}$ and are thus given by 
eqs.~(\ref{linEinsteqsuu-comparb})-(\ref{chieqAcomp}) with $\chi^a = 0$. In addition, the stress-energy tensor does not appear in any equations at the orders relevant to this analysis. It is clear from the arguments that led to \Cref{theoremcharinitprblmgrav} that it must be possible to eliminate $h_{ab}^{(n-2)}$ and $h_{ab}^{(n-1)}$ in favor of $h_{ab}^{(n)}$, but it is useful to have an explicit construction, which we now give. 

First, we can directly invert the angular operator appearing in eq.~(\ref{linEinsteqsuu-comparb}) to solve for $\bar{h}_{uu}^{(n-1)}$ in terms of $\bar{h}_{uu}^{(n)}$. Explicitly, we have 
\begin{equation}
{\bar{h}}_{uu}^{(n-1)} = - (2n-d+2)\big[\mathcal{D}^{2} +(n-1)(n-d+2)\big]^{-1} \partial_{u}{\bar{h}}_{uu}^{(n)}.
\end{equation}
Note that $\bar{h}_{uu}^{(n)}$ appears in this solution only in the form $\partial_u \bar{h}_{uu}^{(n)}$. We then iterate this procedure to obtain $\bar{h}_{uu}^{(n-2)}$ in terms of $\bar{h}_{uu}^{(n-1)}$ and thence $\bar{h}_{uu}^{(n)}$, thereby expressing $\bar{h}_{uu}^{(n-2)}$ in terms of inverse angular operators applied to $\partial^2_u \bar{h}_{uu}^{(n)}$. Next, we eliminate $\mathcal{D}^{A}\bar{h}_{Au}^{(n-1)}$ using \cref{chiequcomp} (with $\chi_{u}^{(n)}=0$) and substitute into eq.~(\ref{linEinsteqsur-comparb}). The resulting equation can then be solved for $\bar{h}_{ur}^{(n-1)}$ in terms of $\partial_u \bar{h}_{ur}^{(n)}$ and $\partial_u \bar{h}_{uu}^{(n)}$. Iterating, we obtain $\bar{h}_{ur}^{(n-2)}$ in terms of $\partial^2_u \bar{h}_{ur}^{(n)}$ and $\partial^2_u \bar{h}_{uu}^{(n)}$. We then similarly invert eq.~(\ref{linEinsteqsuA-comparb}) to solve for $\bar{h}_{uA}^{(n-1)}$ and then $\bar{h}_{uA}^{(n-2)}$. 

Thus far, we have shown how to write the $uu$, $ur$, and $uA$ components of $\bar{h}_{ab}$ at orders $n-2$ and $n-1$ in terms of these components at $n$th order. To proceed further, we note that $\bar{h} \equiv \bar{h}^a_a = -2\bar{h}_{ur} + \bar{h}_{rr} + q^{AB}\bar{h}_{AB}$ satisfies the ordinary scalar wave equation. Hence, we can recursively solve for $\bar{h}^{(n-1)}$ and $\bar{h}^{(n-2)}$ in terms of $\partial_u \bar{h}^{(n)}$ and $\partial^2_u\bar{h}^{(n)}$ respectively. Then one can use eq.~(\ref{linEinsteqsrr-comparb}) and \cref{chieqrcomp} to obtain
\begin{equation}
\big[\mathcal{D}^{2}+(n-d+1)(n-2)]\bar{h}_{rr}^{(n-1)}=2(d-2n+2)\bar{h}_{ur}^{(n-1)}+2\bar{h}^{(n-1)}-(2n-d-2)\partial_{u}\bar{h}_{rr}^{(n)}.
\end{equation}
This equation can be used to solve for $\bar{h}_{rr}^{(n-1)}$ and $\bar{h}_{rr}^{(n-2)}$ in terms of $n$th order quantities. We can then use \cref{chieqAcomp} and eq.~(\ref{linEinsteqsrA-comparb}) to 
solve for $\bar{h}_{rA}^{(n-1)}$ and $\bar{h}_{rA}^{(n-2)}$ in terms of $n$th order quantities. Finally, we solve (\ref{linEinsteqsAB-comparb}) to obtain $\bar{h}_{AB}^{(n-1)}$ and $\bar{h}_{AB}^{(n-2)}$ in terms of $n$th order quantities. 

The above results show explicitly that we can write $h_{\mu \nu}^{(n-2)}$ as an operator (composed of inverses of angular operators and angular derivatives) applied to $\partial^2_u h_{\mu \nu}^{(n)}$. Similarly, we can write  $h_{\mu \nu}^{(n-1)}$ as such an operator applied to $\partial_u h_{\mu \nu}^{(n)}$. Substituting this result in \cref{nthorderweyl}, we see that for all $n < d-3$, the $n$th order Weyl tensor takes the form
\begin{equation}
\label{nthorderweyl2}
    C_{uaub}^{(n)}=O^{(n)}_{ab}{}^{cd}\partial_{u}^{2}\bar{h}_{cd}^{(n)}
\end{equation}
where the operator $O$ is constructed of inverses of angular operators and angular derivatives. 
It follows immediately from \cref{memtensornthord} that for $n < d-3$ the memory tensor takes the form
\begin{equation}
    \Delta_{\mu \nu}^{(n)}=P^{(n)}{}_{\mu \nu}{}^{\rho \sigma}[\Delta \bar{h}_{\rho \sigma}^{(n)}]  \quad \quad \textrm{ for $n<d-3$}
\end{equation}
where
\begin{equation}
    \Delta \bar{h}_{\mu \nu}^{(n)} \equiv \bar{h}_{\mu \nu}^{(n)}(u \to \infty) - \bar{h}_{\mu \nu}^{(n)}(u \to -\infty)
\end{equation}
and $P^{(n)}{}_{\mu \nu}{}^{\rho \sigma}$ is a linear operator constructed from inverses of angular operators and angular derivatives. 
However, as already remarked below \cref{odddstatcond2}, we have $h_{\mu \nu}^{(n)}=0$ for all $n < d-3$ when the metric is stationary at Coulombic order. Thus, the memory tensor vanishes at slower than Coulombic fall-off 
\begin{equation}
    \Delta_{\mu \nu}^{(n)}=0 \quad \quad \textrm{for $n < d-3$}.
\end{equation}
In particular, for $d>4$ the memory tensor vanishes at radiative order \cite{HIW}. 
\par Now consider the case $n= d-3$. The calculation $ \Delta_{\mu \nu}^{(d-3)}$ differs from the above calculation for $n < d-3$ only in that (i) $\Delta \bar{h}_{\mu \nu}^{(d-3)}$ need not vanish and (ii) the recursion relations eqs.~(\ref{linEinsteqsuu-comparb})-(\ref{linEinsteqsAB-comparb}) used to solve for $h_{\mu \nu}^{(d-4)}$ will now contain the additional terms $T_{\mu \nu}^{(d-2)}$ and ${\mathcal G}_{\mu \nu}^{(d-2)}$ (see  \cref{nonlinsub}). With regard to these additional terms the only nonvanishing component of $T_{\mu \nu}^{(d-2)}$ is $T_{uu}^{(d-2)}$. Similarly, it can be seen from \cref{nonlineinstd-2} that all of the components of ${\mathcal G}_{\mu \nu}^{(d-2)}$ except ${\mathcal G}_{uu}^{(d-2)}$ are $u$-derivatives of quantities that vanish in stationary eras. It is not difficult to show that the total $u$-derivative terms do not contribute to $ \Delta_{\mu \nu}^{(d-3)}$ under our stationarity conditions. Thus, the terms involving $T_{\mu \nu}^{(d-2)}$ and ${\mathcal G}_{\mu \nu}^{(d-2)}$ give rise to additional terms in the memory tensor that are proportional to the integral of the total flux, $F$, of matter and gravitational energy to null infinity 
\begin{equation}
\label{nullflux}
    F\equiv  T_{uu}^{(d-2)} + \frac{1}{32\pi}  N^{AB} N_{AB}. 
\end{equation}

Carrying through the calculation of $ \Delta_{\mu \nu}^{(d-3)}$ in the manner described above, we obtain the final formula
\begin{equation}
\label{memsplitordnull}
    \Delta^{(d-3)}_{\mu \nu}=P_{\mu \nu}[\Delta \bar{h}_{\rho \sigma}^{(d-3)}]_{\ell>1}+\int^\infty_{-\infty} du L_{\mu \nu}[F]_{\ell>1}
\end{equation}
where
\begin{align}
\label{ordmemoperator}
   P_{\mu\nu}[\Delta\bar{h}_{\rho\sigma}^{(d-3)}]=&\frac{1}{2}r_{\mu}r_{\nu}\bigg[(d-3)(d-4)^{2}(d-6)\mathcal{D}_{5}^{-2}\mathcal{D}_{4}^{-2}\Delta\bar{h}^{(d-3)}_{uu}-2(d-4)(d-6)\mathcal{D}_{5}^{-2}\Delta \bar{h}_{ru}^{(d-3)}\nonumber \\
    &+\frac{(d-4)^{2}}{d-2}\big((d-5)(d-6)D_{5}^{-2}-2\big)D_{4}^{-2}\Delta\bar{h}^{(d-3)}+\Delta\bar{h}_{rr}^{(d-3)}\bigg]\nonumber \\
    &-(d-4)^{2}(d-6)r_{(\mu}\mathcal{D}_{\nu)}\mathcal{D}_{5}^{-2}\mathcal{D}_{4}^{-2}\Delta\bar{h}_{uu}^{(d-3)}-2(d-3)(d-4)\mathcal{D}_{3}^{-2}r_{(\mu}\mathcal{D}_{\nu)}\mathcal{D}_{4}^{-2}\Delta\bar{h}_{uu}^{(d-3)}\nonumber \\
    &-2(d-3)(d-4)(d-6)\mathcal{D}_{3}^{-2}r_{(\mu}\mathcal{D}_{\nu)}\mathcal{D}_{5}^{-2}\mathcal{D}_{4}^{-2}\Delta\bar{h}_{uu}^{(d-3)} -(d-3)(d-4)r_{(\mu}q_{\nu)}{}^{\rho}\Delta\bar{h}_{\rho u}^{(d-3)}\nonumber \\
    &-\frac{d-4}{d-2}r_{(\mu}\mathcal{D}_{\nu)}\big((d-5)(d-6)\mathcal{D}_{5}^{-2}-1\big)\mathcal{D}_{4}^{-2}\Delta\bar{h}^{(d-3)}+(d-6)r_{(\mu}\mathcal{D}_{\nu)}\mathcal{D}_{5}^{-2}\Delta\bar{h}_{ru}^{(d-3)}\nonumber \\
    &+2(d-6)(d-3)\mathcal{D}_{3}^{-2}r_{(\mu}\mathcal{D}_{\nu)}\mathcal{D}_{5}^{-2}\Delta\bar{h}_{ru}^{(d-3)}+r_{(\mu}q_{\nu)}{}^{\rho}\Delta\bar{h}_{\rho r}^{(d-3)}+\frac{1}{2}(d-4)\big(-(d-6)\mathcal{D}_{\mu}\mathcal{D}_{\nu}\mathcal{D}_{5}^{-2}\nonumber \\
    &+4\mathcal{D}_{(\mu}\mathcal{D}_{3}^{-2}\mathcal{D}_{\nu)}\mathcal{D}_{4}^{-2}-4(d-6)\mathcal{D}_{(\mu}\mathcal{D}_{3}^{-2}\mathcal{D}_{\nu)}\mathcal{D}_{5}^{-2}+q_{\mu\nu}+(d-6)(d-7)q_{\mu\nu}\mathcal{D}_{5}^{-2}\big)\mathcal{D}_{4}^{-2}\Delta\bar{h}_{uu}^{(d-3)}\nonumber \\
    &+\frac{1}{2}\frac{d-4}{d-2}\big(-(d-6)\mathcal{D}_{\mu}\mathcal{D}_{\nu}\mathcal{D}_{5}^{-2}-q_{\mu\nu}+(d-5)(d-6)q_{\mu\nu}\mathcal{D}_{5}^{-2}\big)\mathcal{D}_{4}^{-2}\Delta\bar{h}^{(d-3)}\nonumber \\
    &-\bigg(2(d-6)\mathcal{D}_{(\mu}\mathcal{D}_{3}^{-2}\mathcal{D}_{\nu)}\mathcal{D}_{5}^{-2}-(d-6)q_{\mu\nu}\mathcal{D}_{5}^{-2}-\frac{1}{d-2}q_{\mu\nu}\bigg)\Delta\bar{h}_{ru}^{(d-3)}-\frac{q_{\mu\nu}}{d-2}\Delta\bar{h}_{rr}^{(d-3)} \nonumber \\
    &+(d-4)\mathcal{D}_{(\mu}q_{\nu)}{}^{\rho}\Delta\bar{h}_{\rho u}^{(d-3)}+\frac{1}{2}\bigg(q_{\mu}{}^{\rho}q_{\nu}{}^{\sigma}-\frac{1}{d-2}q_{\mu\nu}q^{\rho\sigma}\bigg)\Delta \bar{h}_{\rho\sigma}^{(d-3)}
\end{align} 
and
\begin{align}
\label{nullmemoperator}
    L_{\mu\nu}=8\pi &\bigg[r_{\mu}r_{\nu}(d-3)(d-4)(d-6)\mathcal{D}_{5}^{-2}\mathcal{D}_{4}^{-2}-2\big((d-4)(d-6)r_{(\mu}\mathcal{D}_{\nu)}\mathcal{D}_{5}^{-2}-2(d-3)\mathcal{D}_{3}^{-2}r_{(\mu}\mathcal{D}_{\nu)}\nonumber \\
    &-2(d-3)(d-6)\mathcal{D}_{3}^{-2}r_{(\mu}\mathcal{D}_{\nu)}\mathcal{D}_{5}^{-2}\big)\mathcal{D}_{4}^{-2}+\big(-(d-6)\mathcal{D}_{\mu}\mathcal{D}_{\nu}\mathcal{D}_{5}^{-2}+4\mathcal{D}_{(\mu}\mathcal{D}_{3}^{-2}\mathcal{D}_{\nu)}\mathcal{D}_{4}^{-2}\nonumber \\
    &-4(d-6)\mathcal{D}_{(\mu}\mathcal{D}_{3}^{-2}\mathcal{D}_{\nu)}\mathcal{D}_{5}^{-2}+q_{\mu\nu}+(d-6)(d-7)q_{\mu\nu}\mathcal{D}_{5}^{-2}\big)\mathcal{D}_{4}^{-2}\bigg]F.
\end{align}
Here, in order to write these equations in a more compact form, we have introduced the notation
\begin{eqnarray}
\mathcal{D}_{3}^{2}\equiv [\mathcal{D}^{2}-(d-3)]\\
\mathcal{D}_{4}^{2}\equiv [\mathcal{D}^{2}-(d-4)]\\
\mathcal{D}_{5}^{2}\equiv [\mathcal{D}^{2}-2(d-5)].
\end{eqnarray}
The notation $[\cdot]_{\ell>1}$ in \cref{memsplitordnull} means that only the $\ell>1$ part of the quantity is to be taken. The memory tensor $\Delta_{\mu\nu}^{(d-3)}$ has only $\ell>1$ spherical harmonic parts (see \cref{lgreat2}). However, $\Delta \bar{h}_{\rho\sigma}^{(d-3)}$ and $F$ have $\ell=0,1$ parts. The $\ell = 0,1$ parts of $\Delta \bar{h}_{\rho\sigma}^{(d-3)}$ and $F$ should be excluded from \cref{ordmemoperator,nullmemoperator} for the computation of ordinary and null memory. 

\Cref{memsplitordnull} naturally splits the memory tensor into a ``null memory'' piece associated with the flux $F$ of stress-energy and/or Bondi news to null infinity, and an ``ordinary memory'' piece associated with the change in the metric in harmonic gauge at Coulombic order. The ordinary memory piece can be rewritten in terms of $\Delta E_{\mu\nu}^{(d-1)}=\Delta C_{\mu u \nu u}^{(d-1)}$ as follows\footnote{It should be possible to derive \cref{nullmemoperator,ordmemgaugeinv} directly from \cref{divofRiem,waveRiem}, bypassing the need to introduce the harmonic gauge. We have shown that such a derivation can be given in linearized gravity with sources.}:
\begin{align}
\label{ordmemgaugeinv}
    P_{\mu\nu}[\Delta \bar{h}_{\rho\sigma}^{(d-3)}]=&-r_{\mu}r_{\nu}(d-4)(d-6)\mathcal{D}_{5}^{-2}\mathcal{D}_{4}^{-2}\Delta E_{rr}^{(d-1)}+2(d-4)r_{(\mu}\mathcal{D}_{\nu)}(\mathcal{D}^{2}-2)\mathcal{D}^{-2}\mathcal{D}_{5}^{-2}\mathcal{D}_{4}^{-2}\Delta E_{rr}^{(d-1)} \nonumber \\
&-2(d-2)(d-4)\mathcal{D}_{3}^{-2}(\mathcal{D}^{2}-1)^{-1}r_{(\mu}[q_{\nu)}{}^{\lambda}-\mathcal{D}_{\nu)}\mathcal{D}^{-2}\mathcal{D}^{\lambda}]\Delta E_{r\lambda}^{(d-1)}\nonumber \\
&+d(d-2)q_{\mu}{}^{\rho}q_{\nu}{}^{\sigma}(\mathcal{D}^{2}-2)^{-1}\mathcal{D}_{-4}^{-2}\Delta E_{\rho\sigma}^{(d-1)}-2d(d-2)(\mathcal{D}^{2}-2)^{-1}\mathcal{D}_{-4}^{-2}\mathcal{D}_{(\mu}\mathcal{D}_{-3}^{-2}\mathcal{D}^{\lambda}\Delta E_{\nu)\lambda}^{(d-1)}\nonumber \\
    &+2d(d-2)(\mathcal{D}^{2}-2)^{-1}\mathcal{D}_{-4}^{-2}\mathcal{D}_{(\mu}\mathcal{D}_{-3}^{-2}\mathcal{D}_{\nu)}\mathcal{D}^{-2}\mathcal{D}^{\lambda}\mathcal{D}^{\kappa}\Delta E_{\lambda\kappa}^{(d-1)} \nonumber \\
    &-\frac{d(d-2)^{2}}{d-3}(\mathcal{D}^{2}-2)^{-1}\mathcal{D}_{-4}^{-2}\bigg(\mathcal{D}_{\mu}\mathcal{D}_{\nu}-\frac{1}{d-2}q_{\mu\nu}\mathcal{D}^{2}\bigg)\mathcal{D}_{-2}^{-2}\mathcal{D}^{-2}\mathcal{D}^{\lambda}\mathcal{D}^{\kappa}\Delta E_{\lambda\kappa}^{(d-1)}\nonumber \\ 
    & +(\mathcal{D}^{2}-2)^{-1}\mathcal{D}_{-4}^{-2}\bigg(2d\mathcal{D}_{(\mu}\mathcal{D}_{-3}^{-2}\mathcal{D}_{\nu)}+d\bigg(\mathcal{D}_{\mu}\mathcal{D}_{\nu}-\frac{1}{d-2}q_{\mu\nu}\mathcal{D}^{2}\bigg)\mathcal{D}_{-2}^{-2}+dq_{\mu\nu}\bigg)\Delta E_{rr}^{(d-1)}\nonumber \\
    &+2(d-2)\mathcal{D}_{(\mu}\mathcal{D}_{-3}^{-2}\mathcal{D}_{-5}^{2}\mathcal{D}_{3}^{-2}(\mathcal{D}^{2}-1)^{-1}[q_{\nu)}{}^{\lambda}-\mathcal{D}_{\nu)}\mathcal{D}^{-2}\mathcal{D}^{\lambda}]\Delta E_{r\lambda}^{(d-1)} \nonumber \\
    &+\frac{1}{d-3} \bigg(\mathcal{D}_{\mu}\mathcal{D}_{\nu}-\frac{1}{d-2}q_{\mu\nu}\mathcal{D}^{2}\bigg)[(d-6)-2(d-4)(\mathcal{D}^{2}-2)\mathcal{D}^{-2}]\mathcal{D}_{5}^{-2}\mathcal{D}_{4}^{-2}\Delta E_{rr}^{(d-1)}\nonumber \\
    &+\frac{(d-4)(d-6)}{d-2}q_{\mu\nu}\mathcal{D}_{5}^{-2}\mathcal{D}_{4}^{-2}\Delta E_{rr}^{(d-1)}
\end{align}
where
\begin{equation}
    \mathcal{D}_{-4}^{2}\equiv[\mathcal{D}^{2}+(d-4)] 
\end{equation}
\begin{equation}
    \mathcal{D}_{-5}^{2}\equiv[\mathcal{D}^{2}+(d-5)]
\end{equation}
\begin{equation}
    \mathcal{D}_{-3}^{2}\equiv[\mathcal{D}^{2}+(d-3)]
\end{equation}
\begin{equation}
   \mathcal{D}_{-2}^{2}\equiv [\mathcal{D}^{2}+(d-2)].
\end{equation}
Again, the The $\ell = 0,1$ parts of $\Delta E_{\rho\sigma}^{(d-1)}$ should be excluded from \cref{ordmemgaugeinv} for the computation of memory.

\par Since $F$ is gauge invariant, null memory is manifestly gauge invariant. Since $C_{\alpha\beta\gamma\delta}^{(d-1)}$ is gauge invariant in stationary eras, ordinary memory is also manifestly gauge invariant when expressed in the form of \cref{ordmemgaugeinv}. We shall see in \Cref{memd=4} that \cref{nullmemoperator,ordmemgaugeinv} also hold in $d=4$.

We now consider the effects on the memory tensor at Coulombic order of placing stronger stationarity conditions than those imposed by \cref{evendstatcond} on the metric at early and late times in even dimensions. Specifically, suppose we were to require that
\begin{equation}
\label{evendstatcondstrong}
\partial_u h^{(k)}_{\mu \nu} \to 0 \quad {\rm as} \quad u \to \pm \infty \textrm{ for $k\leq d-2$},
\end{equation}
i.e., suppose that we require stationarity at one order faster fall-off than Coulombic. Suppose that, in addition, we require
\begin{equation}
\label{statcondstrongstress}
T^{(d-1)}_{\mu \nu} \to 0 \quad {\rm as} \quad u \to \pm \infty  \, .
\end{equation}
In the stationary eras, the nonlinear terms in Einstein's equation are $O(1/r^{2(d-2)})$ and will not enter our analysis to the orders we consider. It then follows from \Cref{lgstat} that $\bar{h}^{(n)}_{\mu \nu}$ can be put in harmonic gauge such that in the stationary eras, we have $\partial_u h^{(n)}_{\mu \nu} = 0$ for all $n \leq d-2$. It further follows from \Cref{statcoulordcorr} that, in the stationary eras, $\bar{h}^{(n)}_{\mu \nu} = 0$ for all $n < d-3$. Furthermore, $h^{(d-3)}_{\mu \nu}$ satisfies 
eqs.~(\ref{linEinsteqsuu-comparb})-(\ref{linEinsteqsAB-comparb}) for $n=d-2$ with all terms involviing $\chi_\mu$, $u$-derivatives, and stress-energy put to zero. In addition, $h^{(d-3)}_{\mu \nu}$ satisfies eqs.~(\ref{cu})-(\ref{cA}) for $n=d-3$ with vanishing stress-energy terms. It is not difficult to show that the unique solution to these equations is 
\begin{equation}
\label{schwarz}
\bar{h}^{(d-3)}_{uu} = \bar{h}^{(d-3)}_{ur} = \bar{h}^{(d-3)}_{rr} = {\textrm const.} 
\end{equation}
with all other components vanishing. This corresponds to the Schwarzschild solution in harmonic gauge at Coulombic order. Thus, with the stronger stationarity conditions \cref{evendstatcondstrong} and \cref{statcondstrongstress}, the solution approaches the Schwarzschild solution (possibly with different masses) at early and late retarded times. Thus, 
$\Delta h_{\mu\nu}^{(d-3)}$ has only an $\ell=0$ part, and cannot contribute to memory by \cref{lgreat2}. Thus, if the stronger stationarity conditions \cref{evendstatcondstrong} and \cref{statcondstrongstress} hold at early and late retarded times, then ordinary memory vanishes (but a nonvanishing null memory effect may still occur).

\subsubsection{$d$ odd}

For $d$ odd, the analysis of the memory effect for $n < d-3$---where $n$ is now half-integral---follows the even dimensional case exactly, and we find that 
\begin{equation}
    \Delta_{\mu \nu}^{(n)}=0 \quad \quad \textrm{ for $n < d-3$}.
\end{equation}

Since $\tilde{h}_{ab}^{(d-3)}$ is the leading order term in the integer power part of the expansion of $h_{ab}$ (see
eq.~(\ref{hdoddan})), the only contribution to $C_{uaub}^{(d-3)}$ is 
\begin{equation}
    C_{u\mu u\nu}^{(d-3)}=\gamma^{(d-3)}{}_{\mu \nu}{}^{\rho \sigma}\partial_{u}^{2} \tilde{h}_{\rho \sigma}^{(d-3)}
\label{weylodd}
\end{equation}
where the $\gamma^{(d-3)}{}_{\mu \nu}{}^{\rho \sigma}$ is given by \cref{gamma} with $n=d-3$. Einstein's equation in harmonic gauge yields 
\begin{equation}
    (d-4)\partial_{u}\bar{\tilde{h}}_{\mu \nu}^{(d-3)}=-16\pi T^{(d-2)}_{\mu \nu} + 2 {\mathcal G}^{(d-2)}_{\mu \nu}.
\end{equation}
However, we have
\begin{equation}
   T^{(d-2)}_{\mu \nu} = T^{(d-2)}_{uu} K_\mu  K_\nu
\end{equation}
and
\begin{equation}
   {\mathcal G}^{(d-2)}_{\mu \nu} = - \frac{1}{4} N^{AB} N_{AB} K_\mu  K_\nu + \partial_u B_{\mu \nu}
\end{equation}
where $B_{\mu \nu}$ vanishes in stationary eras. From \cref{gamma}, it is easily seen that $\gamma^{(d-3)}{}_{\mu \nu}{}^{\rho \sigma} K_\rho K_\sigma =0$. It can also be seen immediately from \cref{memtensornthord} and \cref{weylodd} that $B_{\mu \nu}$ cannot contribute to $\Delta_{\mu \nu}$. 
Thus, we find that for $d$ odd,
\begin{equation}
\label{EinseqCoulorder}
    \Delta_{\mu \nu}^{(d-3)}=0 \quad \quad \textrm{ for $d$ odd.}
\end{equation}
and thus the memory effect vanishes at Coulombic order (as well as slower fall-off) in odd dimensions.

At first sight, it may seem paradoxical that there is a major difference between odd and even dimensions in the memory effect at Coulombic order: First, in odd dimensions there is a flux of energy to null infinity at order $1/r^{d-2}$ in exact parallel with the even dimensional case, so why isn't there a null memory contribution at Coulombic order? Second, if one considers, e.g., the scattering of timelike particles, one would expect that the retarded solution at early and late times should behave like \cref{hcoul} at late and early times, potentially giving rise to a nonvanishing $\Delta h_{\mu \nu}$ at Coulombic order in odd dimensions. Why doesn't this give rise to an ordinary memory effect?

The answer to the first question is that the key difference that occurs in odd dimensions---as compared with even dimensions with $d>4$---is that terms with integer power fall-off slower than $1/r^{d-3}$ are not permitted. In even dimensions with $d > 4$, the possible presence of a nonvanishing $h_{\mu\nu}^{(d-4)}$ and  $h_{\mu\nu}^{(d-5)}$ effectively makes the null and ordinary memory independent. In odd dimensions, there can, indeed, be a null memory effect, but it is always exactly canceled by ordinary memory.

The answer to the second question is more subtle and has to do with the manner in which the retarded solution approaches a solution of the form \cref{hcoul} at late times for particles moving on inertial trajectories.
To see this, it is illuminating to consider the concrete example of the retarded solution, $\phi$, to the scalar wave equation \cref{scalar2} with source corresponding to the creation of a scalar particle with scalar charge $q$ at time $t=0$ at the origin in $5-$dimensional Minkowksi spacetime
\begin{equation}
S = q\theta(t) \delta^{(4)} (\vec{x}).
\end{equation}
The exact retarded solution for such a source is 
\begin{equation}
\label{phicreate}
    \phi =q\frac{\theta(u)}{(2\pi r)^{2}} \frac{r+u}{\sqrt{u(2r+u)}}.
\end{equation}
For $r\gg u$, \cref{phicreate} admits an expansion in half-integer powers of $1/r$ fully consistent with our ansatz eq.~(\ref{doddan})
\begin{equation}
\label{phiseriescreate}
    \phi = \frac{q}{2\sqrt{\pi}(2\pi r)^{3/2}}\frac{\theta(u)}{\sqrt{u}}+\frac{3\sqrt{\pi }q}{4(2\pi r)^{5/2}}\theta(u)\sqrt{u}+O(1/r^{7/2}).
\end{equation}
No integer powers of $1/r$ occur. 
In particular, at all retarded times, the scalar field vanishes at Coulombic order. In addition, as $u \to \infty$, we have $\phi^{(3/2)} \to 0$, so $\phi$ vanishes at late retarded time at Coulombic and slower fall-off. On the other hand, if we fix $r$ and take the limit of the exact solution \cref{phicreate} as $u \to \infty$, we obtain the Coulomb solution
\begin{equation}
    \lim_{u\rightarrow \infty}\phi = \frac{q}{4\pi^{2}r^{2}}.
\label{couls}
\end{equation}
In other words, $\phi$ approaches the Coulomb solution at {\em timelike infinity} ($u \to \infty$ at fixed $r$), but does {\em not} approach the Coulomb solution at {\em null infinity} ($r \to \infty$ at fixed $u$) even if $u$ is then taken to be arbitrarily large. In other words, the Coulomb solution \cref{couls} will not be evident to an observer unless he waits a time much longer than the light travel time to the source.

Similarly, in the gravitational case in any odd dimension $d \geq 5$, consider classical particle scattering wherein the particles move on timelike, inertial trajectories at early and late times. Then at early retarded times, the retarded solution at Coulombic order, $\tilde{h}^{(d-3)}_{\mu \nu}$, will have the multipolar structure corresponding to the incoming particles, as in \cref{hcoul}. However, 
except for $h_{uu}$, this multipolar structure will not change with $u$ and will remain the same as $u \to \infty$, i.e., $\Delta \tilde{h}^{(d-3)}_{\mu \nu} = 0$ except for $\mu = \nu = u$. The ordinary memory effect that may result from a nonvanishing $\Delta \tilde{h}^{(d-3)}_{uu}$ will be exactly canceled by the null memory effect.

Thus, the total memory effect vanishes at Coulombic order in odd dimensions. However, the above considerations suggest that it may be possible to define a notion of a memory effect at timelike infinity that would be nonvanishing.

\subsubsection{$d$ = 4}
\label{memd=4}
In dimension $d=4$, radiative and Coulombic order coincide, since $d/2 - 1 = d-3 = 1$.
Our analysis for $d > 4$ was based upon the imposition of the harmonic gauge, so it cannot be applied\footnote{However, our harmonic gauge analysis can be applied in linearized gravity to the case where $T^{(2)}_{\mu \nu} = 0$, in which case ordinary memory is possible.} in nonlinear gravity when $d=4$ if $T^{(2)}_{\mu \nu} \neq 0$ or $N_{AB} \neq 0$. Thus, we cannot impose the harmonic gauge, i.e., we cannot set $\chi_a = 0$ in eqs.~(\ref{linEinsteqsuu-comparb})-(\ref{chieqAcomp}), nor can we use the corresponding simplifications in calculating the nonlinear terms in Einstein's equation. Nevertheless, the properties of the metric perturbation at radiative order described in the paragraph below \cref{dec} 
still apply. In particular, the Bondi news tensor 
\begin{equation}
    N_{AB} = \left(q_{A}{}^{C}q_{B}{}^{D}-\frac{1}{2}q_{AB}q^{CD}\right)\partial_{u}h^{(1)}_{CD}
\end{equation}
is gauge invariant. Furthermore, at radiative (= Coulombic) order, the only components of the Weyl tensor that can be nonvanishing are the $uAuB$ components, which are given by
\begin{equation}
\label{4dweyl}
C^{(1)}_{uAuB} = - \frac{1}{2} \partial_u N_{AB} = - \frac{1}{2} \partial^2_u h^{(1)}_{AB}.
\end{equation}
Integrating this equation twice, we immediately obtain the following extremely simple formula for the memory tensor:
\begin{equation}
\label{delh1}
\Delta^{(1)}_{AB} =  \frac{1}{2} \Delta h^{(1)}_{AB}
\end{equation}
where $\Delta h^{(1)}_{AB}$ denotes the difference between $h^{(1)}_{AB}$ in the initial and final stationary eras. \Cref{delh1} holds in any gauge compatible with our ansatz.

To proceed further we use Einstein's equations eqs.~(\ref{linEinsteqsuu-comparb})-(\ref{chieqAcomp}) with $\chi_a$ {\em not} put to zero and with $T_{ab}^{(2)}$ replaced by $T_{ab}^{(2)} - \mathcal{G}_{ab}^{(2)}/8 \pi$. These equations can be simplified significantly by restricting consideration to the case $h_{rr}^{(1)} = 0$ (see \Cref{appnonlin}), in which case we can impose
the Bondi gauge conditions $h_{rr} = h_{rA} = 0$ and $\partial_{r}(\det(h_{AB}))=0$.
Einstein's equations do not directly yield an equation for $\partial_u h^{(1)}_{AB}$, but they do yield an equation for $\partial_u {\mathcal D}^B h^{(1)}_{AB}$, which can be integrated to obtain ${\mathcal D}^B\Delta^{(1)}_{AB}$ and thence $\Delta^{(1)}_{AB}$. We will not carry out the analysis here, as it has already been done by many authors\footnote{References \cite{Zeld,BG} worked in the context of linearized gravity whereas \cite{Eanna,Christ} analyzed the memory effect in four dimensions in full, nonlinear general relativity. References \cite{BG,Eanna,Christ} considered contributions to memory from null sources whereas \cite{Zeld} did not. References \cite{Zeld,Christ,Eanna,BG} all considered ordinary memory effects in which $\Delta Q=0$ in \cref{ordmem4}.} \cite{Zeld,Christ,Eanna,BG}. The final result is that the memory 
tensor in $4$ dimensions can be expressed as \cite{Christ,Eanna,BG}
\begin{equation}
\label{mem4d}
    \Delta_{AB}^{(1)}=[P_{AB}]_{\ell>1}+\int_{-\infty}^{\infty}duL_{AB}[F]_{\ell>1}.
\end{equation}
Here the ``ordinary memory'' (the first term in \cref{mem4d}) is given by 
\begin{equation}
\label{ordmem4}
    P_{AB}=-2\bigg(\mathcal{D}_{A}\mathcal{D}_{B}-\frac{1}{2}q_{AB}\mathcal{D}^{2}\bigg)\mathcal{D}^{-2}(\mathcal{D}^{2}+2)^{-1}\Delta P+2\epsilon_{(A}{}^{C}\mathcal{D}_{B)}\mathcal{D}_{C}\mathcal{D}^{-2}(\mathcal{D}^{2}+2)^{-1}\Delta Q
\end{equation}
where 
\begin{equation}
\label{P}
 P \equiv C_{urur}^{(3)},
\end{equation}
\begin{equation}
\label{Q}
    Q\equiv \frac{1}{2}\epsilon^{\mu\nu}C^{(3)}_{\mu\nu ru},
\end{equation}
and $\Delta P$ and $\Delta Q$ correspond to the difference in these quantities at early and late retarded times. Only the $\ell > 1$ parts of $\Delta P$ and $\Delta Q$ enter the formula for memory. The contributions to ordinary memory of $\Delta P$ and $\Delta Q$ are usually referred to as its ``electric parity" and ``magnetic parity" parts\footnote{As we shall see in the next subsection, the ``magnetic parity'' part is the same as the ``vector part'' in a spherical tensor decomposition.}, respectively. The ``null memory'' (the second term in \cref{mem4d}) is given by \cite{Christ,Eanna,BG} 
 \begin{equation}
 \label{nullmem4}
     L_{AB}[F]=16\pi\bigg(\mathcal{D}_{A}\mathcal{D}_{B}-\frac{1}{2}q_{AB}\mathcal{D}^{2}\bigg)\mathcal{D}^{-2}(\mathcal{D}^{2}+2)^{-1}F
 \end{equation}
 where $F$ is the total flux of matter and gravitational energy to null infinity---given by \cref{nullflux} with $d=4$---and only the $\ell > 1$ part is taken. \Cref{ordmem4,nullmem4} agree with \cref{ordmemgaugeinv,nullmemoperator} with $d$ set equal to $4$. 

Finally, suppose that we were to impose the strong stationarity conditions \cref{evendstatcondstrong} and \cref{statcondstrongstress} at early and late retarded times. Our analysis for $d > 4$ used the harmonic gauge, which we cannot assume here. However, the gauge freedom for the metric at order $1/r$ that preserves strong stationarity is given by \cref{radgaugetrans} with $d=4$, with the requirement that $\xi_{a}^{(0)}$ is stationary and $\xi_{a}^{(1)}$ vanishes. We can use up the full gauge freedom of $h_{ab}^{(1)}$ by setting $\eta^{ab}h_{ab}^{(1)} = 0$ and $h_{AB}^{(1)} = 0$. One then can show that Einstein's equations with these gauge conditions imply that 
when \cref{evendstatcondstrong} and \cref{statcondstrongstress} hold
the metric at Coulombic order (i.e. order $1/r$) must be Schwarzschild. The stronger stationarity conditions together with the field equations also imply that $h_{\mu\nu}^{(2)}$ and $h_{\mu\nu}^{(3)}$ do not contribute to $P$ or $Q$ as defined in \cref{P,Q}. Since $h_{\mu\nu}^{(1)}$ is spherically symmetric it follows that $Q=0$ and $P$ is spherically symmetric. Hence, as was the case for $d > 4$, we find that when $d=4$, the ordinary memory vanishes if the stronger stationarity conditions \cref{evendstatcondstrong} and \cref{statcondstrongstress} are imposed.
 
\bigskip

We summarize the main results of this subsection in the following theorem: 
\begin{thm}
\label{memthm}
Suppose $d \geq 4$ and the metric satisfies the stationarity condition \cref{evendstatcond} (for even dimensions) or \cref{odddstatcond1,odddstatcond2} (for odd dimensions) at early and late retarded times. Then the memory tensor, defined by \cref{memtensornthord}, has the following properties:
\begin{enumerate}
    \item In odd dimensions, $\Delta_{ab}^{(n)}=0$ for all $n \leq d-3$.
    \item In even dimensions, $\Delta_{ab}^{(n)}=0$ for all $n < d-3$. For $n=d-3$, the memory tensor can be decomposed into ``ordinary memory'' and ''null memory'' as in \cref{memsplitordnull}. For $d > 4$, the ordinary and null memory are given, respectively, by \cref{ordmemoperator} (or \cref{ordmemgaugeinv}) and \cref{nullmemoperator}. For $d=4$, the ordinary and null memory are given, respectively, by \cref{ordmem4} and \cref{nullmem4}. If one imposes the stronger stationarity conditions \cref{evendstatcondstrong} and \cref{statcondstrongstress} at early and late retarded times, then the ordinary memory vanishes at Coulombic order (but null memory may still be nonvanishing at Coulombic order). 
\end{enumerate}
\end{thm}

\subsection{Non-scalar Memory}
\label{nonscal}

As proven in \cite{IW} (see Propositions 2.1 and 2.2 of that reference), any (co)-vector field, $v_A$, on a sphere in $(d-2)$-dimensions can be decomposed into its vector and scalar parts via
\begin{equation}
    w_A =W_{A}+\mathcal{D}_{A} W
\end{equation}
where $\mathcal{D}^{A}W_{A}=0$. Any symmetric tensor field, $x_{AB}$ on the sphere can be decomposed into its tensor, vector, and scalar parts via
\begin{equation}
    x_{AB}=X_{AB}+\mathcal{D}_{(A}X_{B)}+\bigg(\mathcal{D}_{A}\mathcal{D}_{B}-\frac{1}{d-2}q_{AB}\mathcal{D}^{2}\bigg)X+\frac{1}{d-2}q_{AB}Y
\end{equation}
where $\mathcal{D}^{A}X_{AB}=0=q^{AB}X_{AB}$ and $\mathcal{D}^{A}X_{A}=0$. Any rotationally invariant operator (such as ${\mathcal D}^2$) acting on $w_A$ or $x_{AB}$ maps the scalar, vector, and tensor parts into themselves, i.e., rotationally invariant operations cannot ``mix'' these different parts. 

Thus, the Coulombic order memory tensor $\Delta^{(d-3)}_{\mu \nu}$ may be decomposed into its scalar, vector, and tensor parts via
\begin{equation}
\label{srr}
    \Delta_{rr}^{(d-3)}=-U
\end{equation}
\begin{equation}
    \Delta_{rA}^{(d-3)}=R_{A}+\mathcal{D}_{A}R
\end{equation}
\begin{equation}
\label{tendecom}
    \Delta_{AB}^{(d-3)}=S_{AB}+\mathcal{D}_{(A}S_{B)}+\bigg(\mathcal{D}_{A}\mathcal{D}_{B}-\frac{1}{d-2}q_{AB}\mathcal{D}^{2}\bigg)T+\frac{1}{d-2}q_{AB}U
\end{equation}
where $\mathcal{D}^{A}R_{A}=0=\mathcal{D}^{A}S_{A}$ and $\mathcal{D}^{A}S_{AB}=0=q^{AB}S_{AB}$. Note that the fact that $\Delta^{(d-3)}_{\mu \nu}$ is traceless was used to relate $\Delta_{rr}^{(d-3)}$ to the scalar function $U$ appearing in \cref{tendecom}. In $d=4$ dimensions, the tensor part, $S_{AB}$, in \cref{tendecom} vanishes, since there are no divergence-free, trace-free, symmetric, rank$-2$ tensors on $\mathbb{S}^{2}$. Furthermore, on $\mathbb{S}^{2}$, the vector part $S_{A}$ can always be written as $S_{A}=\epsilon_{AB}\mathcal{D}^{B}S$. Thus, in $d=4$ dimensions, the ``vector part" can be replaced by a ``magnetic parity scalar'' part. In addition, since $\Delta^{(1)}_{r\mu} = 0$ in 4 dimensions (see \cref{4dcomp}), we also have $U=R_A=R=0$ when $d=4$. For $d=6$, we have $U=0$ (see \cref{6drr}).

We shall refer to $U$, $R,$ and $T$ as ``scalar memory'', $R_{A}$ and $S_{A}$ as ``vector memory,'' and $S_{AB}$ as ``tensor memory''. The scalar functions $U$, $R,$ and $T$ are not independent because $\Delta^{(d-3)}_{\mu \nu}$ must satisfy the ``constraint equations'' \cref{Mr} and \cref{MA} with $n=d-3$. This yields
\begin{equation}
\label{Mscalar1}
[\mathcal{D}^{2}-2]U-(d-6)\mathcal{D}^{2}R=0
\end{equation}
and 
\begin{equation}
\label{Mscalar2}
    [\mathcal{D}^{2}+2(d-4)]R+\frac{1}{2}(d-4)[\mathcal{D}^{2}+2(d-3)]T-\frac{d}{d-2}U=0.
\end{equation}
Note that for $d=4$, this implies that $U=R=0$, so scalar memory takes the form
\begin{equation}
\label{scalmem}
[\Delta_{AB}^{(1)}]_{\rm scalar}=\bigg(\mathcal{D}_{A}\mathcal{D}_{B}-\frac{1}{2}q_{AB}\mathcal{D}^{2}\bigg)T \quad \quad \textrm{for $d=4$.}
\end{equation}
The vector part of \cref{Mr} vanishes, but \cref{MA} implies that $R_{A}$ and $S_{A}$ must satisfy
   \begin{equation}
   \label{Mvector}
   [\mathcal{D}^{2}+(d-5)]R_{A}+\frac{1}{2}(d-4)[\mathcal{D}^{2}+(d-3)]S_{A}=0.
   \end{equation}
The constraint equations (\ref{Mr}) and (\ref{MA}) do not give any restrictions on $S_{AB}$. 

We can use \cref{Mscalar1,Mscalar2} to solve for $U$ and $R$ in terms of $T$ and we can use \cref{Mvector} to solve for $R_A$ in terms of $S_A$.  Thus, the memory tensor is fully characterized by $T$, $S_A$, and $S_{AB}$, i.e., the trace-free part of the angle-angle components of the memory tensor.

No other obvious restrictions on $\Delta^{(d-3)}_{\mu \nu}$ arise from Einstein's equations near null infinity for $d$ even---of course, we have already shown that $\Delta^{(d-3)}_{\mu \nu}=0$ for $d$ odd. This suggests that---in addition to scalar memory---magnetic parity memory may be possible for $d=4$, and vector and tensor memory may be possible for $d > 4$ (for $d$ even). We now investigate whether this is possible for physically reasonable solutions.

Consider, first, null memory. The null memory part of $\Delta^{(d-3)}_{\mu \nu}$ is constructed from a rotationally invariant operator $L_{\mu \nu}$ (see \cref{nullmemoperator} and \cref{nullmem4}) acting on the integrated flux $F$. Since $F$ is a scalar function on the sphere, it follows immediately that null memory is always of purely scalar type.

The analysis of ordinary memory requires that we know that the Coulombic order solution $h^{(d-3)}_{\mu \nu}$ at early and late retarded times. This is not feasible in nonlinear general relativity but can be analyzed in linearized gravity. Consider, first, classical particle scattering, as treated in \cite{GHITWeven}. For classical particle scattering, the solution at early and late retarded times is a sum of boosted linearized Schwarzschild solutions. It is easily checked that for boosted, linearized Schwarzschild solutions, $h^{(d-3)}_{\mu \nu}$ is of purely scalar type. Since ordinary memory is obtained by applying a rotationally invariant operator to $\Delta h^{(d-3)}_{\mu \nu}$, it follows that ordinary memory is of purely scalar type for particle scattering in linearized gravity.

However, it is not difficult to show that vector and tensor ordinary memory can occur in linearized gravity for the retarded solution arising from other kinds of ingoing or outgoing matter stress-energy satisfying the dominant energy condition. In particular, magnetic parity (i.e, vector) ordinary memory can be produced in $d=4$. To see this, consider a stress-energy tensor (for $t = u+r > 0$)
\begin{equation}
\label{shellstress}
    T_{ab}=\frac{1}{r^{2}}[\rho u_{a}u_{b}+l_{ab}] \delta(r-vt)
\end{equation}
where $\rho > 0$ is a constant, $u^a$ corresponds to a radially outward $4$-velocity with velocity $1>v>0$, and the components of $l_{ab}$ in a Cartesian basis (or normalized spherical basis) are independent of $t$ and $r$ and, on the unit sphere, are given by
\begin{equation}
\label{magparitypartofstress}
    l_{\mu \nu}=-\epsilon_{(\mu}{}^{\lambda}[u_{\nu)}(\mathcal{D}^{2}+1)-\gamma v\mathcal{D}_{\nu)}]\mathcal{D}_{\lambda}\alpha.
\end{equation}
Here $\alpha$ is a time independent, arbitrary function on the sphere (containing multipoles $l > 1$) and $\gamma\equiv(1-v^{2})^{-1/2}$. For $\alpha = 0$, \cref{shellstress} would correspond to an outgoing spherical dust shell and its stress-energy would be conserved. The $l_{ab}$ term has been constructed so that it is purely of magnetic parity (i.e., vector) type and is conserved by itself, so its addition to the stress-energy tensor does not affect conservation. By choosing $\rho$ sufficiently large, we can ensure that $T_{ab}$ satisfies the dominant energy condition. Thus, we see no principle that would imply that a stress-energy tensor of the form \cref{shellstress} is not physically possible. 

Since we are considering linearized gravity and there is no stress energy flux to null infinity, we may work in the Lorenz gauge. In a Cartesian basis, each component of $\bar{h}_{\mu \nu}$ satisfies the ordinary scalar wave equation with source. At radiative order, the contribution of $l_{ab}$ to the retarded solution for $u > 0$ is independent of $u$ and is given by 
\begin{equation}
    \bar{h}^{(1)}_{\mu \nu} (x^A) = 8\pi\int d \Omega' \frac{l_{\mu \nu} (x'^A)}{1 - v \hat{r}(x^A) \cdot \hat{r}(x'^A)}
\end{equation}
where the integral is taken over a sphere and $\hat{r}$ denotes the unit radial vector (with parallel transport in Euclidean space is understood in taking the dot product of vectors at different points on the sphere). It can be seen that $h^{(1)}_{AB}$ is, in general, nonvanishing. It must be of purely vector type since the source is of purely vector type and the retarded Green's function is rotationally invariant. 

Now suppose one starts in the distant past with a static laboratory and no incoming gravitational radiation. At retarded time $u=0$, a laboratory assistant launches a shell with stress energy of the form
\cref{shellstress}. This shell then continues to move radially outward with velocity $v$ forever. Then, $h^{(1)}_{AB}$ has no magnetic parity part at early retarded times, but it has a nonvanishing magnetic parity part at late times. By \cref{delh1}, this yields a nonvanishing magnetic parity memory tensor.

We note that Madler and Winicour \cite{Mad_Winic} have shown that under the stronger stationarity condition that they impose, magnetic parity memory cannot occur. This result is consistent with our results because, as we have already shown, their stronger stationarity condition rules out all ordinary memory, and null memory is always of scalar type. Bieri \cite{Lydia} has shown that magnetic parity memory cannot occur for vacuum solutions with ``small data'' in nonlinear general relativity. This result also is consistent with our results. 

Finally, we comment that examples with tensor ordinary memory can be obtained for $d > 4$ by choosing a shell stress-energy tensor\footnote{We can, of course, also construct sources with vector memory for $d>4$ in a similar manner to \cref{shellstress,magparitypartofstress}.} 
\begin{equation}
\label{shellstressd}
    T_{ab}=\frac{1}{r^{d-2}}[\rho u_{a}u_{b}+S_{ab}] \delta(r-vt)
\end{equation}
where $S_{ab}$ has vanishing $u$ and $r$ components and its angle-angle components are of purely tensor type. 

In summary, {\em null memory is always of scalar type in linear and nonlinear general relativity. Ordinary memory also is of scalar type for classical particle scattering in linearized gravity. However, ordinary memory need not be of scalar type in general. In particular, we have constructed explicit examples with outgoing shells of matter in linearized gravity that give rise to magnetic parity (= vector) ordinary memory in $4$ dimensions and tensor ordinary memory in higher even dimensions.}

\subsection{Memory as a Diffeomorphism}
\label{memdiff}

In this subsection, we consider the issue of whether the memory tensor up to Coulombic order can be written as an infinitesimal diffeomorphism, i.e., whether there exists a vector field $\xi ^a$ such that 
\begin{equation}
\label{memasdiffeo}
    \Delta^{(n)}_{\mu \nu}=[\nabla_{(\mu}\xi_{\nu)}]^{(n)}
\end{equation}
for all $n \leq d-3$. One reason why this question is of some interest can be seen from the following considerations.

We introduce the following new gauge: For $d>4$, start in the harmonic gauge in the early time stationary era $u<u_0$. For $d=4$, start in an arbitrary gauge compatible with our ansatz and stationarity assumption for $u<u_0$. Then, for $u<u_0$,  we have $h^{(n)}_{\mu \nu} = 0$ for all $n < d-3$ and $\partial_u h^{(d-3)}_{\mu \nu} = 0$. By a further gauge transformation of the form $\psi^a = u f(x^A)/r^{d-3} (\partial/\partial u)^a$, we may, in addition, set $h^{(d-3)}_{uu} = 0$ for $u<u_0$. Now, define coordinates for $u \geq u_0$ by fixing the $(r, x^A)$ coordinates along each geodesic determined by the initial tangent $\partial/\partial u$ and taking the $u$ coordinate to be given by the affine parameter along each geodesic. This agrees with proper time up to and including order $1/r^{d-3}$. Thus, the new coordinates are essentially Gaussian normal coordinates, except that the initial surface $u=u_0$ is not orthogonal to $(\partial/\partial u)^a$. By the same argument as for Gaussian normal coordinates, we have $\partial_u g_{u \mu} = 0$ (and, hence $\partial_u h_{u \mu} = 0$) at all times at Coulombic order and slower fall-off. Note that the new coordinates will {\em not}, in general, be harmonic in the radiative era or the final stationary era.

For $u \geq u_0$, the coordinate vector fields $\partial/\partial r$ and $\partial/\partial x^A$ are deviation vectors for the timelike geodesic congruence with tangent field $u^a = (\partial/\partial u)^a$. We have
\begin{eqnarray}
\frac{\partial^2 h_{\mu \nu}}{\partial u^2} &=& \frac{\partial^2 g_{\mu \nu}}{\partial u^2} = \frac{\partial^2}{\partial u^2}
\left[g_{ab} \left(\frac{\partial}{\partial x^\mu} \right)^a \left(\frac{\partial}{\partial x^\nu} \right)^b \right] \nonumber \\
&=& u^d \nabla_d \, u^c \nabla_c \left[g_{ab} \left(\frac{\partial}{\partial x^\mu} \right)^a \left(\frac{\partial}{\partial x^\nu} \right)^b \right] \nonumber \\
&=& g_{ab} \, u^d \nabla_d \, u^c \nabla_c \left[ \left(\frac{\partial}{\partial x^\mu} \right)^a \left(\frac{\partial}{\partial x^\nu} \right)^b \right].
\label{gcform}
\end{eqnarray}
This equation holds to all orders in $1/r$ in our coordinates. The derivatives of the term in brackets on the right side of eq. (\ref{gcform}) yield terms where $u^d \nabla_d u^c \nabla_c$ acts on a single coordinate vector field and terms where one derivative each acts on each of the two coordinate vector fields. The terms where two derivatives act on a single coordinate vector field can be evaluated from the geodesic deviation equation. The terms where one derivative acts on each of the coordinate vector fields are $O(1/r^{d-2})$. Thus, we obtain in our gauge 
\begin{equation}
\label{gcweyl}
\frac{\partial^2 h^{(n)}_{\mu \nu}}{\partial u^2} = - 2 C^{(n)}_{u \mu u \nu}
\end{equation}
for all $n \leq d-3$.
It follows immediately from the definition, \cref{memtensornthord}, of the memory tensor that in our gauge we have
\begin{equation}
\label{gcmem}
\Delta^{(n)}_{\mu \nu} =  \frac{1}{2}  \Delta h^{(n)}_{\mu \nu}
\end{equation}
for all $n \leq d-3$. Note that the right side of \cref{gcmem} is the full memory tensor, including null memory. This expression is compatible with our previous expression \cref{memsplitordnull} for $d > 4$ because that expression held in harmonic gauge whereas \cref{gcmem} is valid only in the gauge we have defined above. \Cref{gcmem} also is compatible with \cref{delh1} for $d=4$.

Now, suppose we start with an array of geodesic test particles that are initially ``at rest'' at early times and consider their final configuration at late times. If \cref{memasdiffeo} holds, then $\Delta h^{(n)}_{\mu \nu}$ is ``pure gauge'' for all $n \leq d-3$. This means that if we displace the test particles by $\xi^a$ at late times, they will go back to their original relative configuration at Coulombic and slower fall-off. In other words, at Coulombic order, the final spacetime {\em geometry} is the same as the initial geometry. On the other hand, if \cref{memasdiffeo} does not hold, then it is impossible to displace the particles so that they go back to their original relative configuration. A genuine change in the geometry at Coulombic order has occurred.

We now turn to the analysis of whether one can find a $\xi^a$ so that \cref{memasdiffeo} holds. It is clear that in order for $[\nabla_{(\mu}\xi_{\nu)}]^{(n)}$ to vanish for $n<d-3$ and be $u$-independent at $n = d-3$, we must choose $\xi^a$ to be such that
$\xi^{(n)}_\mu = 0$ for $n < d-4$ whereas
\begin{equation}
\label{xiform}
\xi^{(d-4)}_\mu = J_\mu (x^A) \, , \quad \quad \quad \xi^{(d-3)}_\mu = u B_\mu (x^A).
\end{equation}
Decomposing $J_\mu (x^A)$ and $B_\mu (x^A)$ into their scalar, vector, and tensor parts, we see that we have $6$ scalar functions on the sphere, $2$ divergence-free vector fields on the sphere, and no transverse, traceless tensors. On the other hand, the decomposition of a general symmetric tensor, $t_{\mu \nu}$, on the sphere yields $7$ scalar functions, $3$ divergence-free vector fields, and $1$ transverse, traceless tensor (for $d > 4$). Thus, {\em a priori}, we are one free scalar, one free vector, and one free tensor (for $d > 4$) short of being able to express a general tensor on the sphere in the form we seek.
\par However, $\Delta_{\mu \nu}$ is not a general tensor on the sphere. It has vanishing $u$-components, is trace-free, and its scalar and vector parts satisfy the constraint \cref{Mscalar1,Mscalar2,Mvector}. The symmetrized derivative of $\xi_{a}$ at order $1/r^{d-3}$ is 
\begin{equation}
\label{symdxiCoul}
   [\nabla_{(\mu}\xi_{\nu)}]^{(d-3)}=q_{(\mu}{}^{\sigma}\mathcal{D}_{\nu)}J_{\sigma}+(J_{r}-J_{u})q_{\mu\nu}+r_{(\mu}\mathcal{D}_{\nu)}J_{r}-K_{(\mu}\mathcal{D}_{\nu)}J_{u}-q_{(\mu}{}^{\sigma}r_{\nu)}J_{\sigma}-(d-4)r_{(\mu}J_{\nu)}-K_{(\mu}B_{\nu)}.
\end{equation}
It is clear from this equation that we may choose $B_{\nu}$ such that the $u$ components of \cref{symdxiCoul} vanish, so we need only consider whether $J_\mu$ can be chosen so as to make the non-$u$ components of the right side of \cref{symdxiCoul} match $\Delta_{\mu\nu}^{(d-3)}$. We may separately consider the scalar, vector, and tensor parts.
The scalar parts of $J_\mu$ are $J_r$, $J_u$, and $J$, where $J$ denotes the scalar part of $J_A$. Equating the scalar part of 
\cref{symdxiCoul} to the scalar part of $\Delta_{\mu\nu}^{(d-3)}$ (see eqs.~(\ref{srr})-)\ref{tendecom})), 
we obtain the following equations
\begin{equation}
\label{xitodelscalar1}
(d-4)J_{r}=U
\end{equation}
\begin{equation}
\label{xitodelscalar2}
    J_{r}-(d-3)J=2R
\end{equation}
\begin{equation}
\label{xitodelscalar3}
    J=T
\end{equation}
\begin{equation}
\label{xitodelscalar4}
    \mathcal{D}^{2}J+(d-2)(J_{r}-J_{u})=U.
\end{equation}
This is an overdetermined system for $J_r$, $J_u$, and $J$. The necessary and sufficient condition for a solution to exist is that $U$, $R$, and $T$ satisfy
\begin{equation}
\label{UTReq}
\frac{U}{(d-4)}-(d-3)T=2R.
\end{equation}
However, it can be shown that this equation is implied by the constraint equations \cref{Mscalar1,Mscalar2}. Thus, the scalar part of $\Delta_{\mu \nu}$ can always be written in the form \cref{memasdiffeo} for a $\xi^a$ of the form \cref{xiform}. Thus, scalar memory at Coulombic order is always given by a diffeomorphism \cite{Strom:2018}. In particular, as is well known, the scalar memory
\cref{scalmem} for $d=4$ is of the form of a supertranslation. However, a similar calculation shows that no such miracles occur for vector memory, and vector memory can {\em never} be written in the form \cref{memasdiffeo}. Tensor memory, of course, also can never be written in the form \cref{memasdiffeo}.
 
In summary, {\em scalar memory at Coulombic order always can be written as a diffeomorphism, but this never holds for vector and tensor memory.}

\subsection{Charges and Conservation Laws}
\label{chcons}
\subsubsection{Charges and Memory}
\label{chmem}

In $d=4$ dimensions, it is well known \cite{WZ} that all asymptotic symmetries at future null infinity give rise to associated charges and fluxes. In this sub-subsection, we will show that the charges and fluxes associated with supertranslations are intimately related to the memory effect in $4$ dimensions, and, indeed, we will derive the formula for scalar memory in $d=4$ from the supertranslation charges and fluxes. We will then obtain corresponding results for $d > 4$. Since the derivations and formulas of \cite{WZ} apply only to the vacuum case, in the following two paragraphs we will restrict to the case where $T_{ab} = 0$ in a neighborhood of null infinity. We will then restore $T_{ab}$ in our formulas.

Consider a supertranslation, i.e., a diffeomorphism belonging to the gauge equivalence class of
\begin{equation}
\psi^{a} = \alpha(x^A) \bigg(\frac{\partial}{\partial u}\bigg)^{a}-\alpha(x^{A})\bigg(\frac{\partial}{\partial r}\bigg)^{a}-q^{BC}\mathcal{D}_{B}\alpha(x^{A}) \frac{1}{r} \bigg(\frac{\partial}{\partial x^C}\bigg)^{a}+\dots 
\end{equation}
where the $\dots$ stand for a vector field that vanishes as $r\rightarrow \infty$ for fixed $u$ and $x^{A}$. From general considerations \cite{WZ} arising from the Lagrangian formulation of general relativity, a charge ${\mathcal Q}^+_\alpha$, and flux, ${\mathcal F}^+_\alpha$, can be associated with $\psi^a$ such that
for any $u_0, u_1$, we have
\begin{equation}
\label{chfl}
{\mathcal Q}^+_\alpha (u_1) - {\mathcal Q}^+_\alpha (u_0) = \int_{u_0}^{u_1} du \int d \Omega {\mathcal F}^+_{\alpha}.
\end{equation}
An explicit formula for ${\mathcal Q}^+_\alpha$ (originally due to Geroch \cite{Geroch}) is given in eq.(98) of \cite{WZ}, and an explicit formula for ${\mathcal F}^+_\alpha$ is given in eq.(82) of \cite{WZ}. 
Here we have inserted a superscript ``$+$'' to distinguish these charges and fluxes from similar quantities at past null infinity, which will be considered later. The flux is evaluated to be
\begin{equation}
{\mathcal F}^+_\alpha =-\frac{1}{32\pi} (\alpha N^{AB} N_{AB} - 2N^{AB} {\mathcal D}_A {\mathcal D}_B \alpha ).
\end{equation}
The formula for the charge is considerably more complicated, but this formula simplifies considerably in stationary eras, when $N_{AB} = 0$. From eq.(98) of \cite{WZ}, we find that in stationary eras we have
\begin{equation}
\label{Qstat}
{\mathcal Q}^+_\alpha \big\vert_{\rm stationary} =-\frac{1}{8\pi} \int d \Omega \alpha C^{(3)}_{urur}.
\end{equation}
Thus, if we impose the stationarity conditions of \Cref{statcond1} and we let $u_0 \to -\infty$ and $u_1 \to +\infty$ in \cref{chfl}, we obtain
\begin{equation}
\label{chfl2}
{\mathcal Q}^+_\alpha (+\infty) - {\mathcal Q}^+_\alpha (-\infty) = \int_{{\mathscr I}^+} {\mathcal F}^+_\alpha.
\end{equation}
The flux integral can be rewritten as
\begin{eqnarray}
\int_{{\mathscr I}^+} {\mathcal F}^+_\alpha &=& -\int_{{\mathscr I}^+} \alpha F +\frac{1}{16\pi}\int d\Omega {\mathcal D}^A {\mathcal D}^B \alpha \int^\infty_{-\infty} du N_{AB} \nonumber \\
&=&  -\int_{{\mathscr I}^+} \alpha F +\frac{1}{8\pi}\int d \Omega \big({\mathcal D}^A {\mathcal D}^B \alpha\big) \Delta^{(1)}_{AB} \nonumber \\
&=&  -\int_{{\mathscr I}^+} \alpha F + \frac{1}{8\pi}\int d \Omega \alpha {\mathcal D}^A {\mathcal D}^B \Delta^{(1)}_{AB}
\label{flhs5}
\end{eqnarray}
where $F = \frac{1}{32\pi}N^{AB} N_{AB}$ is the Bondi flux, and we used \cref{delh1} in the second line.
The contribution to $\int_{{\mathscr I}^+} {\mathcal F}^+_\alpha$ arising from the term $\alpha F$ is often referred to as the ``hard'' integrated flux (or ``hard charge'') whereas the term involving 
$\Delta^{(1)}_{AB}$ is called the ``soft'' integrated flux (or ``soft charge''). The terms ${\mathcal Q}^+_\alpha (-\infty)$ and ${\mathcal Q}^+_\alpha (+\infty)$ can be viewed as the contributions to ``hard charge'' coming from the asymptotic past (spatial infinity) and future (timelike infinity). From eqs. (\ref{Qstat}) - (\ref{flhs5}), we obtain,
\begin{equation}
\int d \Omega \alpha C^{(3)}_{urur} \big\vert_{+\infty} - \int d \Omega \alpha C^{(3)}_{urur} \big\vert_{-\infty} - 8\pi \int_{{\mathscr I}^+} \alpha F = -\int d \Omega \alpha {\mathcal D}^A {\mathcal D}^B \Delta^{(1)}_{AB}
\label{memsuper}
\end{equation}
which relates the hard charges to the soft charge. Note that if $\alpha$ is an $\ell=0$ or $\ell=1$ spherical harmonic (in which case $\psi^a$ is a translation), the term in $\Delta^{(1)}_{AB}$ does not contribute, and this equation corresponds to the integrated conservation law for Bondi $4$-momentum.

Since \cref{memsuper} holds for all $\alpha$, this equation must hold pointwise on the sphere. Therefore, we obtain
\begin{equation}
-{\mathcal D}^A {\mathcal D}^B \Delta^{(1)}_{AB} = C^{(3)}_{urur} \big\vert_{+\infty} - C^{(3)}_{urur} \big\vert_{-\infty} - 8\pi\int_{-\infty}^{\infty} du F. 
\end{equation}
It is easily seen that vector memory makes no contribution to ${\mathcal D}^A {\mathcal D}^B \Delta^{(1)}_{AB}$. On the other hand, substituting the form \cref{scalmem} of scalar memory, we obtain
\begin{equation}
-\frac{1}{2}{\mathcal D}^2  ({\mathcal D}^2 + 2) T= C^{(3)}_{urur} \big\vert_{+\infty} - C^{(3)}_{urur} \big\vert_{-\infty} - 8\pi\int_{-\infty}^{\infty} du F. 
\end{equation}
Solving for $T$ and substituting back in \cref{scalmem}, we obtain a formula for scalar memory that agrees with the scalar part of \cref{mem4d}.

In the above two paragraphs, we have restricted to the case where $T_{ab}=0$ in a neighborhood of null infinity in order to use the formulas given in \cite{WZ}. However, \cref{mem4d} holds when $T_{ab} \neq 0$. This shows that when $T_{ab} \neq 0$, \cref{memsuper} is modified merely by the simple substitution $F = \frac{1}{32\pi}N^{AB} N_{AB} \to \frac{1}{32\pi}N^{AB} N_{AB} + T^{(2)}_{uu}$. 

We now consider the case $d > 4$.
As we have seen in the previous subsection, in $d > 4$ dimensions, scalar memory is still given by a diffeomorphism. However, this diffeomorphism is now pure gauge, i.e., it has vanishing symplectic product with all asymptotically flat perturbations. Thus, nontrivial charges and fluxes cannot be associated with these diffeomorphisms via the Lagrangian formalism. Nevertheless, our general memory formula \cref{memsplitordnull} can be interpreted as a charge/flux formula. Namely, we may write this formula in the form
\begin{equation}
\label{memform}
P_{\mu \nu}[\bar{h}_{\rho \sigma}^{(d-3)}]\big\vert_{\infty} - P_{\mu \nu}[\bar{h}_{\rho \sigma}^{(d-3)}]\big\vert_{-\infty} +\int^\infty_{-\infty} du L_{\mu \nu}[F] =  \Delta^{(d-3)}_{\mu \nu}.
\end{equation}
Now for arbitrary scalar field $\alpha$ on the sphere, {\em define} the scalar charge, ${\mathcal Q}^+_\alpha$ during a stationary era, by\footnote{An important difference between $d > 4$ and $d=4$ is that the scalar charge for $d > 4$ is defined {\em only} during stationary eras, whereas in $d=4$ a local, gauge invariant scalar charge can be defined at all times (even though we gave the formula \cref{Qstat} for scalar charge in $d=4$ only during a stationary era). The existence of local, gauge invariant charge during radiative eras in $d=4$ traces back, by the considerations of \cite{WZ}, to its association with an asymptotic symmetry. Since there is no such association in $d > 4$, we see no reason to believe that a local, gauge invariant scalar charge corresponding to \cref{Qstatscalar} can be defined during radiative eras for $d > 4$.}
\begin{equation}
\label{Qstatscalar}
{\mathcal Q}^+_\alpha = \int d \Omega P_{AB}[\bar{h}_{\rho \sigma}^{(d-3)}]  \bigg(\mathcal{D}^{A}\mathcal{D}^{B}-\frac{1}{d-2}q^{AB}\mathcal{D}^{2} \bigg) \alpha.
\end{equation}
Using \cref{ordmemgaugeinv}, we can rewrite the right side of \cref{memform} in terms of $\Delta E_{rr}^{(d-1)}$. It then can be seen that \cref{Qstatscalar} corresponds to eq.(5.21) of \cite{Strom:2018}, but with different angular weights, i.e. our $\alpha$ is related to their $f$ by angular operators. Multiplying \cref{memform} by $(\mathcal{D}^{A}\mathcal{D}^{B}-\frac{1}{d-2}q^{AB}\mathcal{D}^{2}) \alpha$ and integrating over a sphere, we obtain
\begin{equation}
\label{memform2}
{\mathcal Q}^+_\alpha \big\vert_{\infty} - {\mathcal Q}^+_\alpha \vert_{-\infty} +\int_{\mathscr{I}^{+}} \alpha \bigg(\mathcal{D}^{A}\mathcal{D}^{B}-\frac{1}{d-2}q^{AB}\mathcal{D}^{2}\bigg) L_{AB}[F] = \int d \Omega \alpha \bigg(\mathcal{D}^{A}\mathcal{D}^{B}-\frac{1}{d-2}q^{AB}\mathcal{D}^{2}\bigg) \Delta^{(d-3)}_{AB}
\end{equation}
which is closely analogous to \cref{memsuper} and can be given an interpretation in terms of ``hard'' and ``soft'' charges.

Similarly, during stationary eras we can define the vector charge, ${\mathcal Q}^+_{\beta_A}$, associated with a divergence free vector field $\beta_A$ on the sphere by the formula
\begin{equation}
\label{Qstatvec}
{\mathcal Q}^+_{\beta_A} = \int d \Omega \beta^{B}\mathcal{D}^{A} P_{AB}[\bar{h}_{\rho \sigma}^{(d-3)}]. 
\end{equation}
We then obtain
\begin{equation}
\label{memform3}
{\mathcal Q}^+_{\beta_A} \big\vert_{\infty} - {\mathcal Q}^+_{\beta_A} \big\vert_{-\infty}  = \int d \Omega \beta^B \mathcal{D}^{A} \Delta^{(d-3)}_{AB}.
\end{equation}
No contribution from $F$ appears in this equation since $L_{AB}[F]$ cannot have a vector part. Finally, for any divergence-free, trace-free tensor field $\gamma_{AB}$ on the sphere, we can define the tensor charge ${\mathcal Q}^+_{\gamma_{AB}}$ during a stationary era by 
\begin{equation}
\label{Qstatten}
{\mathcal Q}^+_{\gamma_{AB}} = \int d \Omega \gamma^{AB} P_{AB}[\bar{h}_{\rho \sigma}^{(d-3)}] 
\end{equation}
and obtain
\begin{equation}
\label{memform4}
{\mathcal Q}^+_{\gamma_{AB}} \big\vert_{\infty} - {\mathcal Q}^+_{\gamma_{AB}} \big\vert_{-\infty} = \int d \Omega \gamma^{AB} \Delta^{(d-3)}_{AB}.
\end{equation}
Of course, there is no information contained in \cref{memform2}, \cref{memform3}, and \cref{memform4} than that which already appeared in \cref{memsplitordnull}.

\subsubsection{Conservation Laws}
\label{claws}

Thus far, the analysis of this paper has been concerned solely with the behavior of fields near future null infinity. Of course, the same analysis could be applied to past null infinity. In this sub-subsection, we wish to consider the relationship between quantities at past and future null infinity. Under the assumptions specified below, we will obtain a conservation law relating past and future null infinity.

Consider, first, the case of a scalar field $\phi$ in Minkowski spacetime with $d$ even and $d \geq 4$, with source $S=0$ in a neighborhood of future null infinity. We restrict attention to solutions, $\phi_{\ell m}$, whose angular dependence is given by a single spherical harmonic, $Y_{\ell m}$. (A general solution, of course, can be expressed as a superposition of such solutions.) Suppose that at Coulombic order, $\phi_{\ell m}$ is stationary at early retarded times, $\partial_u \phi_{\ell m}^{(d-3)} = 0$, so that at early times,
\begin{equation}
\label{phicoul}
\phi_{\ell m}^{(d-3)} = c Y_{\ell m} (x^A)
\end{equation}
where $c$ is a constant.
In the recursion relations \cref{scawavew/osourc(1/r^n+1)}, we may replace ${\mathcal D}^2$ by 
$-\ell(\ell+d-3)$, so we have
\begin{equation}
\label{scarecur}
(2n-d+2)\partial_{u}\phi_{\ell m}^{(n)}=  [\ell(\ell+d-3) - (n-1)(n-d+2)]\phi_{\ell m}^{(n-1)}.
\end{equation}
Thus, as usual, we obtain $\phi_{\ell m}^{(n)} = 0$ for $n < d-3$. For $d-3 \leq n < \ell +d -2$, we see that $\phi_{\ell m}^{(n)}$ is a polynomial, ${\mathcal P}_n(u)$, in $u$ of degree $n - d +3$, with the coefficients of the polynomials at the different orders related by \cref{scarecur}. For $n=\ell+d-2$, we obtain $\partial_u \phi_{\ell m}^{(\ell+d-2)} = 0$, so we may terminate the series by setting $\phi_{\ell m}^{(n)} = 0$ for $n  \geq \ell + d -2$. We thereby obtain an {\em exact solution} of the form 
\begin{equation}
\label{phisol}
\phi_{\ell m} = \sum_{n=\ell + d -3}^{d-3} \frac{{\mathcal P}_n(u)}{r^n} Y_{\ell m} (x^A).
\end{equation}
This solution is of direct physical interest, since it corresponds to the $Y_{\ell m}$ part of the retarded solution with source corresponding to matter in inertial motion (e.g., classical incoming particles on inertial timelike trajectories). The general solution with $Y_{\ell m}$ angular dependence that is stationary at Coulomb order is \cref{phisol} plus a solution with an asymptotic expansion whose slowest fall-off term is at order $1/r^{\ell+d-2}$, and with the coefficients of the higher powers of $1/r^n$ being polynomials in $u$ of degree $n-(\ell+d -2)$. This series cannot terminate.

We consider, now, the exact solution \cref{phisol}.
The highest power of $u$ in \cref{phisol} appears as the term $Cu^{\ell}/r^{\ell+d-3}$, where $C$ is related to the coefficient of the Coulombic order coefficient $c$ by an $\ell$ fold product of the numerical factors 
arising from successively solving \cref{scarecur}. Now, consider the behavior of  the solution \cref{phisol} near past null infinity. We can determine this behavior by
writing $u = v - 2r$ and re-expanding in $1/r$. It is immediately clear that the highest power of $v$ occurring in this solution will be the term $C^{\prime} v^{\ell}/r^{\ell + d - 3}$. The coefficient $C$ is related to the Coulombic order coefficient $C^{\prime}$ at past null infinity by a set of recursion relations. The recursion relations at past null infinity are the same as the recursion relations at future null infinity except for the following important difference: $\partial/\partial r$ is now past directed, which gives rise to a change in the sign of the $\partial/\partial u$ term in each of the recursion relations. Thus, we end up with $\ell$ sign flips by the time we reach Coulombic order. We thereby obtain $C' = (-1)^\ell C $, i.e., we have
\begin{equation}
\label{phicoulminus}
\phi_{\ell m}^{(d-3)} \big\vert_{{\mathscr I}^-} = (-1)^{\ell} C Y_{\ell m} (x^A).
\end{equation}
Since $(-1)^{\ell} Y_{\ell m} (x^A) = Y_{\ell m} (- x^A)$, this means that the solution \cref{phisol} at Coulombic order has an ``antipodal matching'' between ${\mathscr I}^+$ and ${\mathscr I}^-$ \cite{Strom:2017}. 

The antipodal matching \cref{phicoulminus} has been shown only for the exact solutions \cref{phisol} that terminate at order $1/r^{\ell + d -3}$. However,
since the additional terms in the asymptotic series of more general solutions behave no worse than $u^k/r^{k+\ell + d - 2}$ for $k \geq 0$, these individual terms would not contribute at Coulombic order at ${\mathscr I}^-$. Of course, the series composed of these terms is merely an asymptotic series near ${\mathscr I}^+$, and we clearly cannot determine the behavior of solutions near ${\mathscr I}^-$ from an asymptotic expansion near ${\mathscr I}^+$. Nevertheless, it seems not implausible that the antipodal matching may hold for a much more general class of solutions than the exact solutions \cref{phisol}. In any case, since the antipodal matching holds for \cref{phisol}
for all $\ell,m$ and the retarded solution corresponding to incoming inertial particles is a sum of such solutions, the antipodal matching holds for the retarded solution for incoming inertial particles---as can be verified directly from the explicit form of the solution \cite{Strom:2017}. 

Similar antipodal matching results hold for Maxwell's equations and for linearized gravity \cite{Strom1:2014,Strom2:2014,Strom:2017,Kartik_Maxwell}. The situation in nonlinear general relativity is less clear. Even for a solution that is stationary at Coulombic order, nonlinear terms will enter Einstein's equation at order $2(d-2)$. However, even in the linear case above, the behavior at ${\mathscr I}^-$ at Coulombic order depends on the form of the solution at order $n= \ell + d - 3$ near ${\mathscr I}^+$. Thus, for large $l$, the nonlinear terms in Einstein's equation cannot be ignored. Nevertheless, it remains not implausible that the antipodal matching may continue to hold in quite general circumstances.

In any case, we will now {\em assume} that we have a solution to Einstein's equation for which the antipodal matching holds at Coulombic order and consider the consequences. The key point is that the matching of the Coulombic order metrics implies a corresponding matching of the charges of the previous subsection, since the charges are constructed out of the Coulombic order metric. In particular, in $d=4$ dimensions, we have
\begin{equation}
{\mathcal Q}^+_\alpha \vert_{u=-\infty} = {\mathcal Q}^-_{\tilde{\alpha}} \vert_{v=+\infty}
\end{equation}
where ${\mathcal Q}^-_{\tilde{\alpha}}$ denotes the charge at ${\mathscr I}^-$ associated to the supertranslation $\tilde{\psi}^a$ with $\tilde{\alpha}$ antipodally matched to $\alpha$. Since, in analogy to \cref{chfl2}, we have
\begin{equation}
\label{chfl5}
{\mathcal Q}^-_{\tilde{\alpha}} \vert_{v=+\infty} - {\mathcal Q}^-_{\tilde{\alpha}} \vert_{v=-\infty} = \int_{{\mathscr I}^-} {\mathcal F}^-_{\tilde{\alpha}}
\end{equation}
where 
\begin{equation}
\label{fluxpast}
{\mathcal F}^-_{\tilde{\alpha}} =\frac{1}{32\pi} (\tilde{\alpha} N^{AB} N_{AB} + 2N^{AB} {\mathcal D}_A {\mathcal D}_B \tilde{\alpha} )
\end{equation}
we obtain the conservation law \cite{Strom1:2014,Strom2:2014,Strom:2017}
\begin{equation}
{\mathcal Q}^+_\alpha \vert_{u=+\infty} + \int_{{\mathscr I}^+} \alpha F -\frac{1}{8\pi} \int d \Omega \alpha {\mathcal D}^A {\mathcal D}^B \Delta^{(1)}_{AB} \vert_{{\mathscr I}^+}  = {\mathcal Q}^-_{\tilde{\alpha}} \vert_{v=-\infty} + \int_{{\mathscr I}^-} \tilde{\alpha} F + \frac{1}{8\pi}\int d \Omega \tilde{\alpha} {\mathcal D}^A {\mathcal D}^B \Delta^{(1)}_{AB}\vert_{{\mathscr I}^-}. 
\label{hscons}
\end{equation}
This may be interpreted as saying that the ingoing hard charge plus the integrated hard and soft fluxes at ${\mathscr I}^-$ are equal to the corresponding quantities at ${\mathscr I}^+$.

Similarly, in $d > 4$ dimensions, we get a similar antipodal matching of the scalar, vector, and tensor charges defined by \cref{Qstatscalar}, \cref{Qstatvec}, and \cref{Qstatten}, which leads to similar conservation laws.

\subsection{Memory and Infrared Divergences in Quantum Field Theory (``Soft Theorems'')}
\label{softthm}

In $d=4$ dimensions, there is a very close relationship between the memory effect and infrared divergences that occur in quantum field theory. This follows directly from the fact that, by \cref{delh1}, the memory tensor is just the change in $h^{(1)}_{AB}$ between late and early retarded times. Thus, if $\Delta^{(1)}_{AB} \neq 0$, then $h^{(1)}_{AB}(u, x^A)$ cannot vanish at future null infinity at both $u \to -\infty$ and $u \to + \infty$. It follows that the Fourier transform of $h^{(1)}_{AB}$ with respect to $u$ will diverge at small $\omega$ as $1/\omega$. As we shall now explain, this behavior gives rise to infrared divergences in quantum field theory. Exactly similar behavior occurs in the scalar and electromagnetic cases, but we will restrict our discussion here to the gravitational case.

Let $d \geq 4$, with $d$ allowed to be odd as well as even. The Lagrangian formulation of general relativity gives rise to a conserved symplectic current density $w^a$ constructed out of a background solution $g_{ab}$ and two perturbations $h_{ab}$ and $h'_{ab}$. Consider the symplectic flux $(\partial/\partial u)^a w_a = w_u$ near future null infinity. Only the leading order term $w^{(d-2)}_u$ can contribute to this flux. However, only the radiative order parts of $h_{ab}$ and $h'_{ab}$ can contribute to $w^{(d-2)}_u$, and the deviation of $g_{ab}$ from the flat metric $\eta_{ab}$ cannot contribute at all. We obtain
\begin{equation}
\label{sympflux}
w^{(d-2)}_u ({h'}_{AB},h_{CD}) = \frac{1}{32\pi}(C^{AB} {N'}_{AB} - {C'}^{AB} N_{AB})
\end{equation}
where $N_{AB}$ is the Bondi news tensor, \cref{news} and $C_{AB}$ is the trace free part of the projection of $h_{AB}^{(d/2-1)}$ onto the sphere. In writing \cref{sympflux}, we have imposed the gauge conditions $h_{rA}^{(1)} =h_{uu}^{(1)} = \eta^{ab} h_{ab}^{(1)} = 0$ in $d=4$ and we have imposed the harmonic gauge for $d > 4$.
The integrated symplectic flux can be used to define a symplectic form $\Omega ({h'}_{AB},h_{CD})$ at future null infinity
\begin{equation}
\Omega ({h'}_{AB},h_{CD}) = \int_{-\infty}^\infty du \int d \Omega \,  w^{(d-2)}_u ({h'}_{AB},h_{CD}). 
\label{Omega}
\end{equation}

\Cref{Omega} gives us the necessary structure to define a Fock space of ``outgoing graviton" states. We define the ``one-particle outgoing Hilbert space'' ${\mathcal H}_{\rm out}$ as the space of radiative order trace-free $\psi_{AB}$ that are purely positive frequency with respect to $u$, with inner product given by 
\begin{equation}
\langle {\psi'}_{AB} \vert \psi_{CD}\rangle  = -i  \Omega ({\psi}^{\prime*}{}_{AB}, \psi_{CD}) 
\label{KG}
\end{equation}
where ``$*$'' denotes complex conjugation.
More precisely, we define ${\mathcal H}_{\rm out}$ by starting with smooth positive frequency $\psi_{AB}$ with fast fall-off in $u$, defining the inner product \cref{KG} on such $\psi_{AB}$, and taking the Cauchy completion.
The inner product \cref{KG} is positive definite, as can be seen from the fact that in Fourier transform space, it is given by 
\begin{equation}
\langle {\psi'}_{AB} \vert \psi_{CD}\rangle = \frac{1}{16\pi}\int d \Omega \int_0^\infty  d \omega \omega \hat{\psi}^{\prime*}_{AB} \hat{\psi}^{AB} 
\label{KG3}
\end{equation}
where the ``hat'' denotes the Fourier transform. A classical solution $h_{\mu \nu}$ can be associated with a state in ${\mathcal H}_{\rm out}$ via $h_{\mu \nu} \to h^{(1)}_{+AB}$---where the subscript ``$+$'' denotes the positive frequency part---provided, of course, that $h^{(1)}_{+AB} \in {\mathcal H}_{\rm out}$.
Given ${\mathcal H}_{\rm out}$, one may then define the corresponding Fock space ${\mathcal F} ({\mathcal H}_{\rm out})$. A free field operator, ${\bf h}^{\rm out}_{\mu \nu}$, on ${\mathcal F} ({\mathcal H}_{\rm out})$ can then be defined in the usual manner in terms of annihilation and creation operators. Note that this construction is well defined even if the quantum gravity theory has not been defined in the interior spacetime \cite{Ashtekar:1981}.

However, this space, ${\mathcal F} ({\mathcal H}_{\rm out})$, of outgoing graviton states need not be adequate to describe all physically relevant outgoing states. This is most easily seen by considering 
the theory of {\em linearized} quantum gravity (i.e., a massless, spin-2 field) with a {\em classical} stress energy source, i.e., the stress-energy operator is taken to be $T_{ab} {\bf I}$ where $T_{ab}$ is a classical stress energy and $\bf I$ is the identity operator. This is a well defined, mathematically consistent theory that can be solved exactly. 
After analyzing this theory, we will discuss the implications for a full theory in which the stress-energy is fully quantum and the nonlinear effects of gravity are taken into account.

The Heisenberg equations of motion for the field operator ${\bf h}_{\mu \nu}$ for linearized gravity with a classical stress-energy source are easily solved to yield
\begin{equation}
\label{heisfd}
{\bf h}_{\mu \nu} = {\bf h}^{\rm in}_{\mu \nu} + h^{\rm ret}_{\mu \nu} {\bf I}
\end{equation}
where ${\bf h}^{\rm in}_{\mu \nu}$ is the free field operator corresponding to the ``in'' field and $h^{\rm ret}_{\mu \nu}$ is the classical retarded solution with classical source $T_{ab}$. Suppose we consider the state $| 0_{\rm in} \rangle$, corresponding to the vacuum state of ${\bf h}^{\rm in}_{\mu \nu}$. If we {\em assume} that this state corresponds to some state $\Psi \in {\mathcal F} ({\mathcal H}_{\rm out})$, then it follows from \cref{heisfd} that for any one particle state $\psi_{AB}$, we have
\begin{equation}
{\bf a}_{\rm out} (\psi_{AB}) \Psi = - \langle {\psi}_{AB} \vert h^{\rm ret}_{+AB} \rangle \Psi.
\end{equation}
The solution to this equation is the coherent state associated with $h^{\rm ret}_{+AB}$, namely
\begin{equation}
\Psi \propto \exp\left[-{\bf a}^\dagger_{\rm out} (h^{\rm ret}_{+AB})\right] \, \vert 0_{\rm out} \rangle
\label{cohstate}
\end{equation}

\Cref{cohstate} was derived under the assumption that $\Psi \in {\mathcal F} ({\mathcal H}_{\rm out})$. If $h^{\rm ret}_{+AB}$ has finite norm in the inner product \cref{KG}, then the right side of \cref{cohstate} defines a state in ${\mathcal F} ({\mathcal H}_{\rm out})$, and this state corresponds to $| 0_{\rm in} \rangle$. However, if $h^{\rm ret}_{+AB}$ does not have finite norm in the inner product \cref{KG}, then the right side of \cref{cohstate} does not define a state in ${\mathcal F} ({\mathcal H}_{\rm out})$. It follows that $| 0_{\rm in} \rangle$ cannot correspond to a state in 
${\mathcal F} ({\mathcal H}_{\rm out})$. This should not be a cause of any distress. The Heisenberg state $| 0_{\rm in} \rangle$ is well defined everywhere as a state on the algebra of local field observables. It is similarly well defined on the algebra of asymptotic field observables near future null infinity. All of its correlation functions are well defined at future null infinity. If we wish to represent this state as a vector in a Hilbert space, $\tilde{\mathcal F}_{\rm out}$, carrying a representation of the ``out'' field observables,
we may always do so via the GNS construction. However, if $h^{\rm ret}_{+AB}$ does not have finite Klein-Gordon norm, the representation of the field observables on $\tilde{\mathcal F}_{\rm out}$ cannot be unitarily equivalent to its representation on ${\mathcal F} ({\mathcal H}_{\rm out})$ (see \cite{Ashtekar:1981}, Section V.A of \cite{Asht:2018}).

Now let $d=4$ and consider the case where the classical source $T_{ab}$ is such that the corresponding retarded solution $h^{\rm ret}_{ab}$ has a nonvanishing memory tensor $\Delta^{(1)}_{AB} \neq 0$. Then, as already noted in the first paragraph of this subsection, the Fourier transform of of $h^{{\rm ret} \, (1)}_{AB}$ with respect to $u$ will diverge at small $\omega$ as $1/\omega$.
But by \cref{KG3}, we then have  
\begin{equation}
\left\lVert \hat{h}^{{\rm ret} \, (1)}_{+AB} \right\rVert^2   =\frac{1}{16\pi} \int d \Omega \int_0^\infty  d \omega \omega \vert \hat{h}^{{\rm ret} \, (1)}_{+AB}  \vert^2 = \infty
\label{KG2}
\end{equation}
on account of the ``infrared divergence'' as $\omega \to 0$. Thus, the ``out'' state corresponding to $| 0_{\rm in} \rangle$---or, for than matter, any other state in ${\mathcal F} ({\mathcal H}_{\rm in})$---does not live in ${\mathcal F} ({\mathcal H}_{\rm out})$, and one would have to work with a different representation to represent this state as a vector in a Hilbert space.
Exactly analogous results hold in the scalar and electromagnetic cases for $d=4$.

We have just shown that in linearized gravity with a classical source for which a nontrivial memory effect is present in the classical retarded solution---as would occur generically in classical particle scattering---the ``out'' state $\Psi$ is not a state in ${\mathcal F} ({\mathcal H}_{\rm out})$. However, the infrared divergence described in the previous paragraph is sufficiently innocuous that one can, in effect, proceed as though one were dealing with a state in ${\mathcal F} ({\mathcal H}_{\rm out})$. To see this, consider, first, the case where no infrared divergences occur and $h^{\rm ret}_{+AB}$ has finite Klein-Gordon norm,
so \cref{cohstate} defines a state in ${\mathcal F} ({\mathcal H}_{\rm out})$. Choose a frequency $\omega_0 > 0$ and decompose ${\mathcal H}_{\rm out}$ into the direct sum of its ``hard'' and ``soft'' graviton spaces
\begin{equation}
{\mathcal H}_{\rm out} = {\mathcal H}^H_{\rm out} \oplus {\mathcal H}^S_{\rm out} 
\end{equation}
where ${\mathcal H}^H_{\rm out}$ is spanned by trace-free $\psi_{AB}$ composed of frequencies $\omega \geq \omega_0$ and ${\mathcal H}^S_{\rm out}$ is spanned by trace-free $\psi_{AB}$ composed of frequencies $\omega_0 > \omega \geq 0$. The Fock space ${\mathcal F} ({\mathcal H}_{\rm out})$ then factorizes as
\begin{equation}
{\mathcal F} ({\mathcal H}_{\rm out}) = {\mathcal F} ({\mathcal H}^H_{\rm out}) \otimes {\mathcal F} ({\mathcal H}^S_{\rm out}).
\end{equation}
Now decompose $h^{\rm ret}_{+AB}$ into its ``hard'' and ``soft'' parts,
\begin{equation}
h^{\rm ret}_{+AB} = [h^{\rm ret}_{+AB}]^H + [h^{\rm ret}_{+AB}]^S.
\end{equation}
The creation operator ${\bf a}^\dagger_{\rm out} (h^{\rm ret}_{+AB})$ appearing in \cref{cohstate} and be written as the sum of creation operators for the hard and soft parts of $h^{\rm ret}_{+AB}$. Since these operators commute, $\Psi$ factorizes as 
\begin{equation}
\Psi = \Psi^H \otimes \Psi^S
\label{psifac}
\end{equation}
where $\Psi^H \in {\mathcal F} ({\mathcal H}^H_{\rm out})$ is the coherent state associated with 
$[h^{\rm ret}_{+AB}]^H$ and $\Psi^S \in {\mathcal F} ({\mathcal H}^S_{\rm out})$ is the coherent state associated with 
$[h^{\rm ret}_{+AB}]^S$. The factorization in \cref{psifac} implies that if we are interested solely in the ``hard part'' of the outgoing state, we may effectively put in an ``infrared cutoff'' at $\omega = \omega_0$ and work with the state $\Psi^H$ in the Fock space ${\mathcal F} ({\mathcal H}^H_{\rm out})$. In particular, the probability that $\Psi^H \in {\mathcal F} ({\mathcal H}^H_{\rm out})$ contains a specified number of ``hard gravitons'' in specified modes is the same as the sum of the probabilities that $\Psi \in {\mathcal F} ({\mathcal H}_{\rm out})$ contains these ``hard gravitons'' and any number of ``soft gravitons.'' This is the essential content of the ``soft theorems'' \cite{Weinberg:1965}. In perturbation theory, the fact that inclusion of the effects of ``soft gravitons'' does not affect the calculation of ``hard graviton'' probabilities manifests itself in a cancelation of the contributions of ``real soft gravitons'' and ``virtual soft gravitons.'' 

The above discussion assumed that $h^{{\rm ret} \, (1)}_{AB}$ does not have infrared divergences, in which case there is no need to decompose the ``out'' state into ``hard'' and ``soft'' parts. Now consider the case where a memory effect is present and $h^{{\rm ret} \, (1)}_{AB}$ does have an infrared divergence. Then, as discussed above, $\Psi \notin {\mathcal F} ({\mathcal H}_{\rm out})$. Nevertheless, we may still write
\begin{equation}
\Psi = \Psi^H \otimes \tilde{\Psi}^S
\label{psifac2}
\end{equation}
where $\Psi^H \in {\mathcal F} ({\mathcal H}^H_{\rm out})$ is the coherent state associated with 
$[h^{\rm ret}_{+AB}]^H$ and $\tilde{\Psi}^S \in \tilde{\mathcal F}^S_{\rm out}$ is the ``soft graviton'' state written as a vector in the Hilbert space $\tilde{\mathcal F}^S_{\rm out}$ in the representation to which it belongs. Although the ``soft graviton'' content of $\Psi$ near future null infinity is ill defined (since $\Psi^S \notin {\mathcal F} ({\mathcal H}^S_{\rm out})$), the ``hard graviton'' content of $\Psi$ is well defined (since $\Psi^H \in {\mathcal F} ({\mathcal H}^H_{\rm out})$). Thus, if we are interested only in the ``hard particle'' content of $\Psi$ near future null infinity, we may, in effect, put in an infrared cutoff and treat $\Psi$ as an ordinary Fock space state $\Psi^H \in {\mathcal F} ({\mathcal H}^H_{\rm out})$ \cite{Block&Nordsieck:1937}.

All of the above discussion beginning with \cref{heisfd} holds for the rather trivial theory of linearized gravity with a classical stress-energy source. It is quite a leap to go from this theory to the case of quantum, interacting matter and quantum, nonlinear general relativity, especially since a quantum theory of nonlinear general relativity is not in hand. Nevertheless, let us consider a scattering situation where, by assumption, we have non-interacting ingoing ``hard'' particles at early times and non-interacting outgoing ``hard'' particles late times. Consider the ``soft'' content of the outgoing state, associated with $\omega < \omega_0$, where $\omega_0$ is much less than any inverse length or time scale associated with the interaction. Then, it seems plausible that the dominant contributions to this ``soft'' content will come from asymptotically early and late times, where the ``hard'' particles are non-interacting and effectively can be treated classically. If so, then a factorization similar to \cref{psifac2} should occur, but with the following important difference: If we fix an ``in'' state consisting of ``hard'' particles in momentum eigenstates, then the hard content of the ``out'' state should have a nonvanishing amplitude for ``hard'' particles in many different momentum eigenstates. (Of course, total energy-momentum is conserved.) But this means that there also should be nonvanishing amplitudes for different memory tensors. Presumably, one must take the ``out'' Hilbert space for the soft sector to be 
\begin{equation}
\tilde{\mathcal F}^S_{\rm out} = \bigoplus_{\Delta} \, {\mathcal F}^S_{\Delta \, {\rm out}}
\end{equation}
where ${\mathcal F}^S_{\Delta \, {\rm out}}$ describes Hilbert space of soft ``out'' states with memory tensor $\Delta^{(1)}_{AB}$ and the direct sum is taken over the (uncountably infinite) collection of all $\Delta^{(1)}_{AB}$. Instead of \cref{psifac2}, the ``out'' state should be of the form
\begin{equation}
\Psi = \sum_\Delta \Psi_\Delta^H \otimes \Psi_\Delta^S
\end{equation}
where $\Psi_\Delta^H \in {\mathcal F} ({\mathcal H}^H_{\rm out})$ (i.e., the ``hard'' factor lies in the usual Fock space of for all $\Delta^{(1)}_{AB})$, but $\Psi_\Delta^S \in {\mathcal F}^S_{\Delta \, {\rm out}}$ (i.e., the ``soft'' factor lies in different Hilbert spaces for different $\Delta^{(1)}_{AB}$). Thus, the soft gravitons should produce a complete decoherence \cite{Semenoff:2018,Semenoff:2017} of the ``hard'' final states with different memory, although they will not affect the probability of producing any specified ``hard'' final state with particles in momentum eigenstates. Further investigation of this issue is well beyond the scope of this paper.

Finally, we note that essentially all of our discussion above also applies to $d>4$. For $d$ even, the memory tensor is first nonvanishing only at Coulombic order. However, since $C^{(d/2 -1)}_{u \mu u \nu}$ can be expressed as inverse angular operators acting on $\partial^{d/2 -2}C^{(d -3)}_{u \mu u \nu}/\partial u^{d/2 - 2}$, it can be seen that the Fourier transform of $h^{(d/2 -1)}_{AB}(u, x^A)$ behaves as $\omega^{d/2 - 3}$ as $\omega \to 0$. Thus, there are no infrared divergences for $d > 4$. This result holds in odd dimensions as well. Thus, although one can still factorize states into ``hard'' and ``soft'' parts, there is no necessity to do so in order to describe the ``out'' state as a Fock space state.\footnote{However, there may be other considerations that indicate the utility of factorization of the out state into ``hard'' and ``soft'' parts (see \cite{Strom:2017Infradiv,Gomez}).}

\bigskip

\noindent
{\bf Acknowledgements} This research was supported in part by NSF grants PHY 15-05124 and PHY18-04216 to the University of Chicago.

\newpage

\appendix 
\section{Relationship of our Ansatz to Smoothness at $\mathscr{I}^{+}$ in $d=4$}
\label{smoothscri}

As noted in the body of the paper, it is easily seen that in $d=4$, smoothness of $A_a$ 
at ${\mathscr I}^+$ implies that 
our ansatz (\ref{Adevenan}) holds, and smoothness of $\Omega^2 h_{ab}$ at ${\mathscr I}^+$ implies that our ansatz (\ref{hdevenan}) holds. However, for $d=4$ the ansatz (\ref{Adevenan}) implies smoothness of $A_a$ at ${\mathscr I}^+$ only under the additional condition that $A_{r}^{(1)}=0$, and the ansatz (\ref{hdevenan}) implies smoothness of $\Omega^2 h_{ab}$ at ${\mathscr I}^+$ only under the additional condition that $h_{rr}^{(1)}=0$. In this Appendix, we investigate the conditions under which these additional restrictions can be imposed as gauge conditions. We show that this is possible in electromagnetism when $j_{r}^{(3)}=0$ and in linearized gravity when $T_{ur}^{(3)} =T_{rr}^{(3)} = T_{rA}^{(3)} = 0$. However, when these quantities are nonvanishing, there are solutions within our ansatz that are not smooth at ${\mathscr I}^+$. Nevertheless, in nonlinear gravity, we show that $h_{rr}^{(1)}=0$ if the Bondi news is nonvanishing everywhere on one cross-section, in which case our ansatz in $d=4$ is equivalent to smoothness at ${\mathscr I}^+$.

\subsection{Electromagnetism}
\label{appelec}
The $\ell \neq 0$ part of $A_{r}^{(1)}$ is gauge invariant within our ansatz, so if it is nonvanishing, it cannot be set to zero by a gauge transformation. By \cref{psid/2-1}, 
we have $\psi^{(1)} = 0$. \Cref{psiatorder1/r^n} with $n=1$ then yields $\partial_u A_{r}^{(1)} =0$, so $A_{r}^{(1)}$ is independent of $u$.
The $r$-component of Maxwell's equations given by \cref{Maxwellseqsr-comparb} in four dimensions with $n=3$ gives that 
\begin{equation}
\label{Arjr3}
    \mathcal{D}^{2}A_{r}^{(1)}=-4\pi j_{r}^{(3)}.
\end{equation}
This equation implies that the $\ell = 0$ part of $j_{r}^{(3)}$ must vanish. It also implies that $\partial_{u}j_{r}^{(3)} = 0$, as also can be proven directly from current conservation and $\psi^{(1)} = 0$. However, if the $\ell \neq 0$ part of $j_{r}^{(3)}$ is nonvanishing, we will obtain solutions within our ansatz such that $A_{r}^{(1)} \neq 0$. Such solutions are not smooth at ${\mathscr I}^+$ in any gauge.

Conversely, if $j_{r}^{(3)} = 0$, then the $\ell \neq 0$ part of $A_{r}^{(1)}$ vanishes by \cref{Arjr3}. The $\ell = 0$ part of 
$A_{r}^{(1)}$ can then be set to zero within our ansatz by a gauge transformation of the form $\phi = c\ln(r)$. Thus, if $j_{r}^{(3)} = 0$, all solutions within our ansatz are smooth at ${\mathscr I}^+$ in some gauge.

\subsection{Linearized Gravity}
\label{applingrav}

The $\ell >1$ part of $h_{rr}^{(1)}$ is gauge invariant within our ansatz, so if it is nonvanishing, it cannot be set to zero by a gauge transformation.
From $\chi_r^{(1)} = 0$ and \cref{chieqrcomp}, we obtain $\partial_u h_{rr}^{(1)} =0$.
The $ur$ and $rr$ components of the linearized Einstein's equation given by \cref{linEinsteqsur-comparb,linEinsteqsrr-comparb} with $n=2$ yield, respectively, 
\begin{equation}
\label{ur3}
    \mathcal{D}^{2}\bar{h}_{ur}^{(1)}-\mathcal{D}^{A}\chi_{A}^{(2)}=-16\pi T_{ur}^{(3)}
\end{equation}
\begin{equation}
\label{rr3}
    [\mathcal{D}^{2}-2]\bar{h}_{rr}^{(1)}+2\bar{h}_{ur}^{(1)}-2\mathcal{D}^{A}\bar{h}_{Ar}^{(1)}+2\chi_{r}^{(2)}=-16\pi T_{rr}^{(3)}.
\end{equation}
The angular divergence of the $rA$ component, \cref{linEinsteqsrA-comparb}, yields
\begin{equation}
\label{rA3}
    \mathcal{D}^{2}\mathcal{D}^{A}\bar{h}_{rA}^{(1)}-2\mathcal{D}^{2}\bar{h}_{ur}^{(1)}+2\mathcal{D}^{2}\bar{h}_{rr}^{(1)}-\mathcal{D}^{2}\chi_{r}^{(2)}+\mathcal{D}^{A}\chi_{A}^{(2)}=-16\pi \mathcal{D}^{A}T_{rA}^{(3)}.
\end{equation}
Applying $\mathcal{D}^{2}$ to \cref{rr3} and taking a linear combination of the above equations, we obtain
\begin{equation}
\label{hrrTr}
    \mathcal{D}^{2}[\mathcal{D}^{2}+2]h_{rr}^{(1)}=-16\pi (\mathcal{D}^{2}T_{rr}^{(3)}+2\mathcal{D}^{A}T_{rA}^{(3)}+2T_{ur}^{(3)})
\end{equation}
where we used the fact that $\bar{h}_{rr}=h_{rr}$. The $\ell = 0$ and $\ell = 1$ parts of the right side must therefore vanish, and the right side must be stationary. Indeed, using conservation of stress energy and the dominant energy condition it can be shown that $T_{ur}^{(3)},T_{rr}^{(3)}$ and $T_{rA}^{(3)}$ are stationary. However, the $\ell > 1$ part of the right side can be nonvanishing, and, if it is, we obtain a solution within our ansatz such that $h_{rr}^{(1)} \neq 0$. Such solutions are not smooth at ${\mathscr I}^+$ in any gauge.

Conversely, if $T_{ur}^{(3)} = T_{rr}^{(3)} = T_{rA}^{(3)} = 0$, then \cref{hrrTr} implies that $h_{rr}^{(1)}$ is a linear combination of an $l=0$ and an $l=1$ spherical harmonic. Let
\begin{equation}
\label{KVX}
X^{a} = c  \bigg(\frac{\partial}{\partial u}\bigg)^{a} + f(x^A) \bigg(\frac{\partial}{\partial u}\bigg)^{a}-f(x^A)\bigg(\frac{\partial}{\partial r}\bigg)^{a}-q^{BC}\mathcal{D}_{B}f(x^A) \frac{1}{r} \bigg(\frac{\partial}{\partial x^C}\bigg)^{a} 
\end{equation}
where $c$ is a constant and $f(x^{A})$ is a linear combination of $\ell =1$ spherical harmonics, so that $X^a$ is a translational Killing field of the background Minkowksi spacetime. By a gauge transformation of the form 
\begin{equation}
    \xi^{a}=X^{a}\ln(r)
\end{equation}
we can set the $\ell = 0,1$ parts of $h_{rr}^{(1)}$ to zero within our ansatz. Thus, we can set $h_{rr}^{(1)} = 0$ and the solution is smooth at ${\mathscr I}^+$.

\subsection{Nonlinear Gravity}
\label{appnonlin}

Again, we obtain $\partial_u h_{rr}^{(1)} =0$. But there now is a new, nontrivial equation containing $h_{rr}^{(1)} =0$. 
The $AB$-components of the Einstein equation given by \cref{linEinsteqsAB-comparb} with $n=1$ where the right hand side of \cref{linEinsteqsAB-comparb} now picks up an additional nonlinear contribution $\mathcal{G}_{AB}^{(2)}$ given by \cref{nonlineinstd-2} with $d=4$. We obtain
\begin{equation}
\label{ABnonlin}
    -q_{AB}\partial_{u}\chi_{r}^{(2)}=2\partial_{u}(h_{rr}^{(1)}N_{AB}).
\end{equation}
The right side is traceless whereas the left side is pure trace, so the only way this equation can hold is if both sides vanish. Thus, using $\partial_u h_{rr}^{(1)} =0$, we obtain
\begin{equation}
\label{hrrnews}
    h_{rr}^{(1)}  \partial_{u} N_{AB} = 0.
\end{equation}
This equation has no analog in the linearized theory. Since $N_{AB} \to 0$ as $u \to \pm \infty$, it implies that if the Bondi news is nonvanishing at angle $x^A$ at any $u$, then $h_{rr}^{(1)}(x^A) = 0$ at all $u$ (since $h_{rr}^{(1)}$ is independent of $u$). Thus, in particular, if the Bondi news is nonvanishing everywhere on one cross-section of ${\mathscr I}^+$, then $h_{rr}^{(1)}=0$, and our ansatz in $d=4$ is equivalent to smoothness at ${\mathscr I}^+$.

 \section{Applying the Lorenz Gauge with a Slower Fall-Off Ansatz for $d > 4$}

In our ansatz eqs.~(\ref{Adevenan})-(\ref{Adoddan}) for $A_a$ and our ansatz eqs.~(\ref{hdevenan})-(\ref{hdoddan}) for $h_{ab}$, the slowest fall-off term was assumed to be at radiative order, $n = d/2 -1$. However, in even dimensions with $d > 4$, the conditions of smoothness of $A_a$ and $\Omega^2 h_{ab} = r^{-2} h_{ab}$ at ${\mathscr I}^+$ would, {\em a priori}, allow terms with slower fall-off than permitted by our ansatz. This suggests a danger that our ansatz might exclude some solutions of physical interest. In this Appendix, we show that this is not the case by weakening our ansatz to permit slower fall-off, allowing the integer powers in even dimensions to start at order $1/r$ and allowing the half-integer powers in odd dimensions to start at order $1/\sqrt{r}$ for all $d>4$. We will show that the Lorenz gauge can still be imposed within the context of this weaker ansatz. Since the Cartesian components of $A_a$ and $h_{ab}$ satisfy the scalar wave equation in Lorenz gauge, it follows from \Cref{termin} that the only additional solutions allowed by our weaker fall-off ansatz vanish in Lorenz gauge. Thus, the only new solutions allowed by the weaker ansatz are pure gauge. This justifies our stronger choice of ansatz eqs.~(\ref{Adevenan})-(\ref{Adoddan}) and eqs.~(\ref{hdevenan})-(\ref{hdoddan})

\subsection{Electromagnetism}
\label{appelectlorgauge}
We take our slower fall-off ansatz for the vector potential $A_{a}$ for $d > 4$ to be
\begin{eqnarray}
A_a &\sim& \sum_{n=1}^\infty \frac{1}{r^{n}} A_a^{(n)}(u, x^A) \, \quad \quad \quad \quad \quad \quad \quad \quad \quad \quad \quad d \,\, {\rm even} \label{Adevenanslow} \\
A_a &\sim& \sum_{n=1/2}^\infty \frac{1}{r^{n}} A_a^{(n)}(u, x^A) + \sum_{p=d-3}^\infty \frac{1}{r^{p}} \tilde{A}_a^{(p)}(u, x^A) \quad \quad d \,\, {\rm odd.} \label{Adoddanslow} 
\end{eqnarray}
As discussed in \Cref{subsecMaxeqa}, in order to impose the Lorenz gauge we must solve the scalar wave equation (\ref{waveeqphi}) for a gauge scalar field $\phi$ with source $\psi$.

Consider, first, the case of $d$ even. We seek to solve \cref{waveeqphi} with the ansatz
\begin{equation}
\label{phiansslow}
    \phi \sim c \ln r + \sum_{n=0}^{\infty}\frac{1}{r^{n}}\phi^{(n)}(u,x^{A})
\end{equation}
where $c$ is a constant, and we require $\partial_{u}\phi^{(0)}=0$ in order that $\partial_a \phi = O(1/r)$. The recursion relations for $\phi^{(n)}$ are given by \cref{scawavew/sourc(1/r^n+1)} with $\psi$ replacing $S$. Although $S = O(1/r^{d-2})$, {\em a priori} we have $\psi = O(1/r)$. However, an analysis similar to the proof of \Cref{lorenz} shows that $\psi^{(1)}$ vanishes and
\begin{equation}
\label{uphi2}
    \partial_{u}\psi^{(2)}=-(d-4)\partial_{u}A_{u}^{(1)}
\end{equation}

To solve \cref{scawavew/sourc(1/r^n+1)}, we start with the radiative order recursion relation ($n = d/2 -1$ in 
\cref{scawavew/sourc(1/r^n+1)}), which yields
\begin{equation}
\label{phid/2minus2}
\left[\mathcal{D}^{2}-(d/2-2)(d/2-1)\right]\phi^{(d/2-2)}=\psi^{(d/2)}.
\end{equation}
This angular operator is invertible, so we may uniquely solve for $\phi^{(d/2-2)}$. There is no difficulty in solving the recursion relations at faster fall-off, since we may then specify $\phi^{(d/2-1)}$ arbitrarily and solve for $\phi^{(n)}$ with $n > d/2 -1$ as in \Cref{first}. To obtain $\phi^{(n)}$ with $n < d/2 -2$ we proceed iteratively by inverting the angular operators in the slower fall-off recursion relations. This works without any difficulty until we get to \cref{scawavew/sourc(1/r^n+1)} with $n=1$. 
\begin{equation}
\label{solvingforphi(0)}
c + \mathcal{D}^{2}\phi^{(0)}=\psi^{(2)}+(d-4)\partial_{u}\phi^{(1)} 
\end{equation}
If the right side of this equation were not stationary, $\phi^{(0)}$ could not be stationary and the desired gauge transformation would not exist. However, we now shall show that the right side of \cref{solvingforphi(0)} is indeed stationary. 

To show this, let 
\begin{equation}
   \gamma \equiv A_{u}-\partial_{u}\phi.
\end{equation}
By Maxwell's equations, when $j_a = 0$, we have 
\begin{equation}
    \Box A_{u}=\partial_{u}\psi.
 \label{umax}   
\end{equation}
Thus, if $\phi$ satisfies \cref{waveeqphi} and if $j_a = 0$, then $\gamma$ satisfies $\Box \gamma = 0$. Of course, $j_a$ need not be zero and we have not yet obtained a solution, $\phi$, to \cref{waveeqphi}. However, we have $j_a^{(n)} = 0$ for all $n < d-2$, and we have constructed above a solution to the recursion relations \cref{scawavew/sourc(1/r^n+1)} to solve for $\phi^{(n)}$ for all $n > 0$. Therefore, we obtain quantities $\gamma^{(n)}$ that satisfy the homogeneous recursion relations \cref{scawavew/osourc(1/r^n+1)} for all $1 < n < d-2$. In parallel with the argument of the previous paragraph, at radiative order, $n = d/2 -1$, these relations imply that $\gamma^{(d/2 -2)} = 0$. It then follows that $\gamma^{(n)} = 0$ for all $1 \leq n \leq d/2-2$. For $n=1$, we obtain
\begin{equation}
\partial_{u}\phi^{(1)} = A_{u}^{(1)}.
\end{equation}
But the Maxwell equation \cref{umax} yields
\begin{equation}
(d-4) \partial_u A_{u}^{(1)} = - \partial_{u}\psi^{(2)}.
\end{equation}
Thus the right side of \cref{solvingforphi(0)} is indeed stationary, as we desired to show.
In parallel with solving \cref{phi0} when $d=4$, we can choose $c$ so as to cancel the $l=0$ part of the right side. We may then 
invert \cref{solvingforphi(0)} to obtain $\phi^{(0)}$. Thus, in even dimensions, the Lorenz gauge can be imposed within the weakened ansatz (\ref{Adevenanslow}).

We now turn to the odd dimensional case. We take the scalar field $\phi$ to have the following expansion in powers of $1/r$ 
\begin{equation}
    \phi \sim \sum_{n= -1/2}^{\infty}\frac{1}{r^{n}}\phi^{(n)}(u,x^{A}) + \sum_{p=d-3}^{\infty}\frac{1}{r^{p}}\tilde{\phi}^{(p)}(u,x^{A}).
\end{equation}
Note that we allow a term, $\phi^{(-1/2)}$, that {\em grows} with $r$ as $r^{1/2}$. In order that $\partial_a \phi$ be consistent with our ansatz (\ref{Adoddanslow}), it is necessary and sufficient that $\partial_u \phi^{(-1/2)} = 0$.

There is no difficulty in solving the recursion relations for $\tilde{\phi}^{(p)}$. There also is no difficulty in solving the recursion relations for $\phi^{(n)}$ for $n \geq 1/2$ in the manner specified in \Cref{second}. However, there is a potential difficulty that arises when one attempts to solve the recursion relation for $\phi^{(-1/2)}$
\begin{equation}
\label{phi-1/2}
    [\mathcal{D}^{2}+\frac{1}{4}(2d-5)]\phi^{(-1/2)}= (d-3)\partial_{u}\phi^{(1/2)} + \psi^{(3/2)}.
\end{equation}
This equation can be uniquely solved for $\phi^{(-1/2)}$, but $\phi^{(-1/2)}$ will be stationary as required
if and only if the right side be stationary. However, the stationarity of the right side can be proven in the same manner as done above for the even dimensional case. Thus, in odd dimensions, the Lorenz gauge can be imposed within the weakened ansatz (\ref{Adoddanslow}).

\subsection{Linearized Gravity}
\label{applingravlorgauge}
We take the slower fall-off ansatz for the metric perturbation $h_{ab}$ to be
\begin{eqnarray}
h_{ab} &\sim& \sum_{n=1}^\infty \frac{1}{r^{n}} h_{ab}^{(n)}(u, x^A) \, \quad \quad \quad \quad \quad \quad \quad \quad \quad \quad \quad \quad d \,\, {\rm even} \label{hdevenanslow} \\
h_{ab} &\sim& \sum_{n=1/2}^\infty \frac{1}{r^{n}} h_{ab}^{(n)}(u, x^A) + \sum_{p=d-3}^\infty \frac{1}{r^{p}} \tilde{h}_{ab}^{(p)}(u, x^A) \quad \quad \, d \,\, {\rm odd.} \label{hdoddanslow} 
\end{eqnarray}
We seek a gauge vector field, $\xi_a$, satisfying \cref{xigaugeeq}.
We take our ansatz for $\xi_a$ to be 
\begin{eqnarray}
\xi_a &\sim& c(\partial/\partial u)_a \ln r + \sum_{n=0}^\infty \frac{1}{r^{n}} \xi_a^{(n)}(u, x^A) \, \quad \quad \quad \quad \quad \quad d \,\, {\rm even} \label{xidevenanslow} \\
\xi_a &\sim& \sum_{n=-1/2}^\infty \frac{1}{r^{n}} \xi_a^{(n)}(u, x^A) + \sum_{p=d-3}^\infty \frac{1}{r^{p}} \tilde{\xi}_a^{(p)}(u, x^A) \quad \quad \quad d \,\, {\rm odd} \label{xidoddanslow} 
\end{eqnarray}
where, in even dimensions, $\partial_{u}\xi_{a}^{(0)}=0$, and, in odd dimensions, $\partial_{u}\xi_{a}^{(-1/2)}=0$.

In even dimensions, we can solve the recursion relations in parallel with the electromagnetic case. The only potential difficulty arises showing that $\partial_u \xi_{a}^{(0)} = 0$.
This requires showing that in the recursion relation for $\xi_{u}^{(0)}$
\begin{equation}
\label{xiu0}
 c +  \mathcal{D}^{2}\xi_{u}^{(0)} = -2\chi_{u}^{(2)}+ (d-4)\partial_{u}\xi_{u}^{(1)},
\end{equation}
the right side must be stationary. However, stationarity can be proven in close parallel with the electromagnetic case by defining
\begin{equation}
   \Gamma \equiv -\bar{h}_{uu} + \frac{1}{d-2}\bar{h}-\partial_{u}\xi_u
\end{equation}
and showing $\Gamma^{(n)} = 0$ for all $1\leq n\leq d/2-2$, from which it can then can be shown that the right side of \cref{xiu0} is stationary. We then can solve \cref{xiu0} to obtain a stationary $\xi_{u}^{(0)}$. The equations for 
$\xi_{r}^{(0)}$ and $\xi_{A}^{(0)}$ can then be solved, and these quantities are stationary. Thus, in even dimensions, the Lorenz gauge can be imposed within the weakened ansatz (\ref{hdevenanslow}). 

The odd dimensional case mirrors the analysis of the electromagnetic case in odd dimensions, with the substitution of the argument of the previous paragraph to prove stationarity of $\xi_{a}^{(-1/2)} = 0$

\bibliographystyle{JHEP}

\end{document}